\UseRawInputEncoding
\documentclass[twocolumn,           
               showpacs,            
               nopreprintnumbers,     
               aps,                 
               prd,          	    
               letterpaper,             
              groupeaddress,      
               nofootinbib,         
               tightenlines,        
               floats,floatfix,      
               showkeys
               ]{revtex4-1}
               
\usepackage[toc,page]{appendix}
\usepackage{graphicx}
\usepackage{dcolumn}
\usepackage{bm}
\usepackage{amsmath}
\usepackage{amsfonts,amssymb}
\usepackage{soul}
\usepackage{color}
\definecolor{v}{rgb}{0.6, 0.2, 0.8} 
\definecolor{MAGA}{rgb}{0.1, 0.43, 0.75}
\definecolor{jm}{rgb}{0.13, 0.48, 0.64}

\usepackage{xcolor}
\usepackage{enumerate}
\usepackage{float}
\usepackage{subfigure}
\usepackage{multirow,tabularx, booktabs}

\usepackage{orcidlink}

\usepackage{silence}
\begin{document}

\title{A comparative study of emergent dark energy models}

\newcommand{\orcidauthorA}{0000-0003-0405-9344} 
\newcommand{\orcidauthorB}{0000-0002-6356-8870} 
\newcommand{\orcidauthorD}{0009-0004-4170-5856} 
\newcommand{\orcidauthorE}{0000-0003-4446-7465} 

\author{Andr\'es Garc\'ia-Rivera$^1$\orcidlink{\orcidauthorD}}
\email{a2578238@correo.uia.mx}

\author{A.  Hern\'andez-Almada$^2$\orcidlink{\orcidauthorA}}
\email{ahalmada@uaq.mx}

\author{Miguel A. Garc\'ia-Aspeitia$^1$\orcidlink{\orcidauthorB}}
\email{angel.garcia@ibero.mx, corresponding author}

\author{V. Motta$^3$\orcidlink{\orcidauthorE}}
\email{veronica.motta@uv.cl}

\affiliation{$^1$ Depto. de F\'isica y Matem\'aticas, Universidad Iberoamericana Ciudad de M\'exico, Prolongaci\'on Paseo \\ de la Reforma 880, M\'exico D. F. 01219, M\'exico}

\affiliation{$^2$ Facultad de Ingenier\'ia, Universidad Aut\'onoma de
Quer\'etaro, Centro Universitario Cerro de las Campanas, 76010, Santiago de 
Quer\'etaro, M\'exico}

\affiliation{$^3$Instituto de F\'isica y Astronom\'ia, Universidad de Valpara\'iso, Avda. Gran Breta\~na 1111, Valpara\'iso, 2360102, Chile}

\begin{abstract}
Late (emergent) dark energy models that involve a hyperbolic tangent in their dynamic equations have achieved success due to their simplicity, the possibility of resolving the Hubble tension, and, in some cases their consistency with the Dark Energy Spectroscopic Instrument (DESI) results. Among the most studied are the Phenomenological and General Emergent Dark Energy Models (PEDE and GEDE, respectively), the variable curvature model, and the graduated dark energy model. Our goal is to include the curvature term in the models and perform a robust comparative study using diverse data samples, in particular, Cosmic Chronometers, Type Ia supernovae, Baryon Acoustic Oscillations from DESI-DR1, and Cosmic Microwave Background Radiation. This study presents the Universe age, the deceleration, the $w_{eff}$ parameter reconstruction, the redshift transition, and the best fit for the free parameters of each model.  Our collected evidence from Akaike Information Criterion and Bayesian Information Criterion indicates that the best candidate is the PEDE model. Nevertheless, the variable-curvature model is the only model that exhibits a decelerated phase near $z=0$, consistent with the results from DESI.
\end{abstract}
\pacs{Dark energy, emergent dark energy, cosmology.}
\maketitle

\section{Introduction}

One of the biggest challenges to cosmology today is understanding how dark energy behaves. This understanding of the evolution of our Universe, thanks to dark energy, has been a subject of constant study through different models that propose different kinds of mathematical tools \cite{Di_Valentino_2021, Motta:2021hvl,CosmoVerseNetwork:2025alb}. Today, the most used model for understanding the evolution of the Universe is the well known $\Lambda$-Cold Dark Matter model ($\Lambda$CDM). Nevertheless, with current observations we are discovering several cracks in the foundations of the model. The first and oldest problem is associated with the value of the energy density of the cosmological constant, where the value obtained through a quantum field theory differs $\sim120$ order of magnitude with the observed value \cite{Weinberg,Zeldovich}.  Another problem is the current tension when we measure the Hubble constant $H_0$ from local redshift observations and high redshift observation \cite{Di_Valentino_2021,CosmoVerseNetwork:2025alb}, right now the tension is around $\sim5\sigma$ presenting an interesting challenge for the standard paradigm. Additionally, the Dark Energy Spectroscopic Instrument (DESI) \cite{DESI:2024hhd,DESI:2025zgx} is observing a discrepancy with the cosmological constant, indicating a probably dynamic behavior, specifically an evolution through phantom, cosmological constant, and finally quintessence dark energy \cite{DESI:2025zgx}, resulting in a reduction of the Universe's acceleration in its final stages.

In this vein, the community is searching for new theories beyond $\Lambda$CDM, and specifically, emergent models for dark energy have had unprecedented success. For example, the Phenomenological Emergent Dark Energy (PEDE) \cite{PEDE:2019ApJ}, Generalized Emergent Dark Energy (GEDE) \cite{Li:2020ybr}, Variable Curvature (VC) \cite{Esteban-Gutierrez:2024rpz}, and Graduated Dark Energy Model (GDE) \cite{Akarsu:2019hmw} are among the most studied. They have a common characteristic, which is a function that in its term acts like dark energy; in this case, the four models have a hyperbolic tangent ($\tanh x$) that allows the transition between a non-accelerated and accelerated Universe. In PEDE and GEDE models, it is possible to infer that the term causing the acceleration is a scalar field; meanwhile, in the VC model, the acceleration is induced by a curvature change, and in GDE, it is produced by a transition of the sign in the $\Lambda$ term. PEDE, GEDE, and GDE are constrained using Cosmic Chronometers and Supernovae of Type Ia, Baryon Acoustic Oscillations, and Cosmic Microwave Background Radiation; meanwhile, for the VC model, we only use low-redshift observations\footnote{The reason for using only low-redshift observations is that the VC model has convergence issues.}, in this case Cosmic Chronometers and Supernovae of Type Ia, with the aim of exploring the current dynamics of the Universe.
Additionally, to ensure a clear comparison, the four models are compared with the $\Lambda$CDM model that includes the curvature term  ($\Omega_k$-$\Lambda$CDM). Similarly, all the models contain a curvature term in their dynamical equations.

The outline of the paper is as follows: Section \ref{sec:cosmology} is devoted to study the mathematical background associated with variable curvature, phenomenological emergent dark energy, general emergent dark energy and dynamical lambda models. Sec. \ref{sec:constraints} presents the datasets used to constrain the free parameters of each model. Sec. \ref{section:Results} is dedicated to present our results and the comparison among models. Finally in Sec. \ref{section:CD} we give our conclusions and discussion. We henceforth use units in which $\hbar=c=k_{B}=1$.

\section{The cosmological models} \label{sec:cosmology}
To characterize the Universe, we need to develop various mathematical models that align with the observations; thus, in this paper, we will explore four distinct models. These models may appear similar mathematically, since they use hyperbolic functions, but each one has a distinct physical interpretation. Certain parameters are shared among the different models; these parameters are: the matter density, $\Omega_{0m}=\Omega_{0b}+\Omega_{0dm}$ that contains baryons and dark matter parameters, the curvature density, $\Omega_k$, and the dark energy density, $\Omega_{0DE}$. We henceforth the subindex $0$ in $\Omega_{0i}$ represents the density parameter at $z=0$.
The radiation term is neglected for the late Universe. The purpose of comparing these models is to understand the effect that their differences have on predicting observations, and to understand which model is closer to reality and why.

To obtain all the cosmography for the four models, we present the dimensionless Friedmann equation 

\begin{equation}
    E(z)\equiv \frac{H(z)}{H_0}=\frac{\dot{a}}{H_0a},
\end{equation}
where $a$ is the scale factor, $z$ the redshift and $H_0$ the Hubble constant. The deceleration parameter is given by the equation
\begin{equation}
    q(z)=-\frac{\ddot{a}}{aH^2}=\frac{(z+1)}{2E(z)^2}\frac{dE(z)^2}{dz}-1,
\end{equation}
and the effective equation of state (EoS), given by

\begin{equation}
    w(z)_{eff}=\frac{1}{3}[2q(z)-1],
\end{equation}
with $w(z)_{eff}=P_{eff}(z)/\rho_{eff}(z)$, where $P_{eff}(z)$ and $\rho_{eff}(z)$ are the pressure and energy density,  respectively.

\subsection{Curvature-$\Lambda$-Cold Dark Matter} \label{subsec:LCDM}

The standard $\Lambda$CDM model, with the addition of a curvature term, is the simplest model constructed under the hypotheses of homogeneity and isotropy. Its dynamical equation takes the form

\begin{eqnarray}
    E(z)^2=\Omega_{0m}(z+1)^3+\Omega_k(z+1)^2+(1-\Omega_{0m}-\Omega_k), 
\end{eqnarray}
where $\Omega_{0i}$ are the density parameters of matter (dark matter and baryons) and curvature, respectively. 

Just for consistency with the other models, the $\Omega_k$ retains the negative sign. The same applies to PEDE, GEDE, and GDE, except for VC, where the sign is not absorbed by the density parameter, in order to avoid confusions and maintain concordance with \cite{Esteban-Gutierrez:2024rpz}. Thus, models can be studied and compared without problem, only having in mind the existence of a minus sign in VC model. 

\subsection{Variable Curvature Model} \label{subsec:cv}

The Variable Curvature Model, studied in \cite{Esteban-Gutierrez:2024rpz}, follows the Friedmann-Lemaitre-Robertson-Walker (FLRW) metric almost perfectly, with the subtlety that the curvature is redshift-dependent. The curvature can be defined as $\kappa(z) = H_0^2 \mathcal{H}(z)$, where $\mathcal{H}(z)$ is the curvature parameter (see Eq. \ref{eq:Hzcurvature}). This model aims to retain all the benefits of the FLRW metric while adding a term that introduces sufficient variability to observe different behavior. This model can be understood by the following dimensionless Friedmann equation, which is expressed as \cite{Esteban-Gutierrez:2024rpz} 

\begin{equation*}
    E(z)^2\approx \Big\{1 + \frac{\gamma(z+1)}{24 \mathcal{H}(z)} (\Delta\Omega_k){\rm sech}^2\Big[\frac{\gamma(z-\alpha)}{2}\Big] \Big\}^{-1}
\end{equation*}
\begin{equation}
    \times [\Omega_{0m}(z+1)^3-\mathcal{H}(z)(z+1)^2],
\end{equation}
where $\gamma$, $\alpha$ are free parameters of the model.

In particular, $\mathcal{H}(z)$ is a curvature parameter defined as 

\begin{equation}
   \mathcal{H}(z) = \frac{1}{2}\Big\{ (\Omega^1_k+\Omega^2_k)  + (\Omega^1_k-\Omega^2_k)\tanh\Big[\frac{\gamma(z-\alpha)}{2}\Big]\Big\},
   \label{eq:Hzcurvature}
\end{equation}
being $\Omega^1_k$, and $\Omega^2_k$ the past and present curvatures, respectively. Since we are using the FLRW metric, it is important that, when we consider the same conditions of constant curvature, we obtain the same dimensionless Friedmann equation. Considering a variable curvature allows us to have a more complex behavior of the evolution of the Universe while keeping similar values for the parameters that are shared with the FLRW metric.

\subsection{Phenomenological Emergent Dark Energy Model} \label{subsec:PEDE}

The Phenomenological Emergent Dark Energy Model (PEDE) studied in \cite{PEDE:2019ApJ,HERNANDEZALMADA2024101668}  follows the FLRW metric, but in contrast with the Variable Curvature model, without modifying the metric. Instead, it has a variable dark energy density parameter, described as \cite{PEDE:2019ApJ,HERNANDEZALMADA2024101668}

\begin{eqnarray}
    \Omega_{DE}(z)=\Omega_{0DE}[1-\tanh(\log_{10}(z+1))].
\end{eqnarray}
The dimensionless Friedmann equation follows 

\begin{eqnarray}
    &&E(z)^2 = \Omega_{0m}(z+1)^3+\Omega_k(z+1)^2+\Omega_{DE}(z),
\end{eqnarray}
under Friedmann constriction $\Omega_{0DE}=1-\Omega_{0m}-\Omega_k$. This model is characterized by having the same free parameters as $\Lambda$CDM cosmology, which allows it to produce similar values.

\subsection{General Emergent Dark Energy Model}
The General Emergent Dark Energy Model (GEDE) studied in \cite{Li:2020ybr,Hernandez-Almada:2020uyr} further generalizes the PEDE model. It introduces more variability using more free parameters, making the behavior it predicts even more complex. Following  \cite{Li:2020ybr,Hernandez-Almada:2020uyr},

\begin{eqnarray}
    &&E(z)^2 = \tilde\Omega_{DE}(z) + \Omega_{0m}(z+1)^3+\Omega_k(z+1)^2, 
\end{eqnarray}
where
\begin{equation}
    \tilde\Omega_{DE}(z)= \Omega_{0DE} \frac{1-\tanh\Big(\Delta \log_{10}\frac{1+z}{1+z_t}\Big)}{1+\tanh(\Delta\log_{10}(1+z_t))},
\end{equation}
and $\Omega_{0DE} = 1-\Omega_{0m}-\Omega_k$ through the Friedmann constraint.  Additionally, $\Delta$ is a free parameter which allows us to recover the $\Lambda$CDM model if $\Delta=0$, or the PEDE model if $\Delta=1$. Moreover, $z_t$ is the transition redshift, $\tilde\Omega(z_t)=\Omega_{0m}(1+z_t)^3$ between $z\sim3$ and $z\sim0$ (see Fig. \ref{fig:ode_comparison}).

\subsection{Graduated Dark Energy Model} \label{subsec:ELambda}
The graduated Dark Energy Model (GDE) studied in \cite{Akarsu:2019hmw} is another emergent dark energy model. The base of GDE is the hypothesis that the inertial mass density exhibits the form 
\begin{equation}
    \rho_{inertial}=\epsilon\rho_0\left(\frac{\rho}{\rho_0}\right)^{\lambda},
\end{equation}
where $\epsilon$ and $\lambda$ are appropriate constants, implying the relation $\omega=-1+\epsilon=const$. This idea allows the construction of a step function written in the form ${\rm sgn}(x)\vert x\vert^y$ that models the DE dynamics. 
Thus, the dimensionless Friedmann equation in this context can be written as \cite{Akarsu:2019hmw}

\begin{eqnarray}
    &&E(z)^2=\Omega_{0m}(z+1)^3+\Omega_k(z+1)^2+\nonumber\\&&\Omega_{0DE} {\rm sgn}[1+\Psi \ln (z+1)] \Big|1+\Psi \ln(z+1)\Big|^{\frac{1}{1-\lambda}},
\end{eqnarray}
where $\rm{sgn}$ is the signum function that can take the values $\rm{sgn}=-1,0,1$ for $x<0$, $x=0$ and $x>0$, respectively and $\Psi$ can be written as $\Psi \equiv -3\epsilon(\lambda-1)<0$ where it follows that $\lambda < 1$ and $\epsilon < 0$, according to \cite{Akarsu:2019hmw}. 
Notice that the Friedmann constraint allows the expression $\Omega_{0DE}=1-\Omega_{0m}-\Omega_k$, in order to reduce the free parameters of the theory.

\section{Datasets and constraints} \label{sec:constraints}

The parameter space of the cosmological models $\boldsymbol{\Theta}$ is as follows:
\begin{itemize}
    \item Variable curvature model  $\boldsymbol{\Theta_{\rm VC}}: (h, \Omega_{0m}, \Omega_k^1, \Omega_k^2, \alpha, \gamma)$,
    \item PEDE model $\boldsymbol{\Theta_{\rm PEDE}}: (h, \Omega_{0m}, \Omega_k)$,
    \item GEDE model $\boldsymbol{\Theta_{\rm GEDE}}: (h, \Omega_{0m}, \Omega_k, \Delta)$,
    \item GDE model $\boldsymbol{\Theta_{\rm GDE}}: (h, \Omega_{0m}, \Omega_k, \Psi, \lambda)$,
\end{itemize}
where $h$ is the dimensionless Hubble constant and the rest of the parameters are characteristic parameters of each cosmological model. These spaces are constrained using cosmic chronometers, type Ia Supernovae, Baryon Acoustic Oscillations, and Cosmic Microwave Background Radiation data, in combination, except for VC, which is constrained only by cosmic chronometers and type Ia Supernovae due to convergence problems. A Bayesian analysis is applied using Monte Carlo Markov Chain (MCMC) tools implemented in the \texttt{Emcee} module \cite{Foreman:2013} under the Python 3 environment. The convergence of the chains is monitored through the autocorrelation function to sample the mentioned phase spaces $\boldsymbol{\Theta}$ generating 4000 chains with 300 steps according to the priors presented in Table \ref{tab:model_priors}. We adopt uniform distributions for all the parameters involved in PEDE, GEDE and GDE. Meanwhile for VC we adopt Gaussian priors for $h=0.7304 \pm 0.0104$ \cite{Riess_2022} and for $\Omega_{0m}=0.3111\pm 0.0056$ \cite{Planck:2018} and Uniform distributions for the rest of the parameters.

\begin{table*}[ht]
\centering
\renewcommand{\arraystretch}{1.2}
\begin{tabular}{lccccc}
\hline
Parameter (Prior) & $\Lambda$CDM & VC & PEDE & GEDE & GDE \\
\hline
$h \sim {\rm Gauss}(0.7304, 0.0104)$ & -- & $\checkmark$ & -- & -- & -- \\
$h \sim U(0.6, 0.76)$ & $\checkmark$ & -- & $\checkmark$ & $\checkmark$ & $\checkmark$ \\
$\Omega_{0m} \sim {\rm Gauss}(0.3111, 0.0056)$ & -- & $\checkmark$ & -- & -- & -- \\
$\Omega_{0m} \sim U(0, 1)$ & $\checkmark$ & -- & $\checkmark$ & $\checkmark$ & $\checkmark$ \\
$\Omega_k \sim U(-0.5,0.5)$ & $\checkmark$ & -- & $\checkmark$ & $\checkmark$ & $\checkmark$ \\
$\Omega_k^1 \sim U(-0.1,0.1)$ & -- & $\checkmark$ & -- & -- & -- \\
$\Omega_k^2 \sim U(-0.1,0.1)$ & -- & $\checkmark$ & -- & -- & -- \\
$\alpha \sim U(-0.5,1)$ & -- & $\checkmark$ & -- & -- & -- \\
$\gamma \sim U(1,20)$ & -- & $\checkmark$ & -- & -- & -- \\
$\Delta \sim U(-0.1,10)$ & -- & -- & -- & $\checkmark$ & -- \\
$\Psi \sim U(-2,0.1)$ & -- & -- & -- & -- & $\checkmark$ \\
$\lambda \sim U(-0.5,0)$ & -- & -- & -- & -- & $\checkmark$ \\
\hline
Number of parameters & 3 & 6 & 3 & 4 & 5 \\
\hline
\end{tabular}
\caption{Priors adopted for the parameters of each cosmological model. Notice that the chosen region for the curvature parameters is due to the fact that recent studies \cite{DiValentino:2019qzk} suggest that our Universe is not necessarily flat.}
\label{tab:model_priors}
\end{table*}

In the following, we summarize the datasets.

\begin{itemize}
\item \textit{The cosmic chronometer} (CC) data consist of 33 measurements of the Hubble parameter spanning the redshift range $0.07<z<1.965$. The CC sample includes 18 uncorrelated measurements and a subset of 15 correlated data points $H(z)$ \cite{M_Moresco_2012, Moresco_2015, Moresco_2016, Moresco_2020, Jiao_2023, Tomasetti_2023}. 

For uncorrelated measurements, the contribution to the likelihood is computed using the standard chi-square function
\begin{equation}
\chi^2_{\rm CC,uncorr} = \sum_{i=1}^{18}
\frac{\left[H_{\rm obs}(z_i)-H_{\rm th}(z_i,\boldsymbol{\Theta})\right]^2}
{\sigma_i^2},
\end{equation}
where $H_{\rm obs}(z_i)$ and $H_{\rm th}(z_i,\boldsymbol{\Theta})$ denote the observed and theoretical values of the Hubble parameter at redshift $z_i$, respectively, $\sigma_i$ is the corresponding observational uncertainty.

For the correlated subset of 15 measurements, the covariance matrix formalism is adopted. In this case, the chi-square function is written as
\begin{equation}
\chi^2_{\rm CC,corr} =
\Delta \mathbf{H}^{\,T}
\mathbf{C}^{-1}
\Delta \mathbf{H},
\end{equation}
where
\begin{equation}
\Delta \mathbf{H} =
H_{\rm obs}(z_i)-H_{\rm th}(z_i,\boldsymbol{\Theta}),
\end{equation}
is the residual vector and $\mathbf{C}$ is the covariance matrix associated with the correlated CC data.

The total chi-square contribution from the CC compilation is, therefore, given by
\begin{equation}
\chi^2_{\rm CC} = \chi^2_{\rm CC,uncorr} + \chi^2_{\rm CC,corr}\,.
\end{equation}

\item \textit{Type Ia Supernovae} (SNIa) observations are taken from the Pantheon+ compilation \cite{Scolnic2018-qf, Brout_2022}, which consists of 1701 correlated measurements of the distance modulus spanning the redshift range $0.001<z<2.26$. Since the Pantheon+ sample includes correlated statistical and systematic uncertainties, likelihood analysis is performed using the full covariance matrix formalism. The theoretical distance modulus is defined as
\begin{equation}
\mu_{\rm th}(z,\boldsymbol{\Theta})
=
5\log_{10}\left[\frac{d_L(z,\boldsymbol{\Theta})}{\rm Mpc}\right]+25+\mathcal{M},
\end{equation}
where $\mathcal{M}$ is the nuisance parameter associated with the calibration of the absolute magnitude. The luminosity distance for a spatially flat Universe is given by
\begin{equation}
d_L(z,\boldsymbol{\Theta})
=
(1+z)\,c\int_0^z \frac{d\tilde{z}}{H(\tilde{z},\boldsymbol{\Theta})},
\end{equation}
where $c$ is the speed of light and $H(z,\boldsymbol{\Theta})$ is the Hubble expansion rate predicted by the cosmological model.

To avoid explicit dependence on $\mathcal{M}$, we follow the analytical marginalization procedure described in \cite{Conley2010}. In this approach, the marginalized chi-square function is given by
\begin{equation}
\chi^2_{\rm SNIa}
=
a-\frac{b^2}{e}+\ln\left(\frac{e}{2\pi}\right),
\end{equation}
where
\begin{equation}
a=\mathbf{X}^{T}\mathbf{C}^{-1}\mathbf{X},
\end{equation}
\begin{equation}
b=\mathbf{X}^{T}\mathbf{C}^{-1}\mathbf{1},
\end{equation}
\begin{equation}
e=\mathbf{1}^{T}\mathbf{C}^{-1}\mathbf{1}.
\end{equation}
Here, $\mathbf{C}$ denotes the full covariance matrix of the Pantheon+ dataset, $\mathbf{1}$ is a unit vector, and the residual vector is defined as
\begin{equation}
\mathbf{X}
=
\mu_{\rm obs}(z_i)-\mu_{\rm th}(z_i,\boldsymbol{\Theta};\mathcal{M}=0),
\end{equation}
with $\mu_{\rm obs}$ and $\mu_{\rm th}$ representing the observed and theoretical distance moduli, respectively.

\item Baryon Acoustic Oscillations (BAO). The BAO signal provides measurements of the dilation scale ratio $D_V(z)/r_d$, where $D_V(z)$ is the volume-averaged distance at redshift $z$ and $r_d \equiv r_s(z_d)$ is the sound horizon at the drag epoch, the comoving angular diameter distance ratio $D_M(z)/r_d$, the Hubble distance ratio $D_H(z)/r_d \equiv (c/H(z))/r_d$, where $r_d$ is given by
    \begin{equation}
        r_d = \int_{z_d}^\infty \frac{c_s(z)dz}{H(z)}\,, \label{eq:rd}
    \end{equation}  
    where $c_s(z)$ is the sound speed, and we use $z_d=1089.80\pm0.21$ \cite{Planck:2018}. The dilation scale is defined as \cite{Wigglez:Eisenstein2005}
    \begin{equation}
        D_V(z) = \sqrt[3]{z\,D_H(z)\,D_M^2(z)}\,.
    \end{equation}
    For a flat geometry, we have
    \begin{equation}
        D_M(z) = c\int_0^z\frac{dz'}{H(z')}\,.
    \end{equation}
    Thus, we consider a $\chi^2$-function as
    \begin{equation}
        \chi^2_{\rm BAO} = \sum_i^{12} \left( \frac{O_{th}(z_i)-O_{obs}^i}{\sigma^i}\right)^2,
    \end{equation}
    where $O_{obs}^i \pm \sigma^i$ is the measurement and its corresponding uncertainty at redshift $z^i$, and $O_{th}^i$ is its theoretical counterpart. In this work, 
    we analyze 12 BAO measurements from the first year of the Dark Energy Spectroscopic Instrument (DESI-DR1) data, spanning $0.1 < z < 4.16$ \cite{desicollaboration2024desi}.

\item Cosmic Microwave Background (CMB). For CMB we can employ the compressed distance posterior derived from the acoustic peaks of the CMB, namely the acoustic scale ($\mathbf{l_A}$), the shift parameter ($R$), and the decoupling redshift ($z_\ast$) to constrain cosmological models.
The acoustic scale is defined as
\begin{equation}
l_A = \frac{\pi r(z_\ast)}{r_d(z_\ast)} ,
\end{equation}
where $r_d$ is the sound horizon defined in Eq. \ref{eq:rd} at the redshift of decoupling $z_\ast$ given
by \cite{Hu_1996},
\begin{eqnarray}
&&z_\ast = 1048 \left[1 + 0.00124(\Omega_{b0}h^2)^{-0.738}\right]\times\nonumber\\&&\left[1 + g_1(\Omega_{m0}h^2)^{g_2}\right] ,
\end{eqnarray}
where
\begin{equation}
g_1 = \frac{0.0783(\Omega_b h^2)^{-0.238}}{1 + 39.5(\Omega_{b0}h^2)^{0.763}} , 
\quad 
g_2 = \frac{0.560}{1 + 21.1(\Omega_{b0}h^2)^{1.81}} .
\end{equation}
The shift parameter is defined as \cite{SHIFT_PARAMETER}
\begin{equation}
R = \frac{\sqrt{\Omega_{m0}} \, H_0}{c}\, r(z_\ast) ,
\end{equation}
where $\Omega_{m0}=\Omega_{dm0}+\Omega_{b0}.$ Thus, the $\mathbf{\chi^2}$ for the CMB data is given as
\begin{equation}
\chi^2_{CMB} = X^T \, {Cov}^{-1}_{CMB} \, X ,
\end{equation}
where $\mathbf{{Cov}^{-1}_{CMB}}$ is the inverse covariance matrix of the distance posteriors and
\begin{equation}
\mathbf{X} =
\begin{pmatrix}
l_A^{\rm th} - l_A^{\rm obs} \\
R^{\rm th} - R^{\rm obs} \\
z_\ast^{\rm th} - z_\ast^{\rm obs}
\end{pmatrix} ,
\end{equation}
the superscripts th and obs refer to the theoretical and observational estimations, respectively. For this work we use the distance posteriors (represented as the obs superscript) for the flat and non-flat cases given by \cite{CMB_DATA2}.

\end{itemize}

Finally, we consider a Gaussian log-likelihood to combine the mentioned samples in the form 
\begin{equation}\label{eq:chi2_joint}
    -2\ln(\mathcal{L}) \;\propto\; \chi^2= \sum_i \chi^2_i\,,
\end{equation}
where the sum runs over each statistic $\chi^2$ for the baseline datasets.

\section{Results} \label{section:Results}

To allow a better comparison between the four models and the standard one, we add the curvature term $\Omega_k$ to all of them. Additionally, the constraints are realized using only low-redshift data samples, in this case, CC and SNIa, with the goal of having a redshift-homogeneous region and avoiding significant discrepancy that could cause bias. As we discussed previously, all the models (except for the standard) contain a hyperbolic tangent function, or a function with similar behavior, which we refer to as an emergent DE. As consequence, dark energy emerges only at low redshifts, in contrast to the cosmological constant, whose presence is assumed since the Big Bang. 

\begin{figure*}
   \centering
   \includegraphics[width=0.6\textwidth]{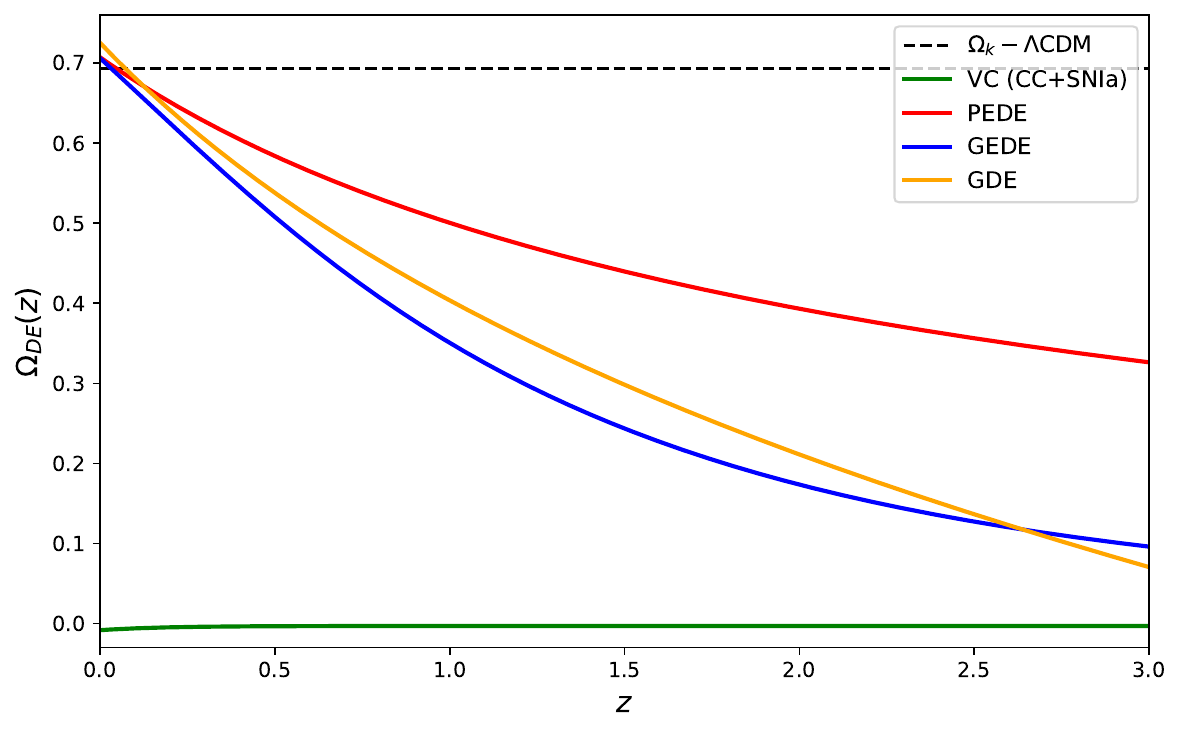}
   \caption{DE density for the cosmological models using best fit values of the combination CC+SNIa+BAO+CMB.}
   \label{fig:ode_comparison}
\end{figure*}

In order to give us an idea of the behavior of the emergent DE term, which involves $\tanh(x)$ or $\rm{sgn}(x)$, we show a comparison between the models in Fig. \ref{fig:ode_comparison}. In particular, note the remarked difference between PEDE, GEDE, GDE, and VC. In the VC case, the universe's acceleration is caused by a change in curvature, whereas in the other models it involves a contribution from DE.

Our results are discussed model by model, in the following sections.

\begin{table*}[ht!]
	\centering
    \footnotesize
	\caption{ Parameter's median values and their $1\sigma$ confidence intervals.}
	\label{tab:bf_model}
	\begin{tabular}{lccccc} 
    \hline
 Parameter & CC & SNIa & CC+SNIa  & CC+SNIa+BAO & CC+SNIa+BAO+CMB\\ [0.9ex]
    \hline
    \multicolumn{6}{c}{$\Omega_k-\Lambda$CDM} \\ [0.9ex]
 $h$  & 	$0.686^{+0.039}_{-0.039}$  & 	$0.680^{+0.054}_{-0.055}$  & 	$0.690^{+0.036}_{-0.289}$  & 	$0.679^{+0.017}_{-0.014}$  & 	$0.680^{+0.006}_{-0.005}$  \\ [0.9ex] 
 $\Omega_{0b}$  & 	$0.042^{+0.008}_{-0.008}$  & 	$0.043^{+0.008}_{-0.009}$  & 	$0.045^{+0.010}_{-0.010}$  & 	$0.048^{+0.002}_{-0.002}$  & 	$0.048^{+0.001}_{-0.001}$  \\ [0.9ex] 
 $\Omega_{0m}$  & 	$0.318^{+0.121}_{-0.118}$  & 	$0.355^{+0.055}_{-0.042}$  & 	$0.390^{+0.608}_{-0.066}$  & 	$0.318^{+0.016}_{-0.015}$  & 	$0.315^{+0.006}_{-0.006}$  \\ [0.9ex] 
 $\Omega_{k}$  & 	$0.026^{+0.312}_{-0.327}$  & 	$-0.150^{+0.101}_{-0.137}$  & 	$-0.075^{+1.071}_{-0.168}$  & 	$-0.047^{+0.031}_{-0.042}$  & 	$-0.008^{+0.006}_{-0.009}$  \\ [0.9ex] 
 $\tau_U \,[\rm{Gyrs}]$  & 	$13.507^{+0.532}_{-0.470}$  & 	$13.860^{+1.217}_{-1.044}$  & 	$13.601^{+0.936}_{-0.591}$  & 	$13.874^{+0.321}_{-0.342}$  & 	$13.723^{+0.024}_{-0.026}$  \\ [0.9ex] 
 $z_T $  & 	$0.615^{+0.158}_{-0.147}$  & 	$0.649^{+0.053}_{-0.048}$  & 	$0.680^{+1.511}_{-0.066}$  & 	$0.663^{+0.030}_{-0.031}$  & 	$0.638^{+0.020}_{-0.018}$  \\ [0.9ex] 
 $q_0 $  & 	$-0.500^{+0.167}_{-0.167}$  & 	$-0.623^{+0.057}_{-0.063}$  & 	$-0.583^{+2.074}_{-0.081}$  & 	$-0.573^{+0.026}_{-0.030}$  & 	$-0.536^{+0.013}_{-0.015}$  \\ [0.9ex] 
$\chi^2$  & $14.60$  & $63.44$  & $72.56$  & $84.93$  & $86.00$  \\ [0.9ex]
\multicolumn{6}{c}{Variable Curvature} \\ [0.9ex]
 $h$  & 	$0.735^{+0.010}_{-0.010}$  & 	$0.730^{+0.011}_{-0.010}$  & 	$0.738^{+0.010}_{-0.010}$ & - & - \\ [0.9ex] 
 $\Omega_{0m}$  & 	$0.312^{+0.005}_{-0.005}$  & 	$0.311^{+0.006}_{-0.006}$  & 	$0.314^{+0.005}_{-0.005}$& - & -  \\ [0.9ex] 
 $\Omega_k^1$  & 	$-0.020^{+0.013}_{-0.020}$  & 	$-0.004^{+0.007}_{-0.005}$  & 	$-0.003^{+0.006}_{-0.004}$ & - & - \\ [0.9ex] 
 $\Omega_k^2$  & 	$-0.079^{+0.027}_{-0.015}$  & 	$-0.057^{+0.124}_{-0.032}$  & 	$-0.066^{+0.117}_{-0.026}$ & - & - \\ [0.9ex] 
 $\tau_U \,[\rm{Gyrs}]$\footnote{For numerical issues, we give only an analytic approximation.}  & 	$13.825$  & 	$18.021$  & 	$16.869$  & - & - \\ [0.9ex] 
 $\alpha$  & 	$0.041^{+0.113}_{-0.223}$  & 	$-0.359^{+0.115}_{-0.094}$  & 	$-0.401^{+0.106}_{-0.070}$  & - & -\\ [0.9ex] 
 $\gamma$  & 	$7.751^{+4.549}_{-2.317}$  & 	$6.378^{+0.638}_{-0.392}$  & 	$5.995^{+0.408}_{-0.247}$ & - & - \\ [0.9ex] 
 $z_T $  & 	$0.533^{+0.142}_{-0.151}$  & 	$0.418^{+0.045}_{-0.043}$  & 	$0.475^{+0.039}_{-0.039}$ & - & - \\ [0.9ex] 
 $q_0 $  & 	$1.047^{+0.362}_{-0.296}$  & 	$-0.056^{+0.191}_{-0.199}$  & 	$0.075^{+0.153}_{-0.152}$ & - & - \\ [0.9ex] 
$\chi^2$  & $17.60$  & $59.44$  & $75.13$ & - & - \\ [0.9ex]
\multicolumn{6}{c}{PEDE} \\ [0.9ex]
 $h$  & 	$0.694^{+0.037}_{-0.041}$  & 	$0.679^{+0.055}_{-0.053}$  & 	$0.694^{+0.025}_{-0.025}$  & 	$0.705^{+0.013}_{-0.011}$  & 	$0.709^{+0.005}_{-0.005}$  \\ [0.9ex] 
 $\Omega_{0b}$  & 	$0.042^{+0.009}_{-0.008}$  & 	$0.043^{+0.009}_{-0.008}$  & 	$0.043^{+0.008}_{-0.009}$  &$0.044^{+0.001}_{-0.002}$  & 	$0.044^{+0.001}_{-0.001}$  \\ [0.9ex] 
 $\Omega_{0m}$  & 	$0.310^{+0.130}_{-0.112}$  & 	$0.391^{+0.049}_{-0.036}$  & 	$0.382^{+0.045}_{-0.032}$  &$0.325^{+0.013}_{-0.013}$  & 	$0.295^{+0.006}_{-0.006}$  \\ [0.9ex] 
 $\Omega_k$  & 	$0.086^{+0.277}_{-0.327}$  & 	$-0.100^{+0.071}_{-0.106}$  & 	$-0.086^{+0.061}_{-0.102}$  & 	$-0.017^{+0.012}_{-0.022}$  & 	$-0.002^{+0.001}_{-0.003}$  \\ [0.9ex] 
 $\tau_U \,[\rm{Gyrs}]$  & 	$13.498^{+0.520}_{-0.465}$  & 	$13.536^{+1.142}_{-1.011}$  & 	$13.288^{+0.438}_{-0.407}$  & 	$13.451^{+0.256}_{-0.261}$  & 	$13.688^{+0.020}_{-0.020}$  \\ [0.9ex] 
 $z_T $  & 	$0.592^{+0.134}_{-0.127}$  & 	$0.549^{+0.038}_{-0.037}$  & 	$0.556^{+0.037}_{-0.035}$  & 	$0.623^{+0.025}_{-0.025}$  & 	$0.678^{+0.014}_{-0.014}$  \\ [0.9ex] 
 $q_0 $  & 	$-0.590^{+0.184}_{-0.197}$  & 	$-0.674^{+0.049}_{-0.057}$  & 	$-0.673^{+0.046}_{-0.056}$  & 	$-0.684^{+0.024}_{-0.024}$  & 	$-0.714^{+0.011}_{-0.011}$  \\ [0.9ex] 
$\chi^2$  & $14.54$  & $60.16$  & $70.10$  & $86.02$  & $97.65$  \\ [0.9ex]
\multicolumn{6}{c}{GEDE} \\ [0.9ex]
 $h$  & 	$0.709^{+0.033}_{-0.041}$  & 	$0.680^{+0.054}_{-0.054}$  & 	$0.706^{+0.027}_{-0.027}$  & 	$0.677^{+0.015}_{-0.014}$  & 	$0.684^{+0.008}_{-0.008}$  \\ [0.9ex] 
 $\Omega_{0b}$  & 	$0.042^{+0.008}_{-0.008}$  & 	$0.043^{+0.008}_{-0.009}$  & 	$0.042^{+0.009}_{-0.008}$  & 	$0.048^{+0.002}_{-0.002}$  & 	$0.048^{+0.001}_{-0.001}$  \\ [0.9ex] 
 $\Omega_{0m}$  & 	$0.344^{+0.054}_{-0.046}$  & 	$0.317^{+0.057}_{-0.047}$  & 	$0.319^{+0.042}_{-0.037}$  & 	$0.312^{+0.013}_{-0.013}$  & 	$0.312^{+0.008}_{-0.008}$  \\ [0.9ex] 
 $\Omega_k$  & 	$-0.062^{+0.360}_{-0.310}$  & 	$-0.179^{+0.123}_{-0.176}$  & 	$-0.175^{+0.121}_{-0.161}$  & 	$-0.029^{+0.021}_{-0.039}$  & 	$-0.018^{+0.013}_{-0.026}$  \\ [0.9ex] 
 $\Delta$  & 	$2.580^{+3.014}_{-1.902}$  & 	$1.049^{+1.404}_{-0.826}$  & 	$1.022^{+1.133}_{-0.759}$  & 	$0.566^{+0.427}_{-0.373}$  & 	$0.301^{+0.233}_{-0.215}$  \\ [0.9ex] 
 $\tau_U \,[\rm{Gyrs}]$  & 	$13.425^{+0.430}_{-0.409}$  & 	$13.957^{+1.225}_{-1.096}$  & 	$13.416^{+0.448}_{-0.411}$  & 	$13.932^{+0.320}_{-0.315}$  & 	$13.730^{+0.022}_{-0.022}$  \\ [0.9ex] 
 $z_T $  & 	$0.544^{+0.110}_{-0.095}$  & 	$0.628^{+0.122}_{-0.109}$  & 	$0.625^{+0.092}_{-0.083}$  & 	$0.640^{+0.031}_{-0.030}$  & 	$0.639^{+0.019}_{-0.019}$  \\ [0.9ex] 
 $q_0 $  & 	$-0.777^{+0.194}_{-0.198}$  & 	$-0.665^{+0.061}_{-0.067}$  & 	$-0.657^{+0.062}_{-0.066}$  & 	$-0.611^{+0.047}_{-0.049}$  & 	$-0.575^{+0.039}_{-0.041}$  \\ [0.9ex] 
$\chi^2$  & $14.85$  & $48.85$  & $63.33$  & $81.91$  & $86.05$  \\ [0.9ex]
\multicolumn{6}{c}{GDE} \\ [0.9ex]
 $h$  & 	$0.701^{+0.036}_{-0.043}$  & 	$0.680^{+0.054}_{-0.054}$  & 	$0.701^{+0.026}_{-0.026}$  & 	$0.686^{+0.017}_{-0.015}$  & 	$0.687^{+0.008}_{-0.007}$  \\ [0.9ex] 
 $\Omega_{0b}$  & 	$0.043^{+0.008}_{-0.009}$  & 	$0.043^{+0.009}_{-0.009}$  & 	$0.043^{+0.008}_{-0.009}$  & 	$0.047^{+0.002}_{-0.002}$  & 	$0.047^{+0.001}_{-0.001}$  \\ [0.9ex] 
 $\Omega_{0m}$  & 	$0.286^{+0.124}_{-0.103}$  & 	$0.375^{+0.057}_{-0.045}$  & 	$0.363^{+0.046}_{-0.036}$  & 	$0.319^{+0.014}_{-0.014}$  & 	$0.309^{+0.007}_{-0.008}$  \\ [0.9ex] 
 $\Omega_k$  & 	$0.137^{+0.250}_{-0.332}$  & 	$-0.094^{+0.068}_{-0.126}$  & 	$-0.095^{+0.069}_{-0.112}$  & 	$-0.036^{+0.024}_{-0.039}$  & 	$-0.006^{+0.004}_{-0.007}$  \\ [0.9ex] 
 $\Psi$  & 	$9.603^{+7.183}_{-7.959}$  & 	$10.492^{+6.713}_{-8.786}$  & 	$11.369^{+6.018}_{-7.996}$  & 	$11.911^{+5.760}_{-7.945}$  & 	$12.758^{+5.180}_{-7.196}$  \\ [0.9ex] 
 $\lambda$  & 	$-0.436^{+0.302}_{-0.361}$  & 	$-0.168^{+0.124}_{-0.245}$  & 	$-0.140^{+0.099}_{-0.190}$  & 	$-0.092^{+0.063}_{-0.112}$  & 	$-0.098^{+0.066}_{-0.100}$  \\ [0.9ex] 
 $\tau_U \,[\rm{Gyrs}]$  & 	$13.556^{+0.500}_{-0.461}$  & 	$13.641^{+1.205}_{-1.042}$  & 	$13.300^{+0.453}_{-0.409}$  & 	$13.763^{+0.316}_{-0.309}$  & 	$13.723^{+0.022}_{-0.024}$  \\ [0.9ex] 
 $z_T $  & 	$0.559^{+0.144}_{-0.124}$  & 	$0.574^{+0.075}_{-0.085}$  & 	$0.584^{+0.062}_{-0.066}$  & 	$0.640^{+0.033}_{-0.032}$  & 	$0.635^{+0.018}_{-0.017}$  \\ [0.9ex] 
 $q_0 $  & 	$-0.813^{+0.293}_{-0.377}$  & 	$-0.737^{+0.100}_{-0.189}$  & 	$-0.735^{+0.100}_{-0.160}$  & 	$-0.665^{+0.068}_{-0.115}$  & 	$-0.645^{+0.074}_{-0.111}$  \\ [0.9ex] 
$\chi^2$  & $14.99$  & $59.76$  & $70.60$  & $87.80$  & $92.80$  \\ [0.9ex]
\hline
	\end{tabular}
\end{table*}

\begin{figure*}
   \centering
   \includegraphics[width=0.6\textwidth]{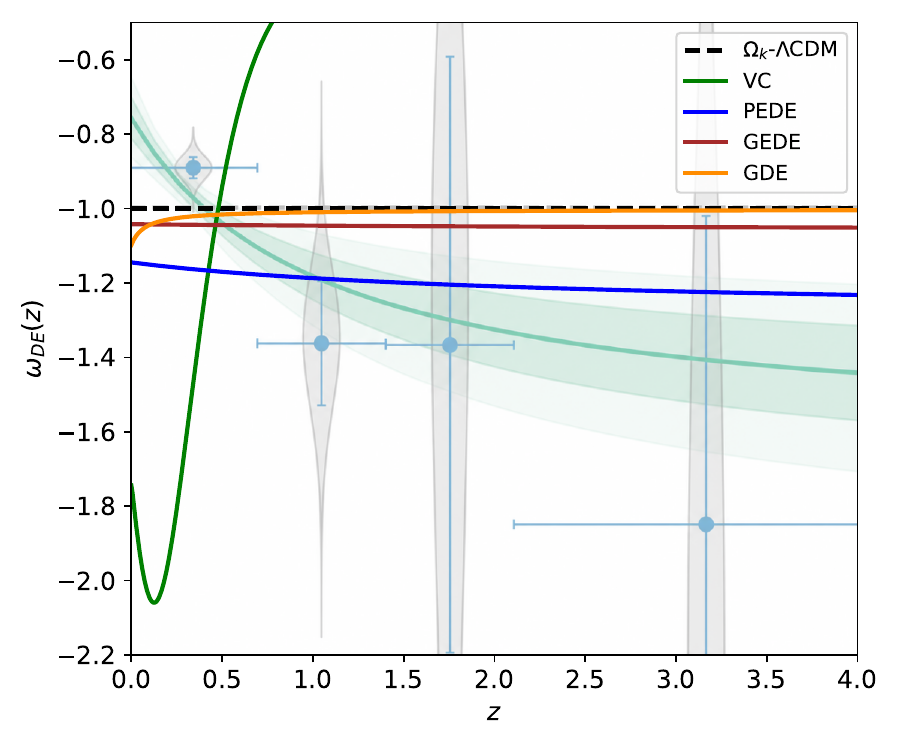}
   \caption{DE equation of state for the cosmological models using best fit values of the combination CC+SNIa+BAO+CMB, meanwhile for VC is used CC+SNIa. Background image corresponds to the Figure 12 in \cite{DESI:2025zgx}. The solid light-green line shows the best-fit $w(z)$ based on $w_0$ and $w_a$ from Chevalier-Polarski-Linder model \cite{Chevallier:2000qy,Linder:2002et} inference and the green contours around it represent the $68\%$ and $95\%$ confidence intervals, according to \cite{DESI:2025zgx}. The blue points are the constraints from the binning approach \cite{DESI:2025fii}, with the horizontal bars indicating the bin width and the vertical bars representing the $1\sigma$ error (see \cite{DESI:2025zgx} for details).}
   \label{fig:eos_comparison}
\end{figure*}

\subsection{Variable curvature}

\begin{figure*}
   \centering
   \includegraphics[width=0.6\textwidth]{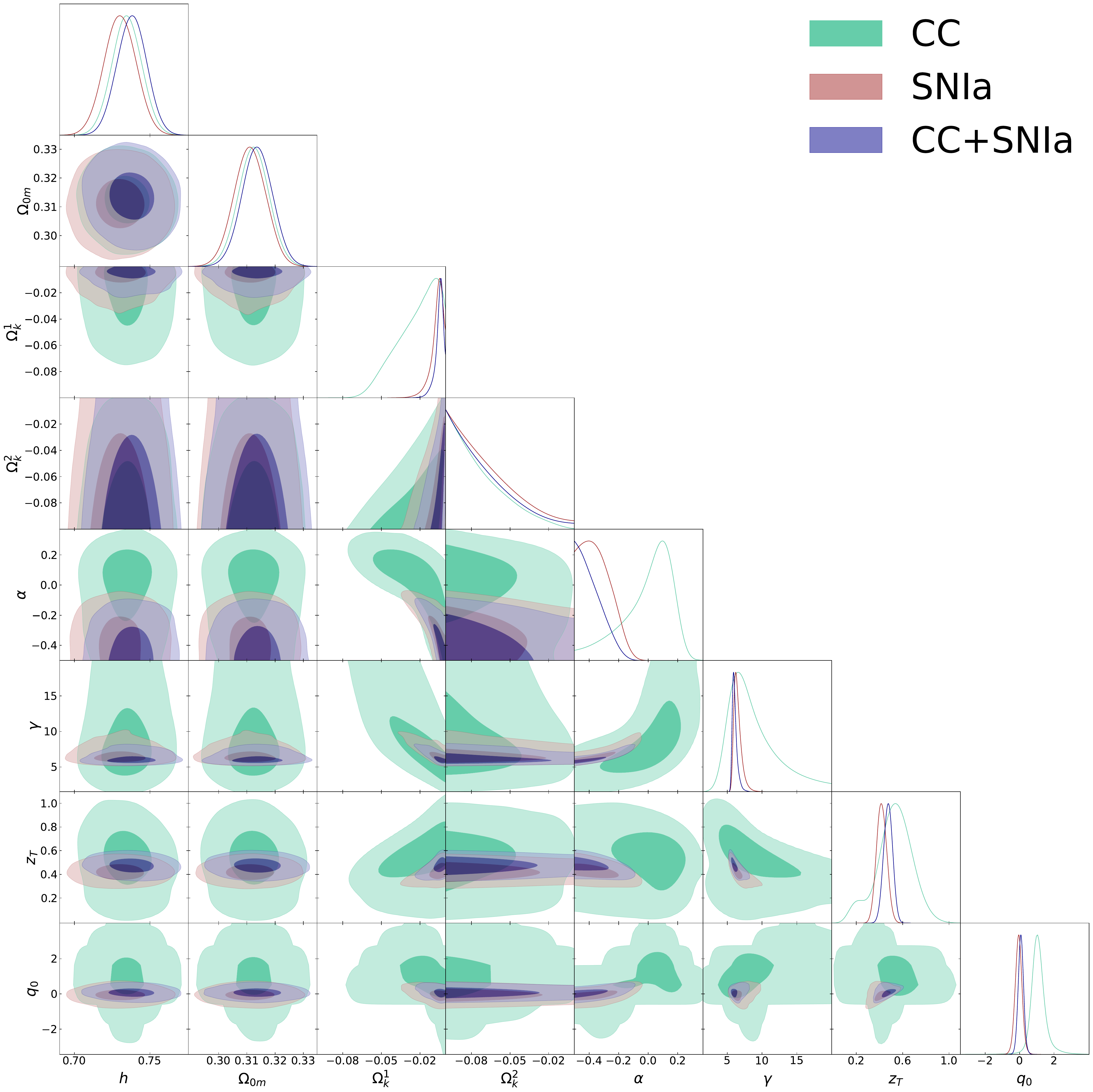}
   \caption{Posterior distributions for the variable curvature model, with parameters $\boldsymbol{\Theta}=(h,\Omega_{0m},\Omega_k^1,\Omega_k^2,\alpha,\gamma)$, obtained from the MCMC analysis. Diagonal panels show the marginalized 1D posteriors, while off-diagonal panels display the joint 2D regions at $1\sigma$  and $3\sigma$ confidence levels.}
   \label{fig:contours_curvar}
\end{figure*}

\begin{figure*}
   \centering
   \includegraphics[width=0.32\textwidth]{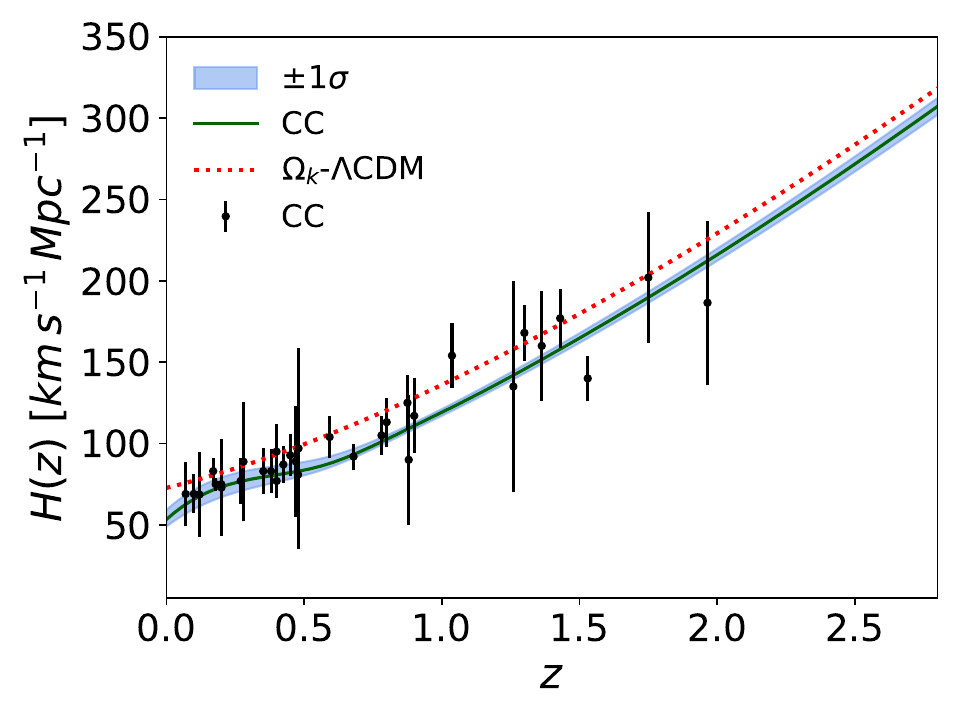}
   \includegraphics[width=0.32\textwidth]{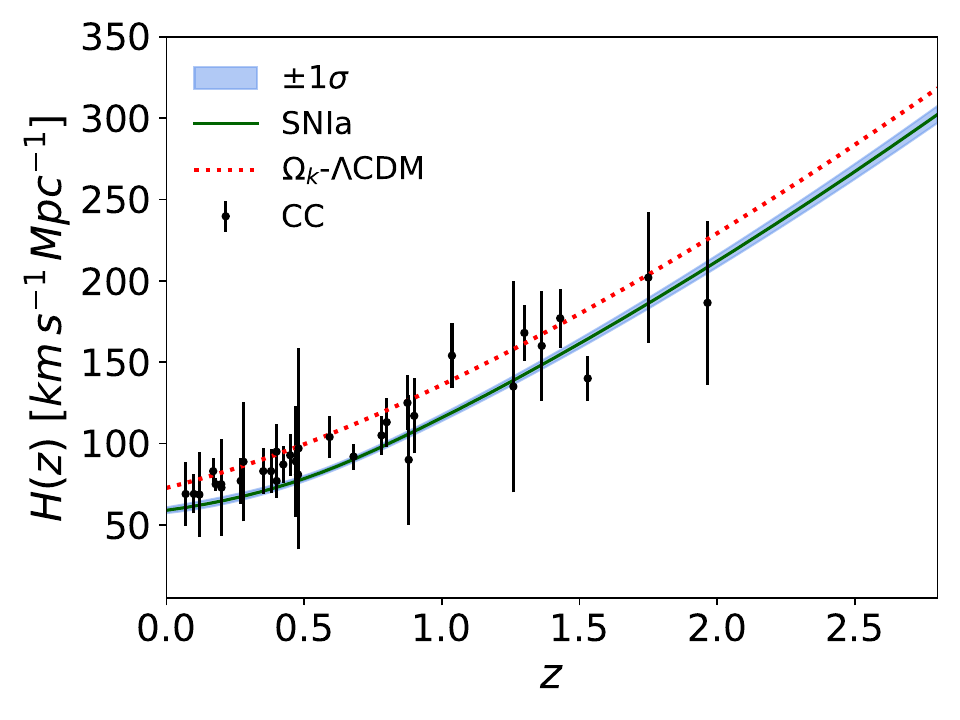}
   \includegraphics[width=0.32\textwidth]{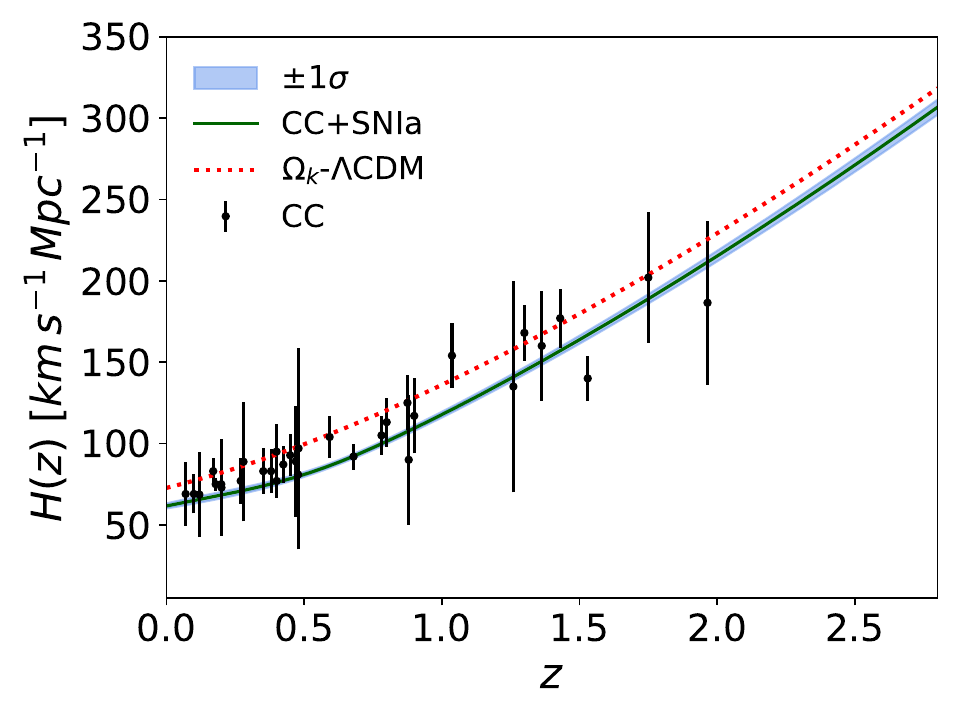}\\
   \includegraphics[width=0.32\textwidth]{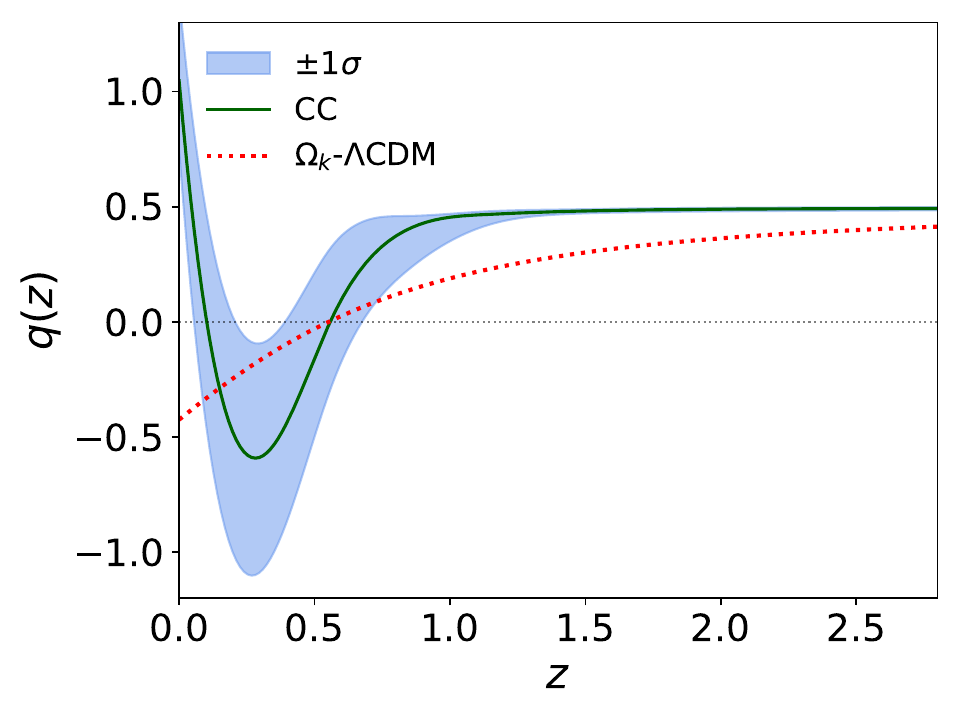}
   \includegraphics[width=0.32\textwidth]{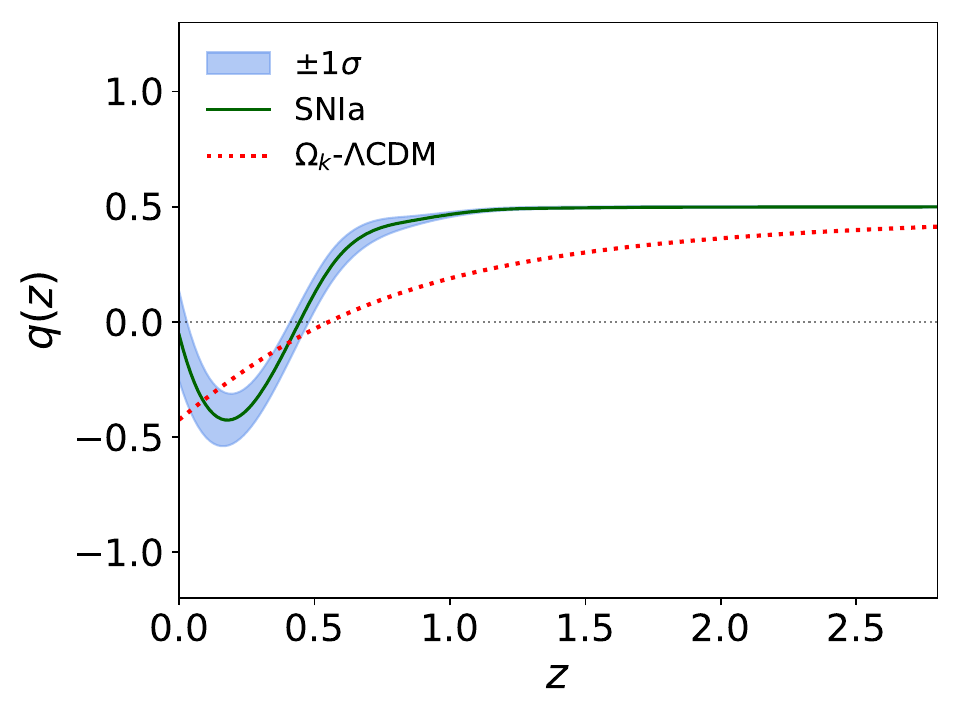}
   \includegraphics[width=0.32\textwidth]{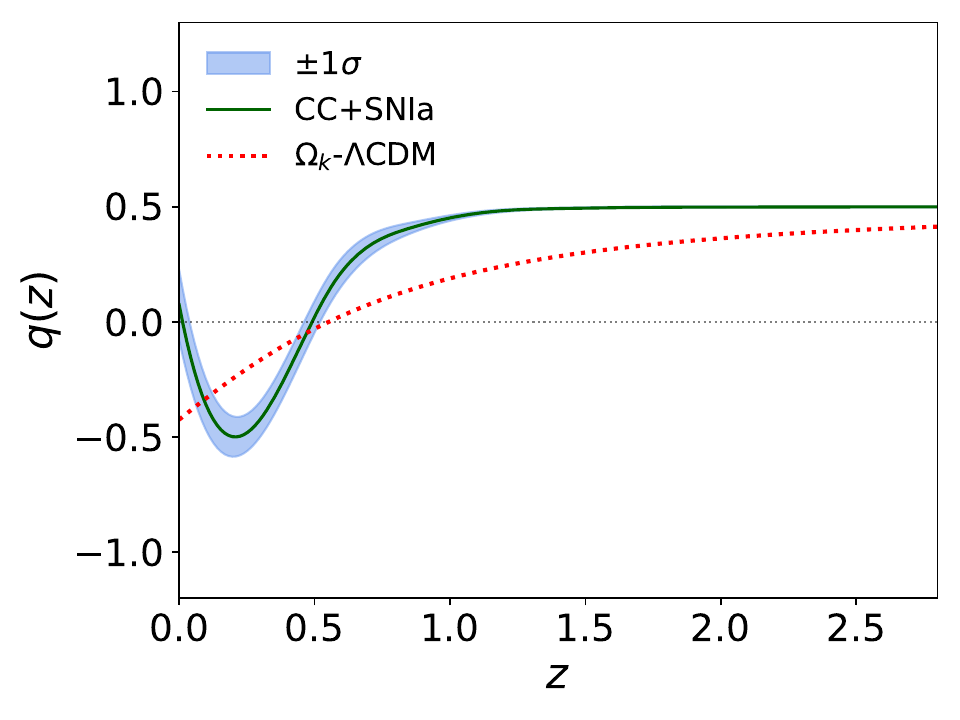} \\
   \includegraphics[width=0.32\textwidth]{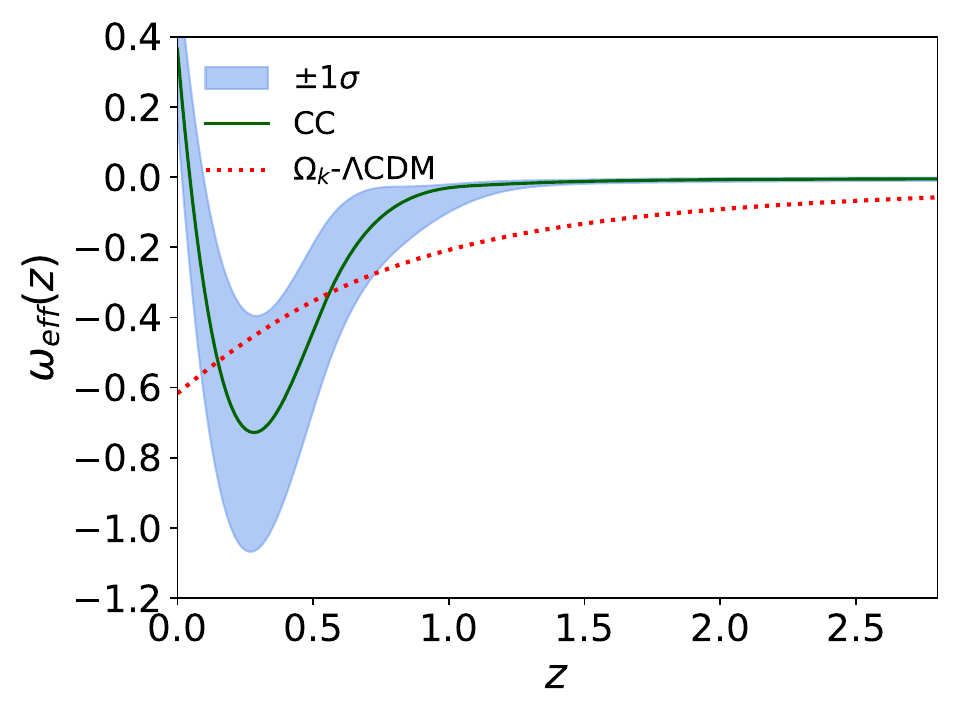}
   \includegraphics[width=0.32\textwidth]{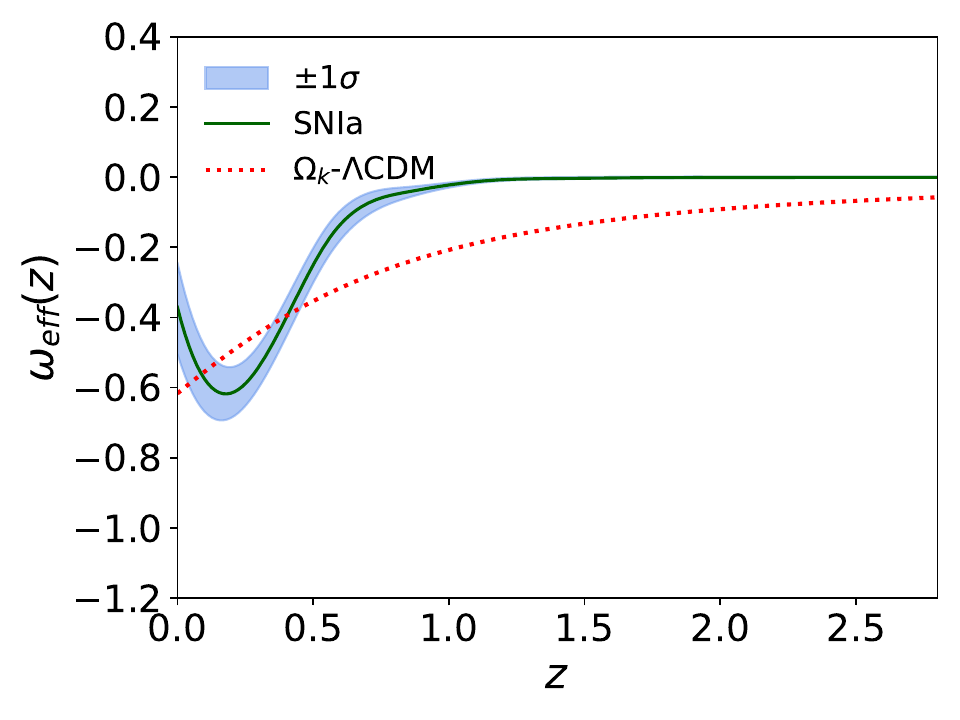}
   \includegraphics[width=0.32\textwidth]{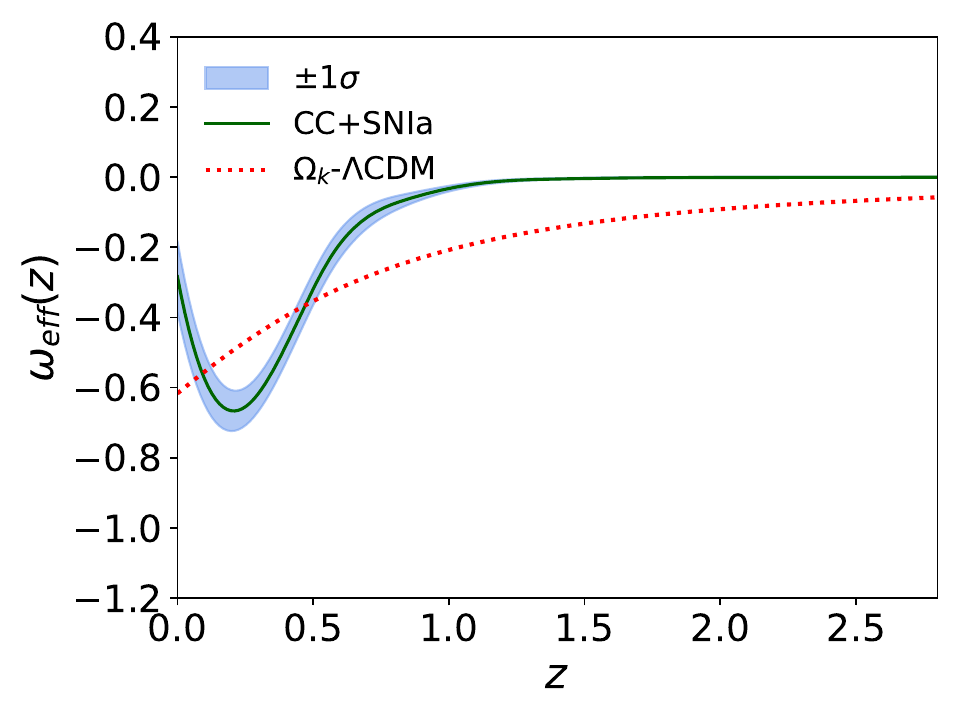}
   \caption{Reconstructions of the expansion history for variable curvature model  using CC, SNIa and CC+SNIa data combinations (left to right). Top panel shows the Hubble parameter $H(z)$, middle panel displays the deceleration parameter $q(z)$ and the bottom panel is the effective EoS $\omega_{\rm eff}(z)$. Solid paths represent to the posterior median reconstructions and their the $1\sigma$ intervals. The reference $\Omega_k$-$\Lambda$CDM prediction is drawn as a red dashed line.}
   \label{fig:cosmo_curvar}
\end{figure*}

The first model reported in this paper is the VC model, introduced in \cite{Esteban-Gutierrez:2024rpz}, whose main characteristic is that the observed acceleration of the Universe is caused by a transition in its curvature. The MCMC results are summarized in \ref{tab:bf_model}, together with its posterior distribution shown in Fig. \ref{fig:contours_curvar}. Moreover, we present the cosmography of the model, shown in Fig. \ref{fig:cosmo_curvar}. Notice that there exists a remarkable difference in the evolution of $H(z)$ compared to $\Omega_k$-$\Lambda$CDM. The deceleration parameter shows a transition in the same epoch as $\Lambda$CDM transits, but a deceleration is also observed at $z=0$. The $w_{eff}$ behavior shows a sudden transition to $w<-1/3$, allowing an accelerated Universe, according to GR, and a later $w>-1/3$ transition towards a decelerated Universe. The age of the Universe was calculated analytically and gives a Universe age above the expected value in $\Omega_k$-$\Lambda$CDM, alleviating the tension with the oldest stars in our Universe (see \cite{OldestStar,Valcin:2021,Valcin_2020} but in specific \cite{deAndres:2024vmr}).

\begin{figure}
   \centering
   \includegraphics[width=0.42\textwidth]{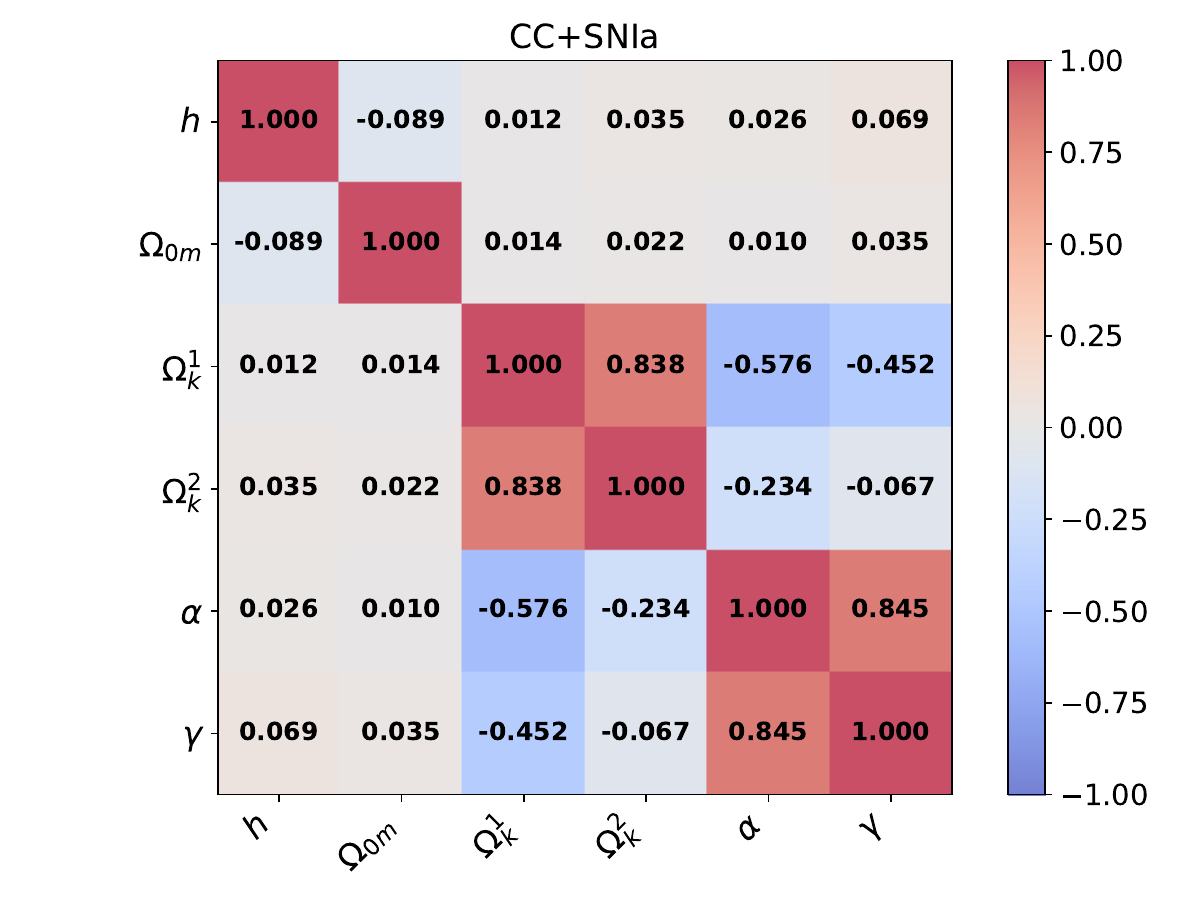}
   \caption{Correlation matrices for the parameters of the VC model for CC+SNIa dataset.}
   \label{fig:corr_curvar}
\end{figure}

Figure~\ref{fig:corr_curvar} presents the correlation matrix for the VC model obtained from the CC+SNIa dataset. The standard cosmological parameters, namely the Hubble constant and the present matter density, exhibit negligible correlations with the additional VC parameters, indicating that the low-redshift observations constrain the background cosmology largely independently of the variable-curvature sector. Instead, the dominant degeneracies are confined to the model-specific parameters. In particular, the two curvature amplitudes, $\Omega_{k}^{1}$ and $\Omega_{k}^{2}$, are strongly positively correlated ($r=0.838$), suggesting that both parameters contribute in a similar manner to the effective curvature evolution and can partially compensate each other. Likewise, the parameters $\alpha$ and $\gamma$ exhibit a strong positive correlation ($r=0.845$), indicating that different combinations of these quantities produce nearly equivalent expansion histories within the redshift range probed by CC and SNIa observations. Moderate anti-correlations are also present between $\Omega_{k}^{1}$ and both $\alpha$ and $\gamma$, reflecting the interplay between the curvature amplitudes and the parameters governing their evolution. Overall, Fig.~\ref{fig:corr_curvar} suggests that the principal parameter degeneracies of the VC model arise within its intrinsic curvature sector rather than between the new parameters and the standard cosmological quantities. This behavior indicates that low-redshift observations are sufficient to constrain the background expansion but provide only limited leverage to disentangle the individual contributions of the variable-curvature parameters, highlighting the potential importance of complementary high-redshift probes for breaking these internal degeneracies.

Finally, the posterior distributions indicate that all VC parameters are statistically constrained by the available data. However, the correlation analysis shows that the additional curvature parameters are not completely independent, exhibiting significant internal degeneracies while remaining only weakly coupled to the standard cosmological parameters. This suggests that several combinations of the VC parameters produce nearly equivalent background expansion histories over the redshift range probed by CC and SNIa observations. As a result, although the VC model provides a viable description of the data, its larger parameter space is only partially exploited by the current low-redshift observations, leading to the stronger complexity penalty reflected by the BIC.

\subsection{PEDE}

\begin{figure*}
   \centering
   \includegraphics[width=0.6\textwidth]{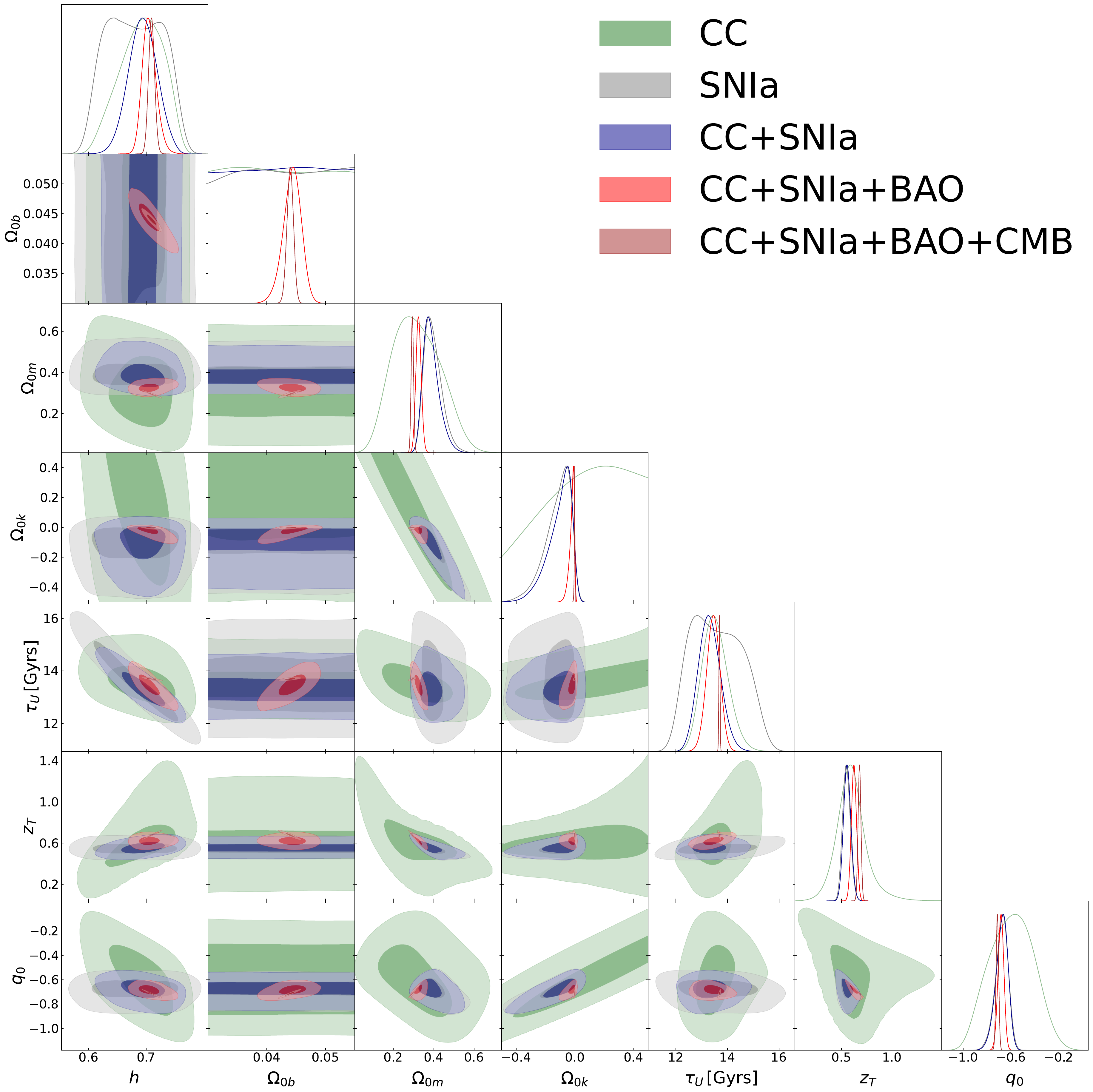}
   \caption{Posterior distributions for the PEDE model parameters $\boldsymbol{\Theta}=(h,\Omega_{0m},\Omega_k)$, obtained from the MCMC analysis. Diagonal panels show the marginalized 1D posteriors, while off-diagonal panels display the joint 2D regions at $1\sigma$  and $3\sigma$ confidence levels.}
   \label{fig:contours_pede}
\end{figure*}

\begin{figure*}
   \centering
   \includegraphics[width=0.32\textwidth]{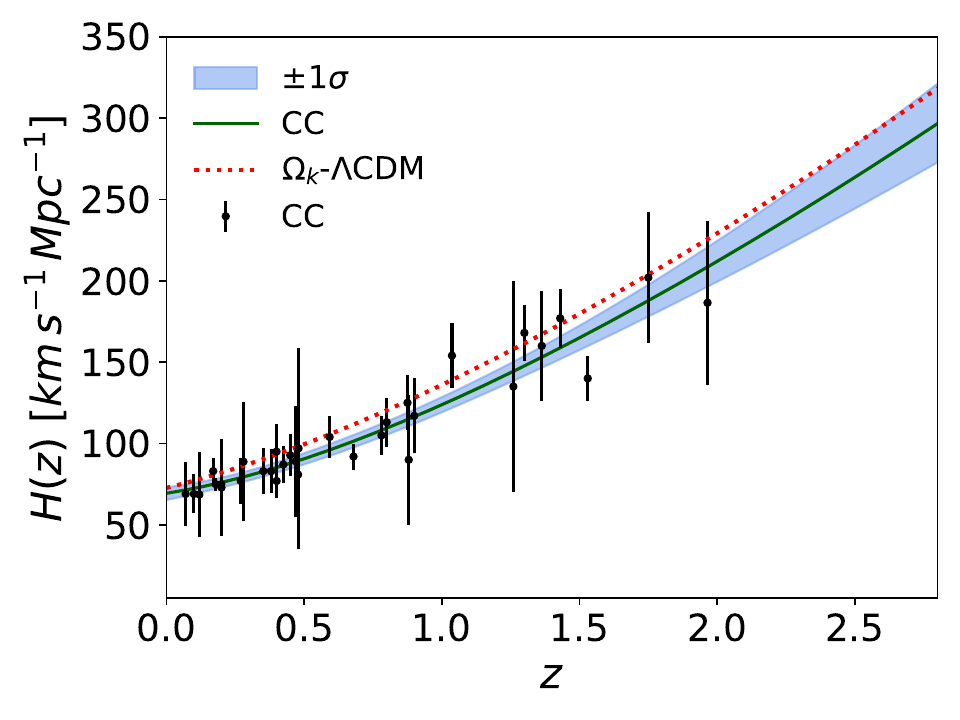}
   \includegraphics[width=0.32\textwidth]{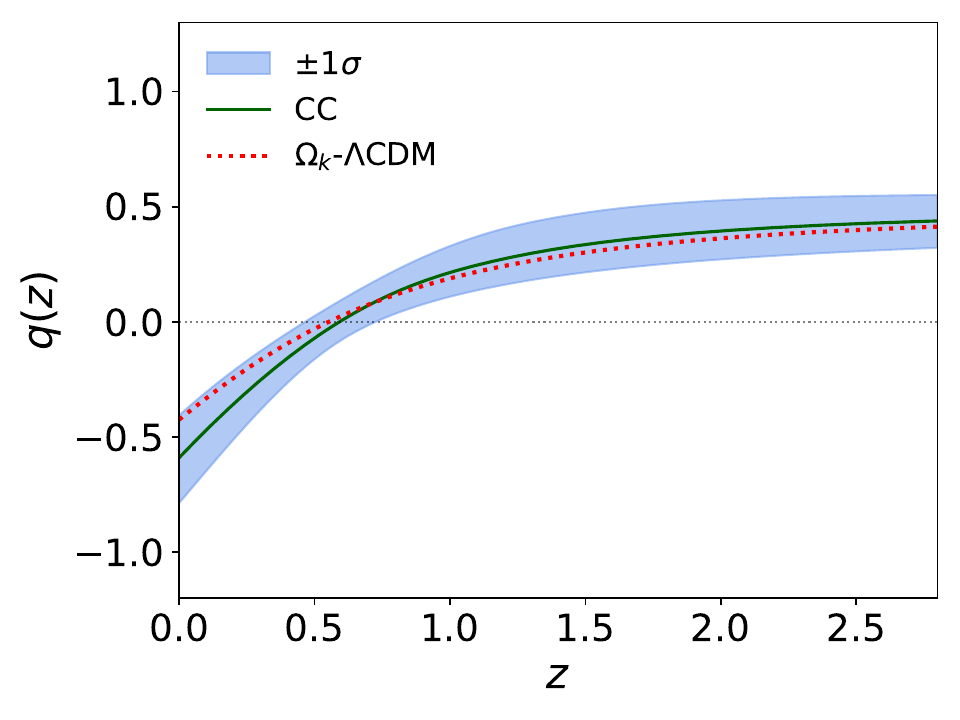}
   \includegraphics[width=0.32\textwidth]{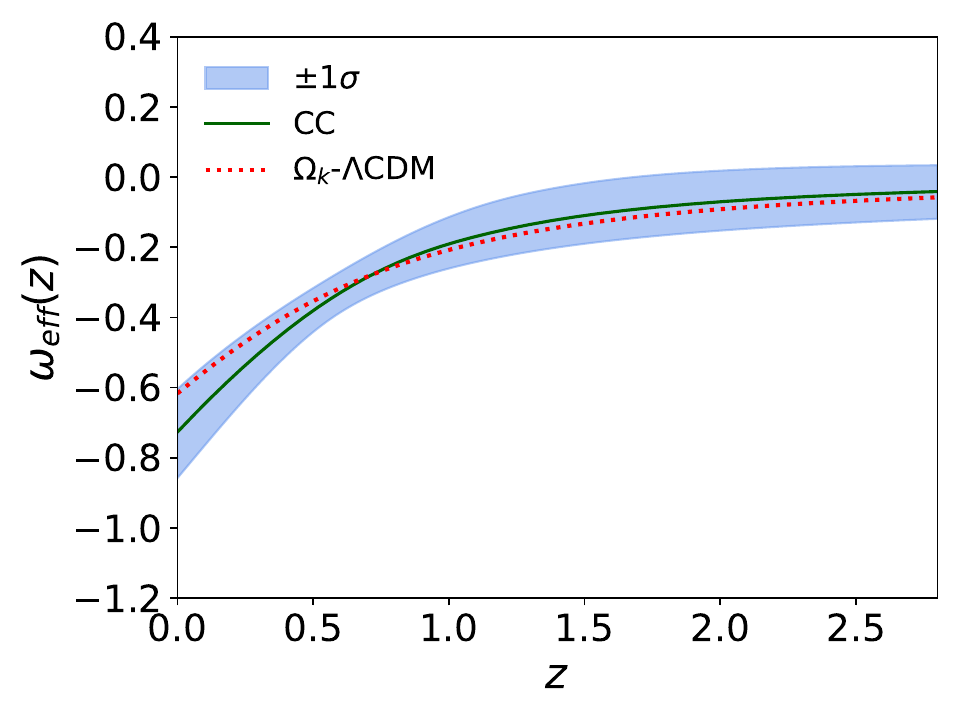}\\
   \includegraphics[width=0.32\textwidth]{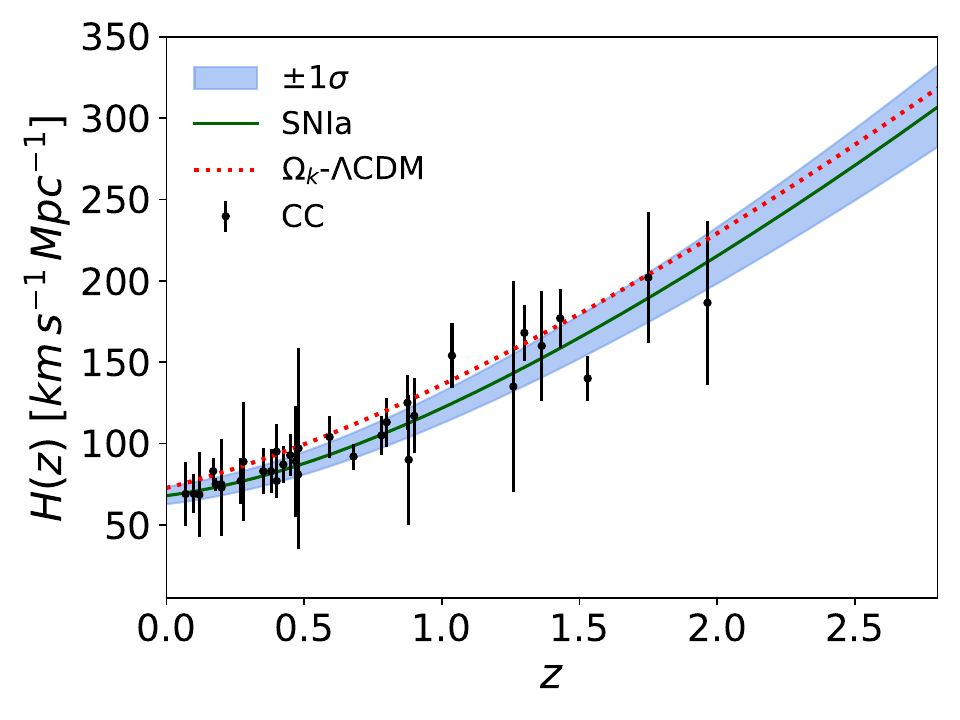}
   \includegraphics[width=0.32\textwidth]{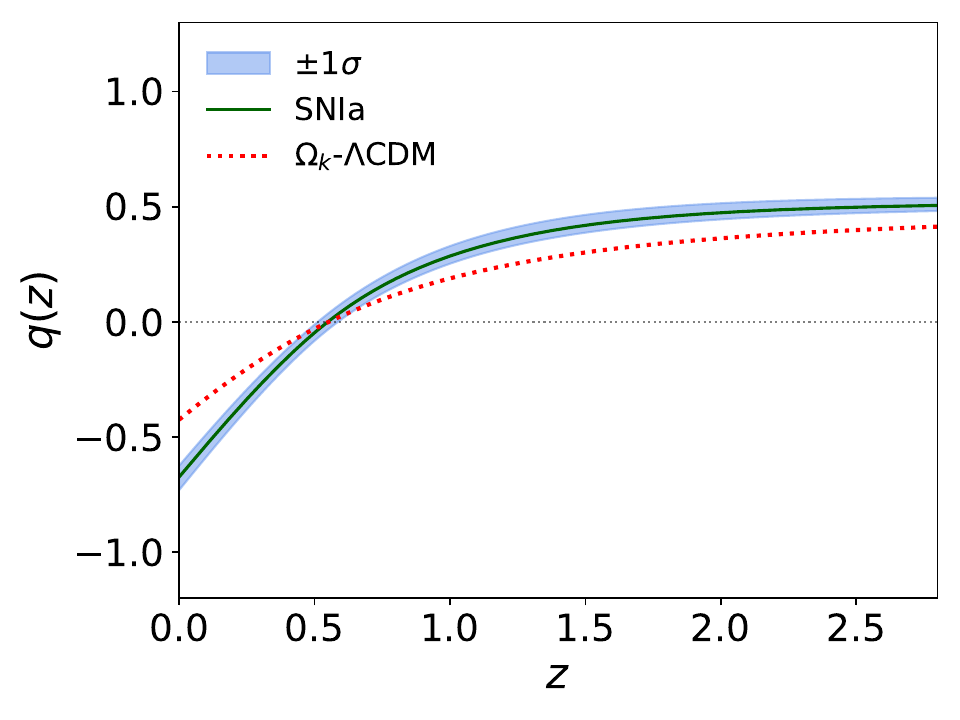}
   \includegraphics[width=0.32\textwidth]{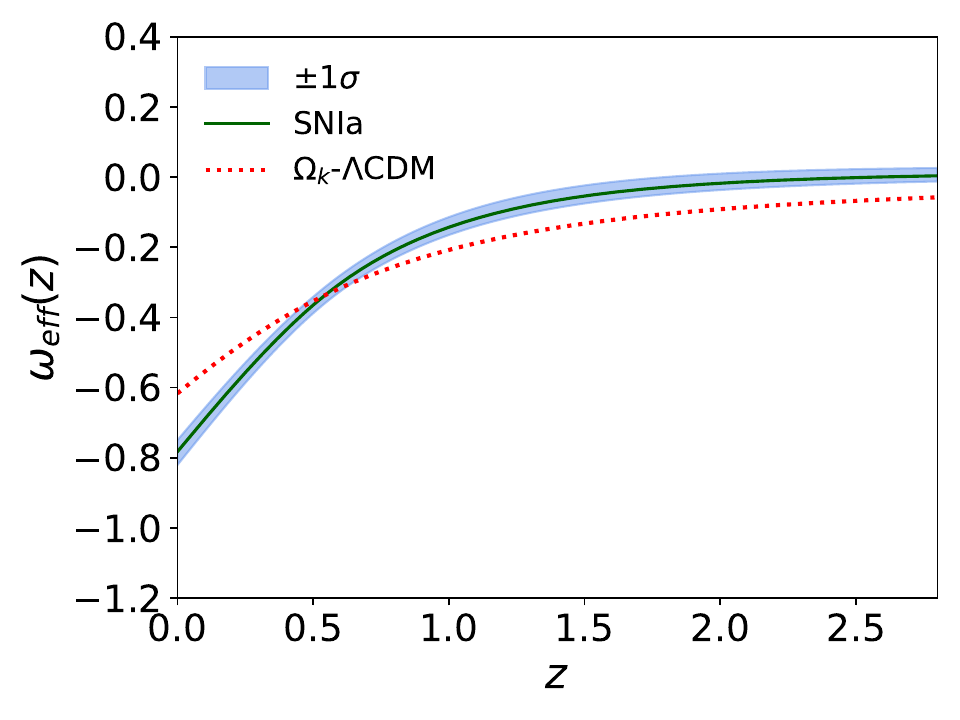} \\
   \includegraphics[width=0.32\textwidth]{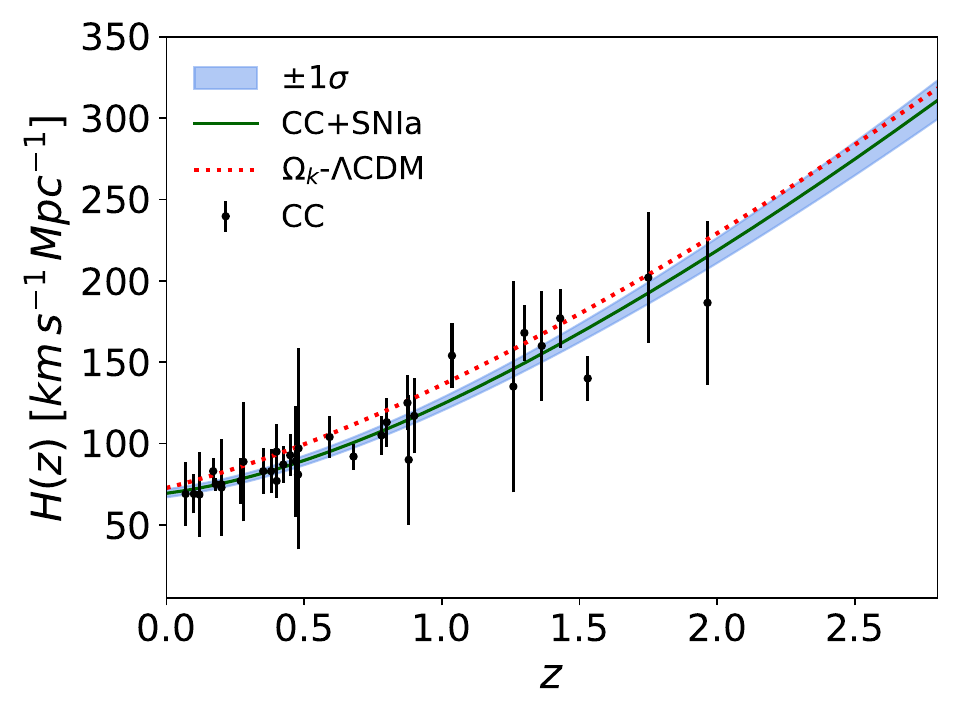}
   \includegraphics[width=0.32\textwidth]{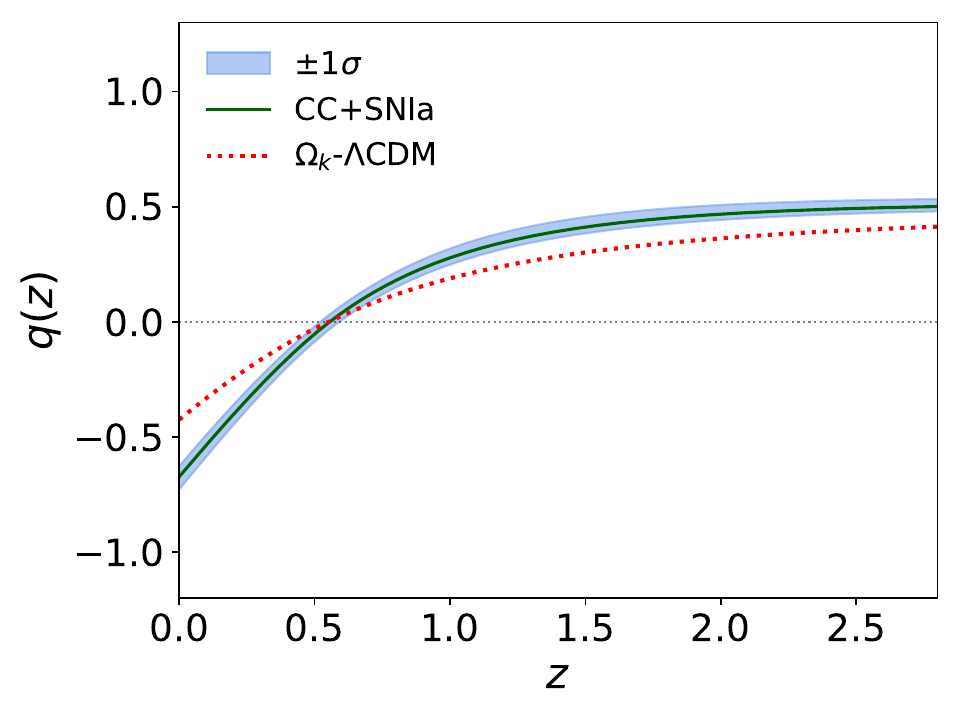}
   \includegraphics[width=0.32\textwidth]{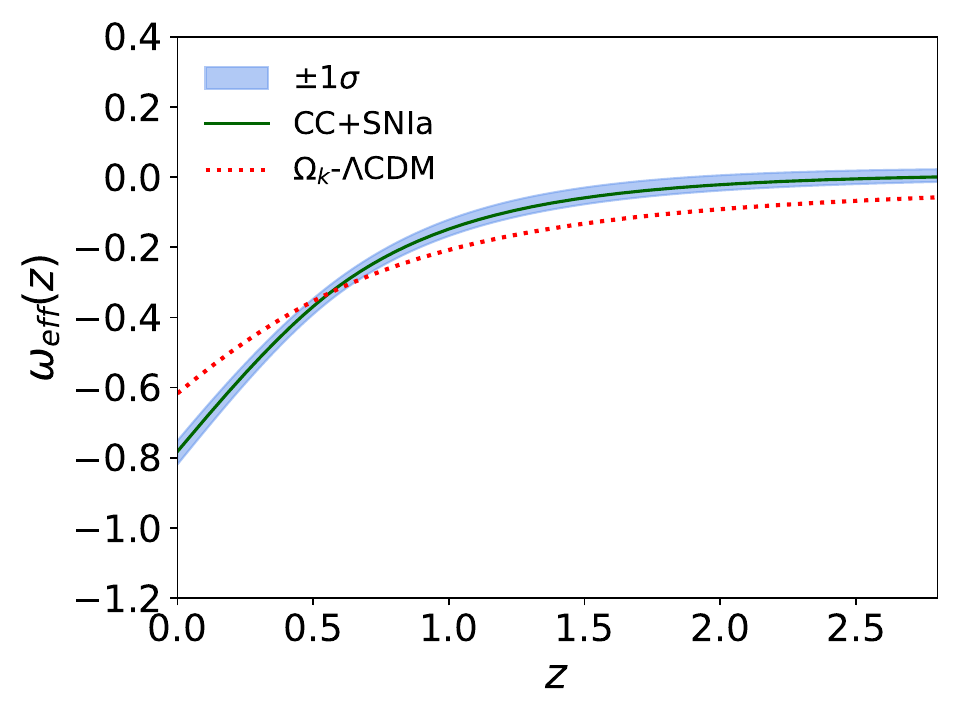} \\
   \includegraphics[width=0.32\textwidth]{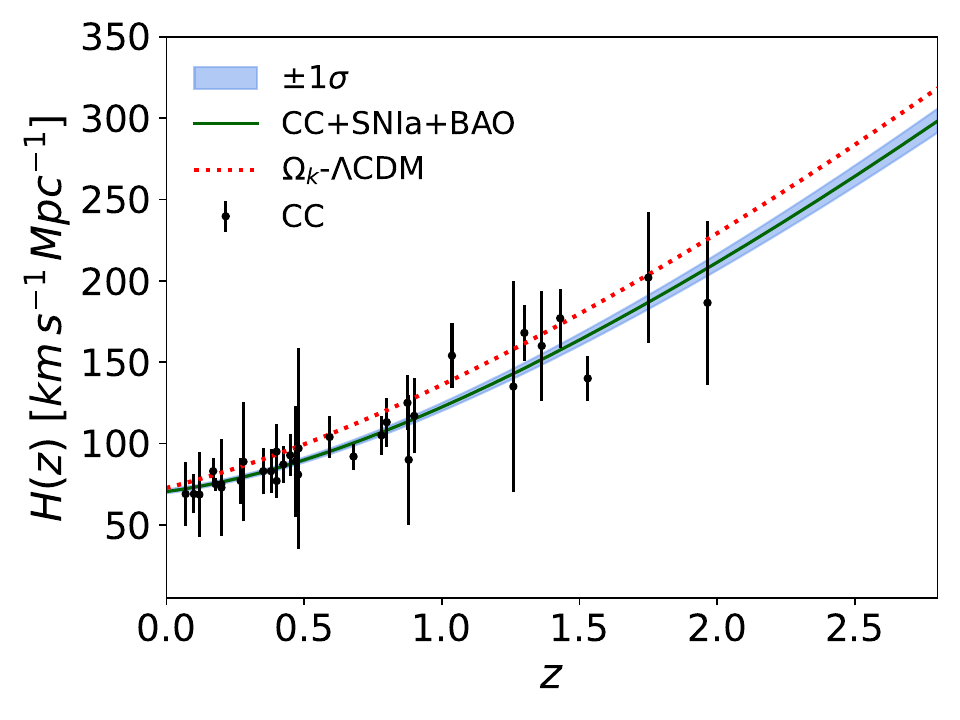}
   \includegraphics[width=0.32\textwidth]{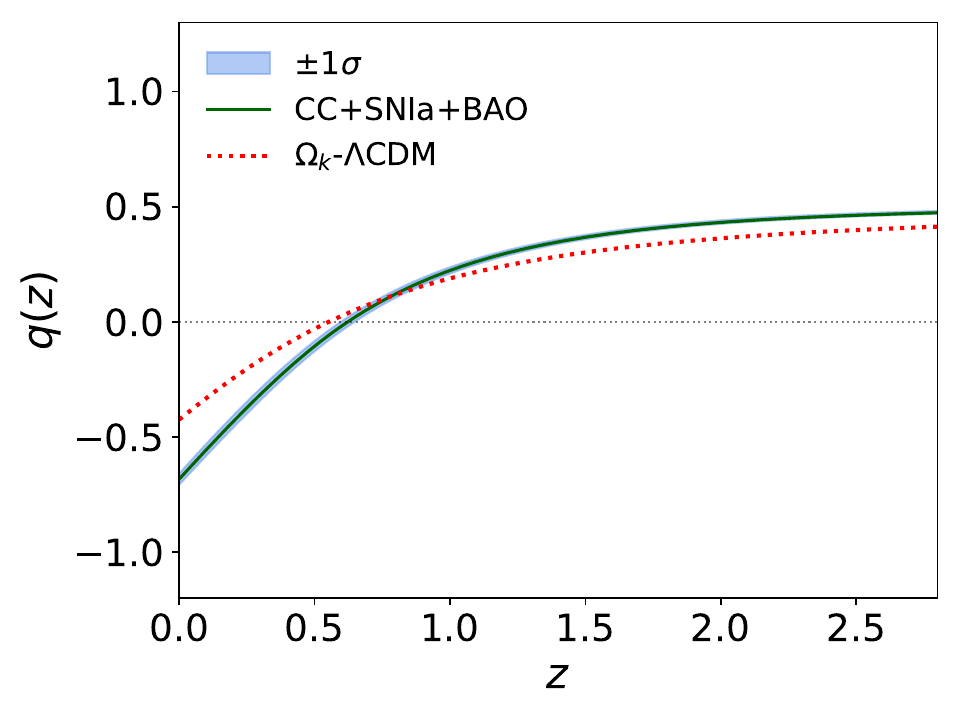}
   \includegraphics[width=0.32\textwidth]{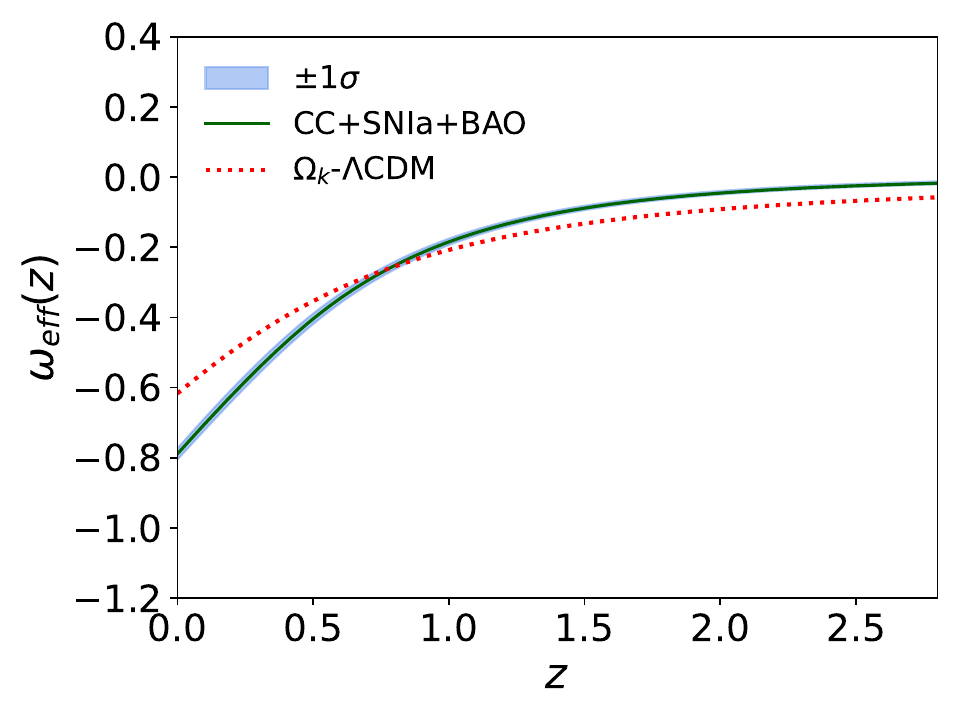} \\
   \includegraphics[width=0.32\textwidth]{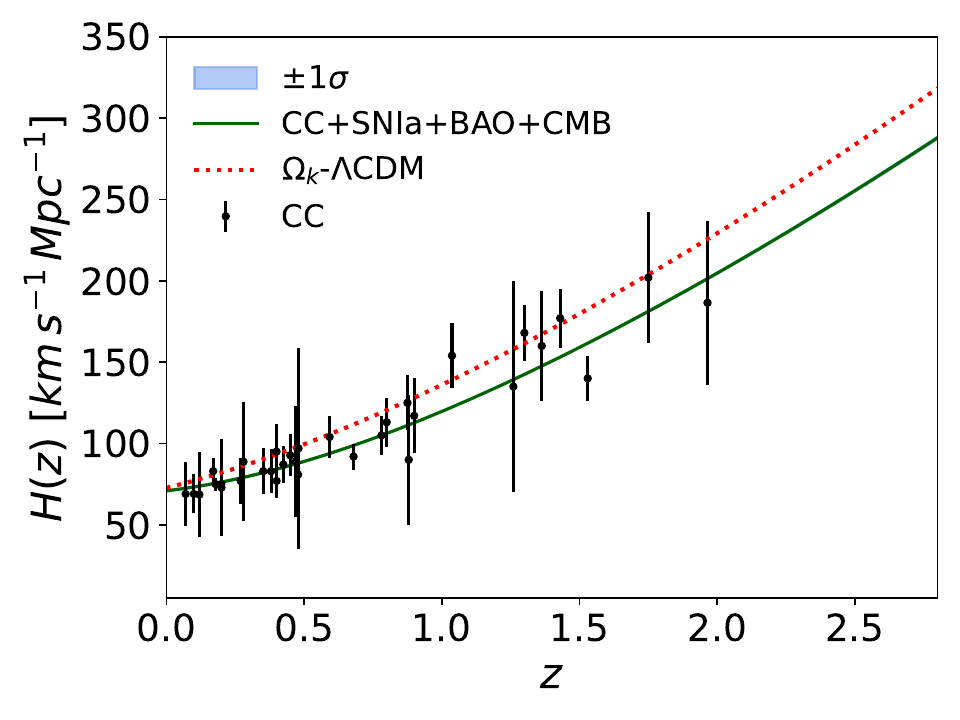}
   \includegraphics[width=0.32\textwidth]{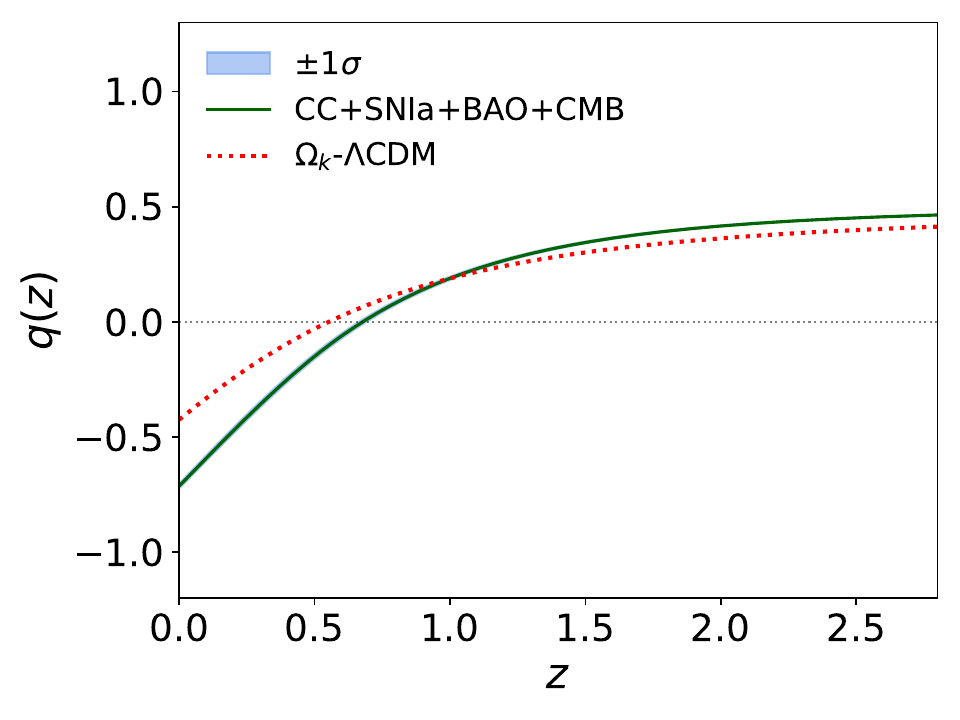}
   \includegraphics[width=0.32\textwidth]{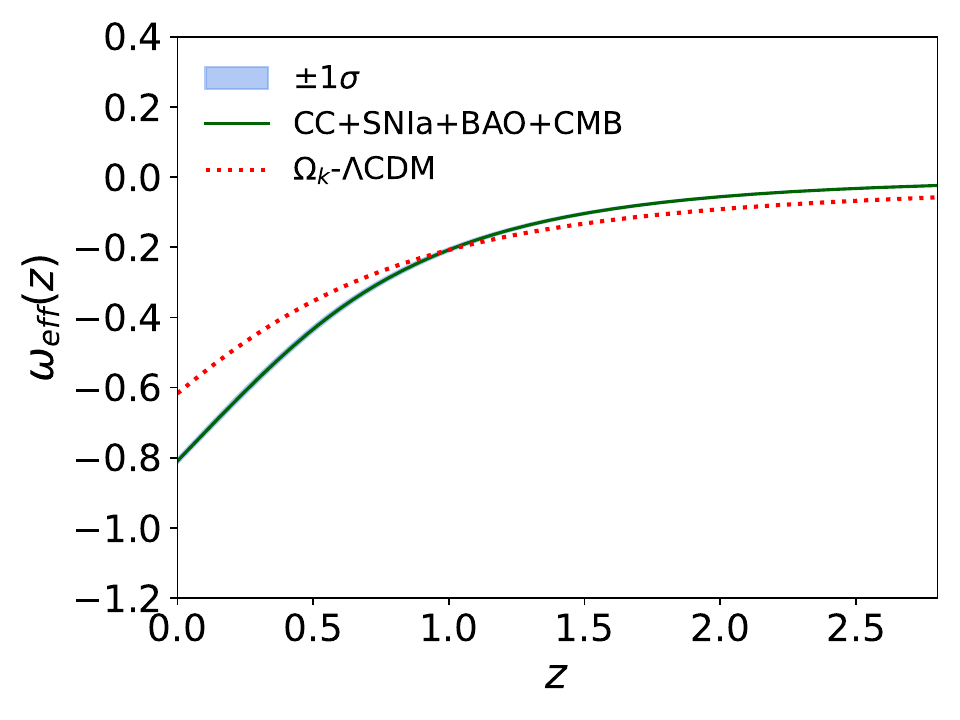}
   \caption{Reconstructions of the expansion history for the PEDE model using CC, SNIa, CC+SNIa, CC+SNIa+BAO, and CC+SNIa+BAO+CMB data combinations (from top to bottom). We show the Hubble parameter $H(z)$ (left panel), the deceleration parameter $q(z)$ (middle panel), and the effective EoS $\omega_{\rm eff}(z)$ (right panel). Solid paths represent the posterior median reconstructions and their $ 1\sigma$ intervals. The reference $\Omega_k$-$\Lambda$CDM prediction is drawn as a red dashed line.}
   \label{fig:cosmo_pede}
\end{figure*}

The next model is one of the most studied in the literature \cite{PEDE:2019ApJ}, where the term $\Omega_{DE}(z)$ is introduced phenomenologically. However, \cite{PhysRevD.80.067301} show that the term $\Omega_{DE}(z)$ could be deduced from the assumption of a scalar field. The achievements of the PEDE model stem from the fact that it has the same degrees of freedom as the $\Omega_k$-$\Lambda$CDM and from the modeling of emergent dark energy. Our results are summarized in Table \ref{tab:bf_model}, together with its posterior distribution shown in Fig. \ref{fig:contours_pede}. From \ref{fig:cosmo_pede} we observe a subtle difference when compared with $\Omega_k$-$\Lambda$CDM. For example, the deceleration parameter indicates that the transition is consistent ($z_T=0.678$) (joint), compared with $\Omega_k$-$\Lambda$CDM, with a similar trend seen for the $w_{eff}$ case. Notice also that, in this case, no deceleration is expected at $z=0$, in contrast to the VC model. Additionally, for this model, the Universe age is consistent with the $\Omega_k$-$\Lambda$CDM but a little bit younger.

\begin{figure*}
   \centering
   \includegraphics[width=0.42\textwidth]{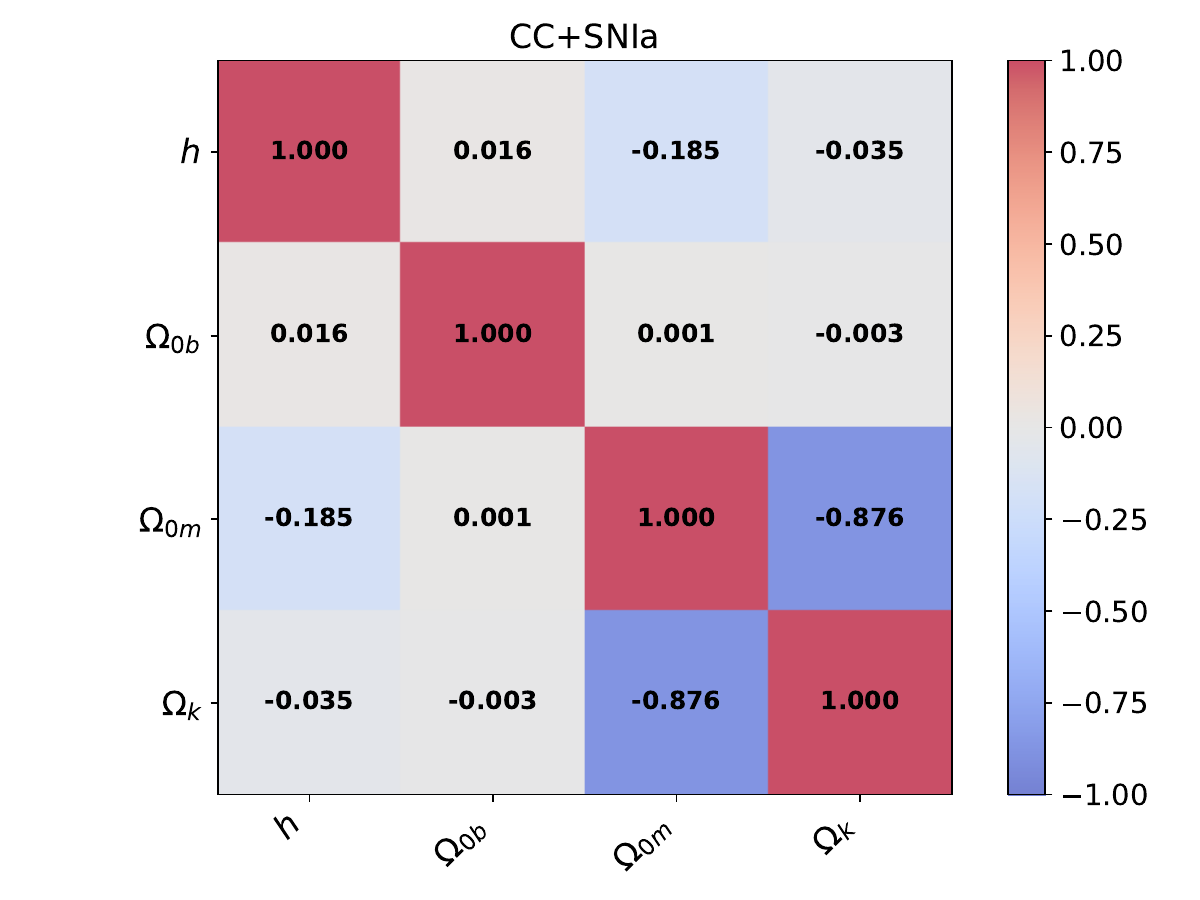}
   \includegraphics[width=0.42\textwidth]{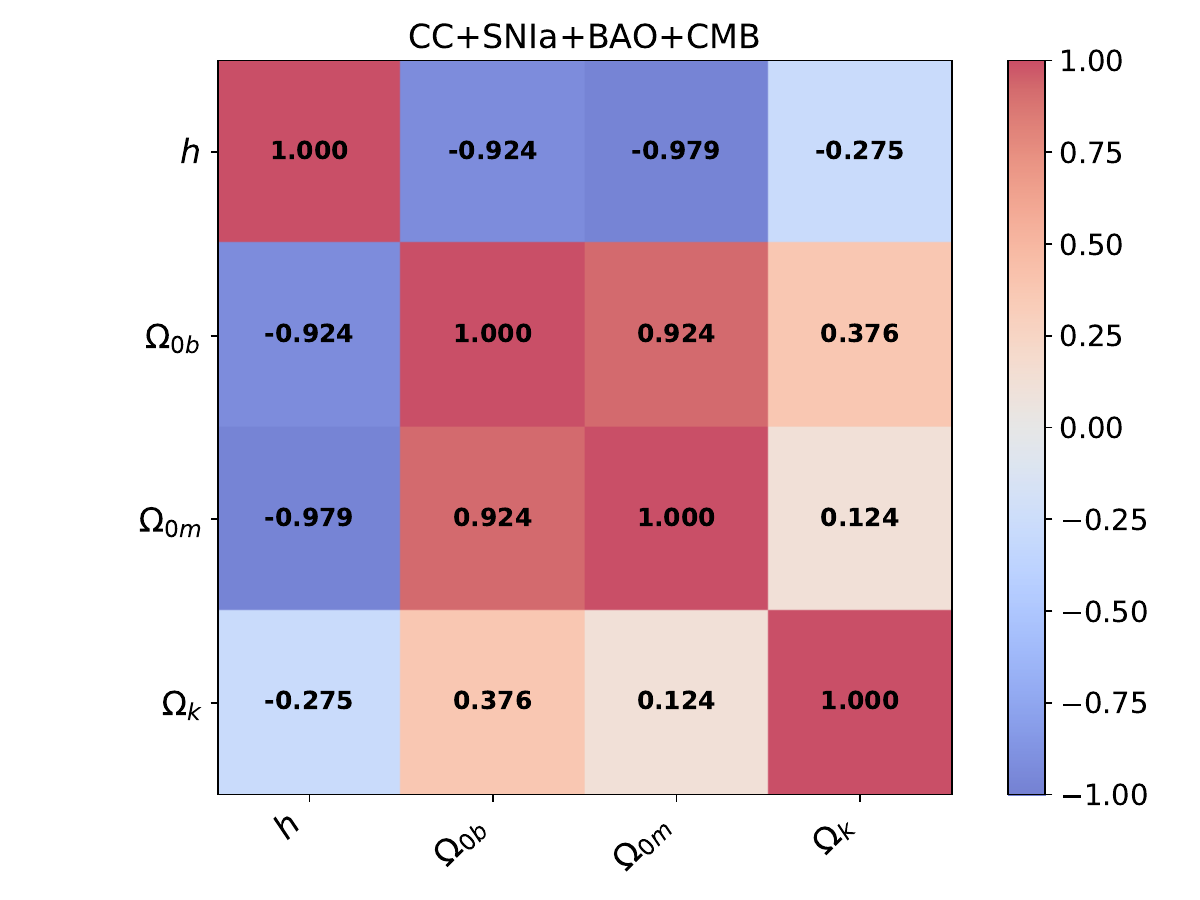}

   \caption{Correlation matrices for the parameters of the PEDE model for CC+SNIa (left panel) and CC+SNIa+BAO+CMB (right panel) datasets.}
   \label{fig:corr_pede}
\end{figure*}

Figure~\ref{fig:corr_pede} shows the correlation matrices of the cosmological parameters for the PEDE model obtained from two different observational combinations. The left panel (CC+SNIa) reveals that most parameters are only weakly correlated, indicating that low-redshift probes provide largely independent constraints on the Hubble constant, baryon density, and spatial curvature. The only notable exception is the strong anti-correlation between $\Omega_{0m}$ and $\Omega_k$ ($r=-0.876$), reflecting the well-known geometric degeneracy whereby variations in the matter density can be partially compensated by changes in the curvature while preserving the low-redshift expansion history. The weak correlations involving $h$ and $\Omega_{0b}$ suggest that these parameters are only marginally affected by degeneracies within the low-redshift dataset. In contrast, the inclusion of BAO and CMB measurements (right panel) substantially modifies the correlation structure. The addition of high-redshift information introduces strong anti-correlations between $h$ and both $\Omega_{0m}$ ($r=-0.979$) and $\Omega_{0b}$ ($r=-0.924$), together with a strong positive correlation between $\Omega_{0m}$ and $\Omega_{0b}$ ($r=0.924$). These correlations arise because BAO and especially CMB observations tightly constrain characteristic physical scales, forcing simultaneous adjustments of the Hubble constant and matter densities to reproduce the observed acoustic scale. Meanwhile, the correlation of $\Omega_k$ with the remaining parameters becomes relatively weak, indicating that the combination of low- and high-redshift observations efficiently breaks the curvature degeneracy present in the CC+SNIa analysis. Overall, Fig.~\ref{fig:corr_pede} demonstrates that the addition of BAO and CMB data not only tightens the parameter constraints but also reshapes the covariance structure, transferring the dominant degeneracies from the matter--curvature sector to the interplay between the Hubble parameter and the matter densities.

\subsection{GEDE}

\begin{figure*}
   \centering
   \includegraphics[width=0.6\textwidth]{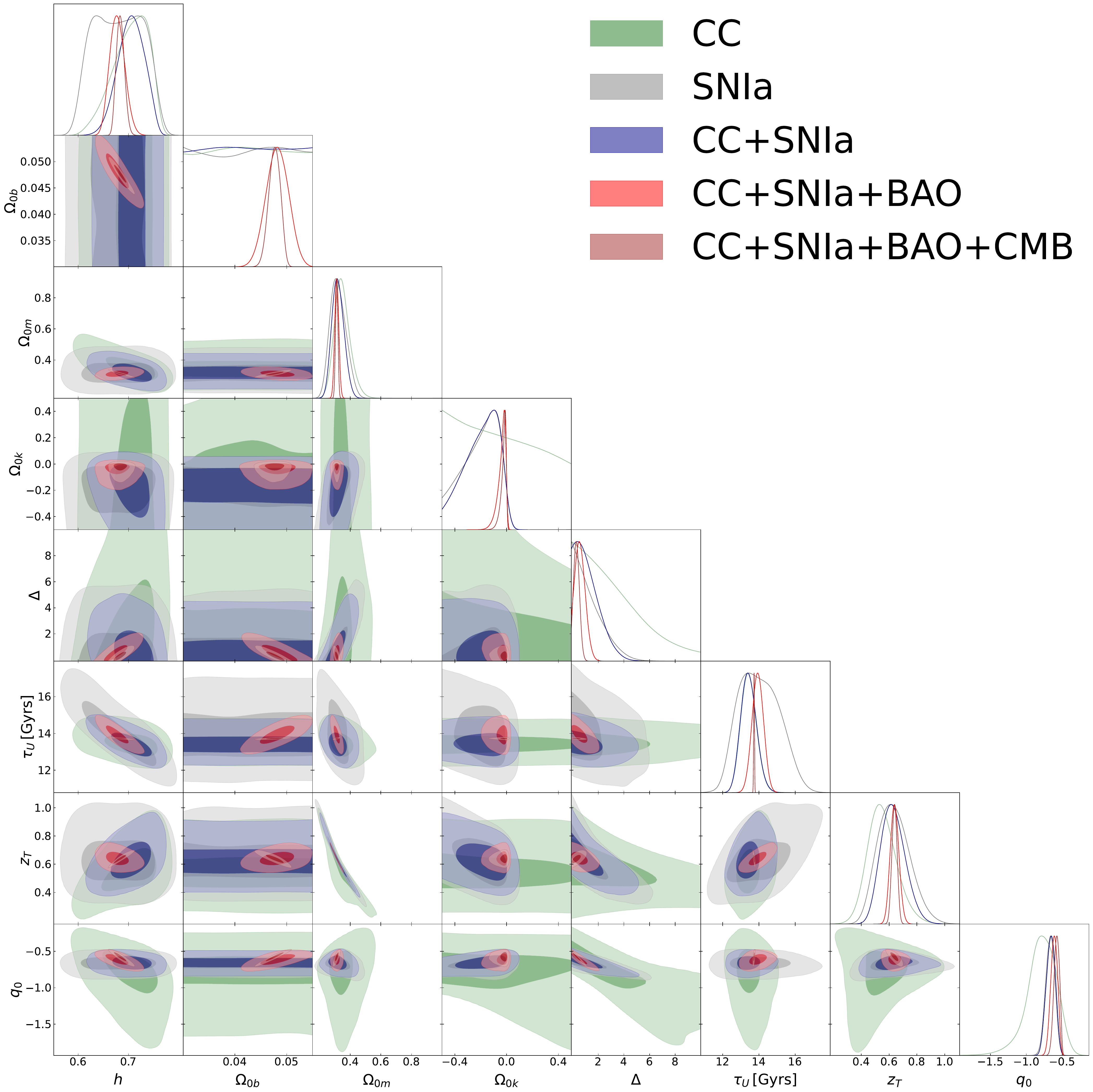}
   \caption{Posterior distributions for the GEDE model parameters $\boldsymbol{\Theta}=(h,\Omega_{0m},\Omega_k,\Delta)$, obtained from the MCMC analysis. Diagonal panels show the marginalized 1D posteriors, while off-diagonal panels display the joint 2D regions at $1\sigma$  and $3\sigma$ confidence levels.}
   \label{fig:contours_gede}
\end{figure*}

A natural extension of PEDE is the GEDE model \cite{Li:2020ybr,Hernandez-Almada:2020uyr}, adding $\Delta$ as an extra free parameter. Again, Table \ref{tab:bf_model} and \ref{fig:contours_gede} contain a summary of the MCMC results, together with their posterior distributions. In this case, a subtle difference is also observed when compared with the $\Omega_k$-$\Lambda$CDM. Among the most interesting features is that the transition to an accelerated Universe happens almost at the same time $q(z)=0.639$ (joint) and the effective EoS shows a tendency towards $-0.7$ at $z=0$, which is lower than the value predicted by the $\Omega_k$-$\Lambda$CDM. In contrast to VC model, there is no observed deceleration phase for $z<0.6$. The age of the Universe is consistent with $\Omega_k$-$\Lambda$CDM, but with a slightly higher value.

\begin{figure*}
   \centering
   \includegraphics[width=0.32\textwidth]{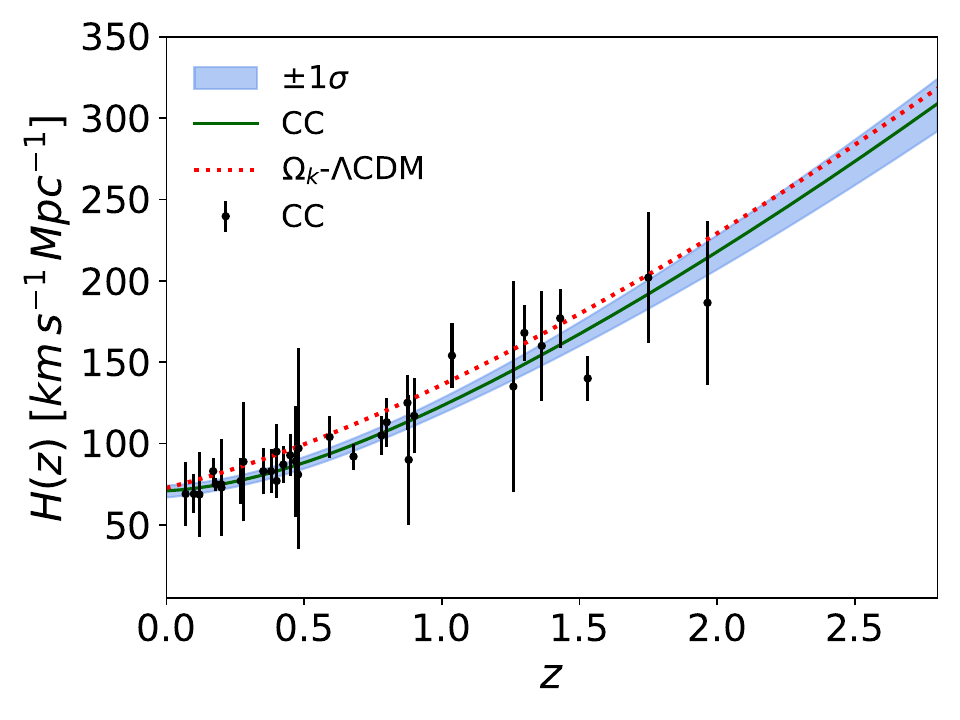}
   \includegraphics[width=0.32\textwidth]{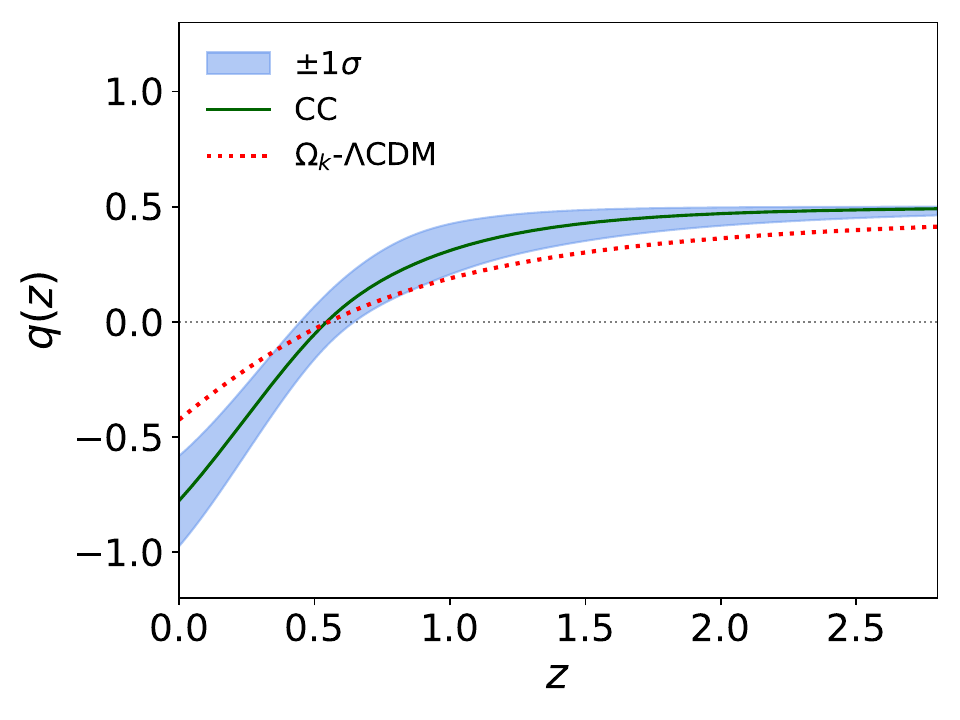}
   \includegraphics[width=0.32\textwidth]{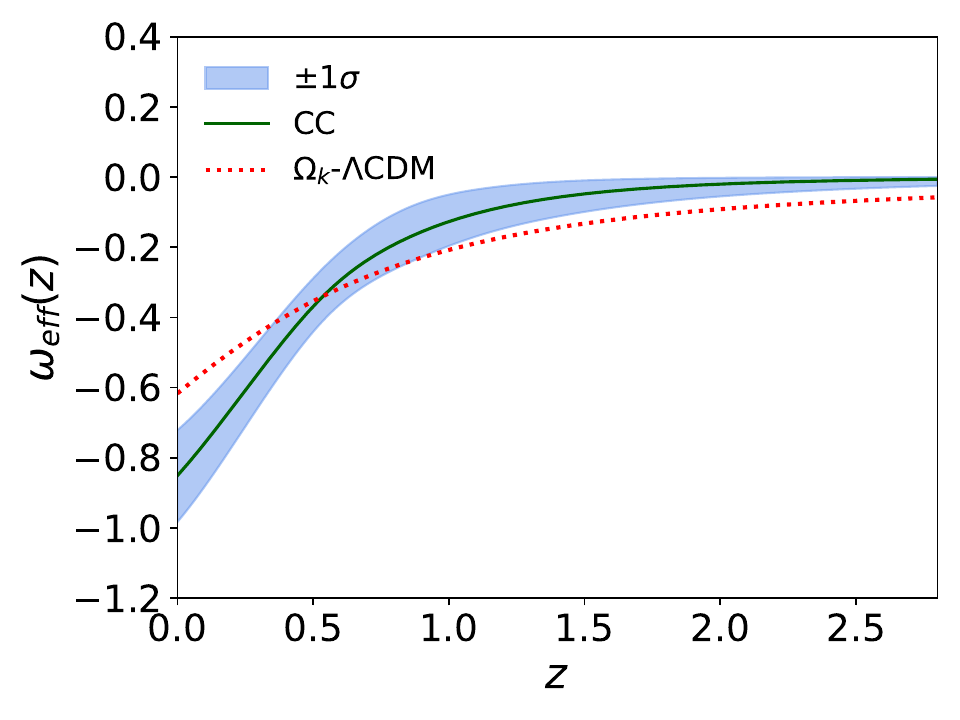}\\
   \includegraphics[width=0.32\textwidth]{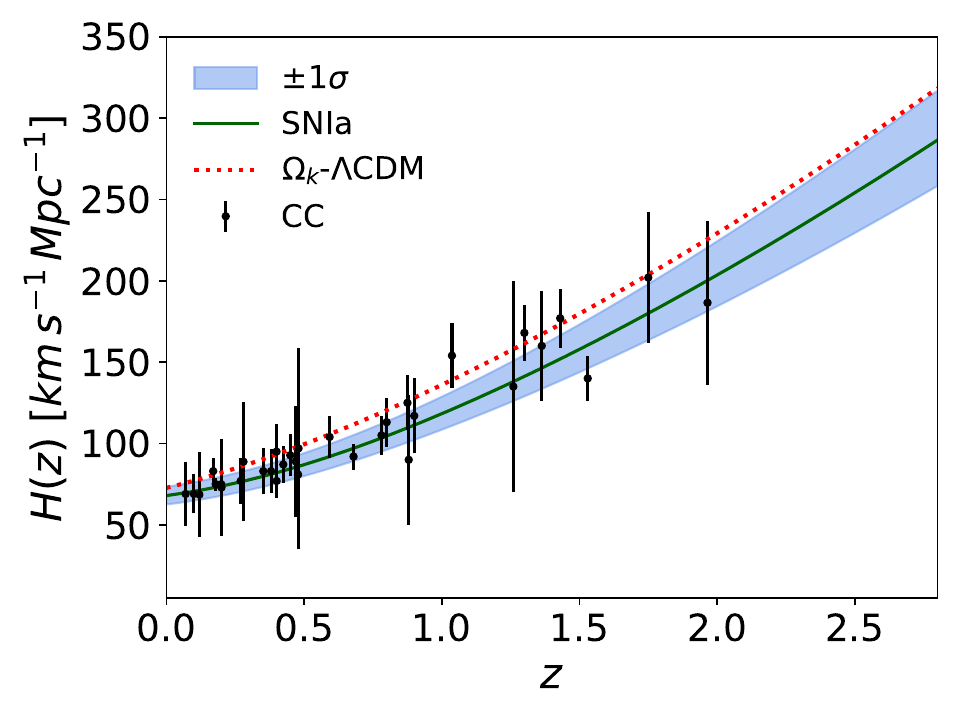}
   \includegraphics[width=0.32\textwidth]{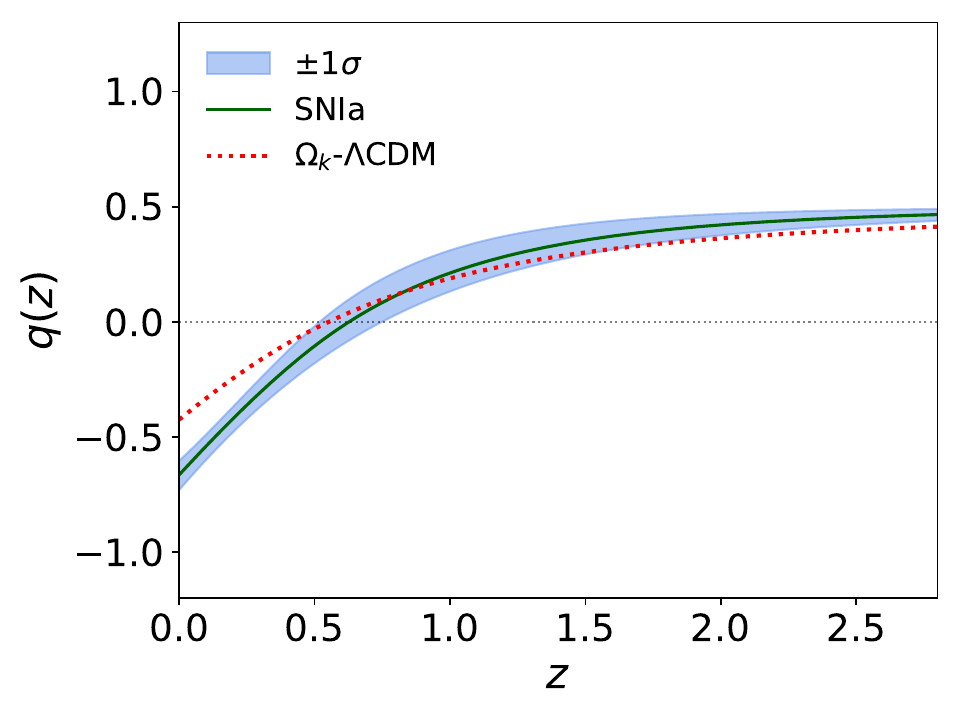}
   \includegraphics[width=0.32\textwidth]{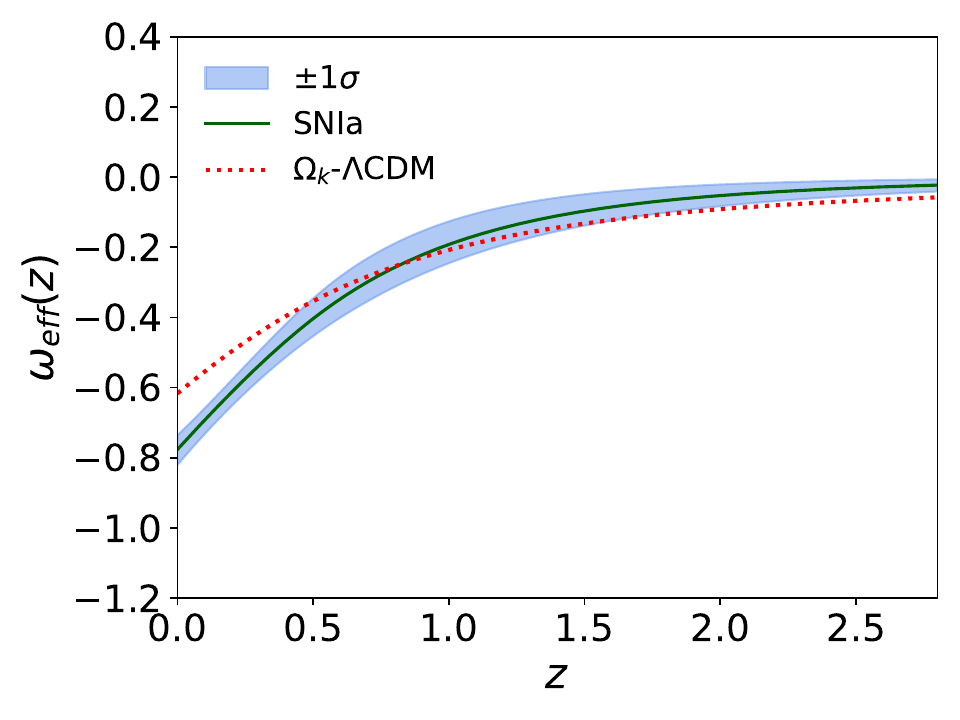} \\
   \includegraphics[width=0.32\textwidth]{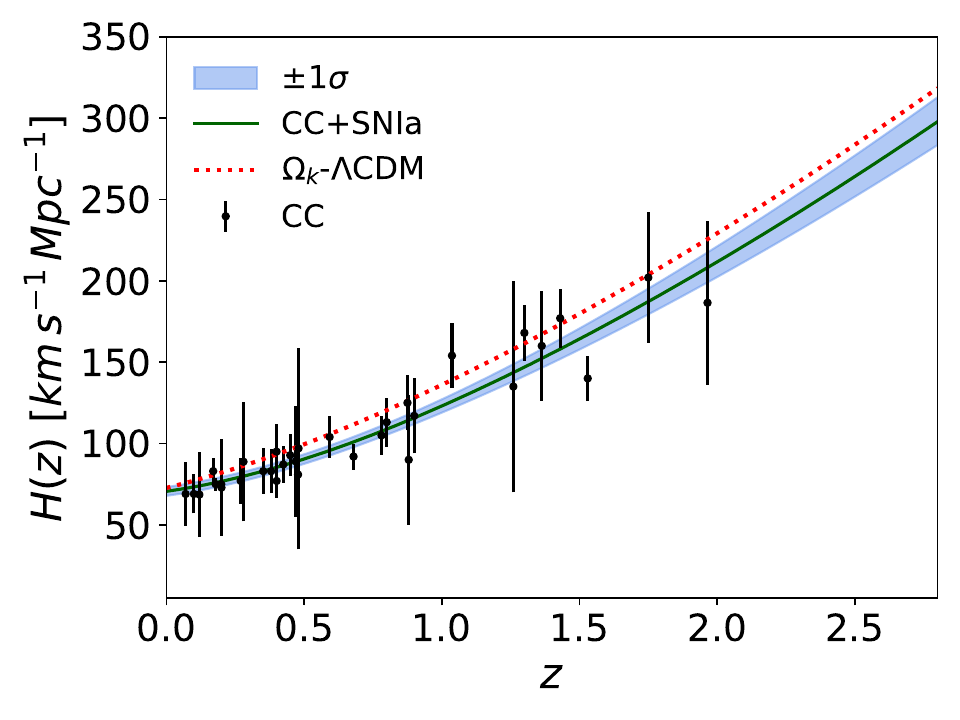}
   \includegraphics[width=0.32\textwidth]{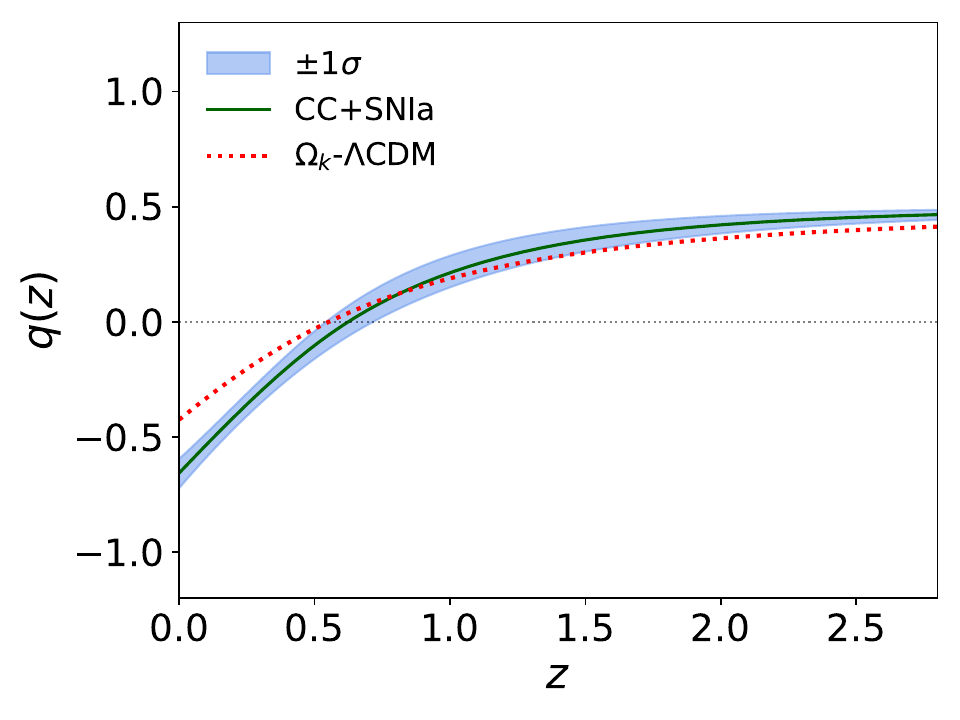}
   \includegraphics[width=0.32\textwidth]{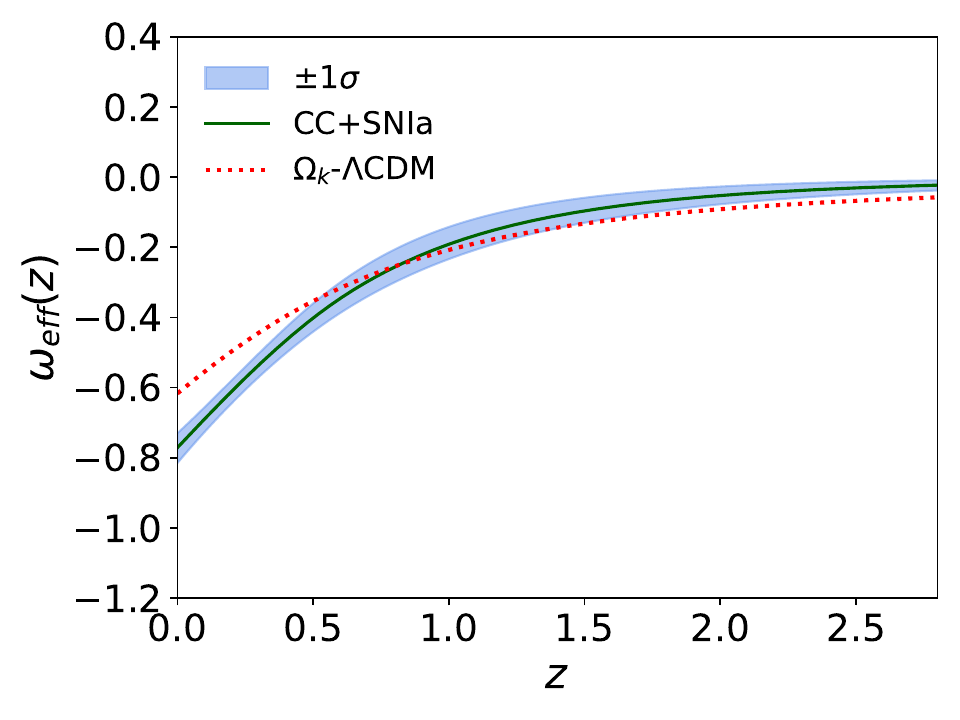} \\
   \includegraphics[width=0.32\textwidth]{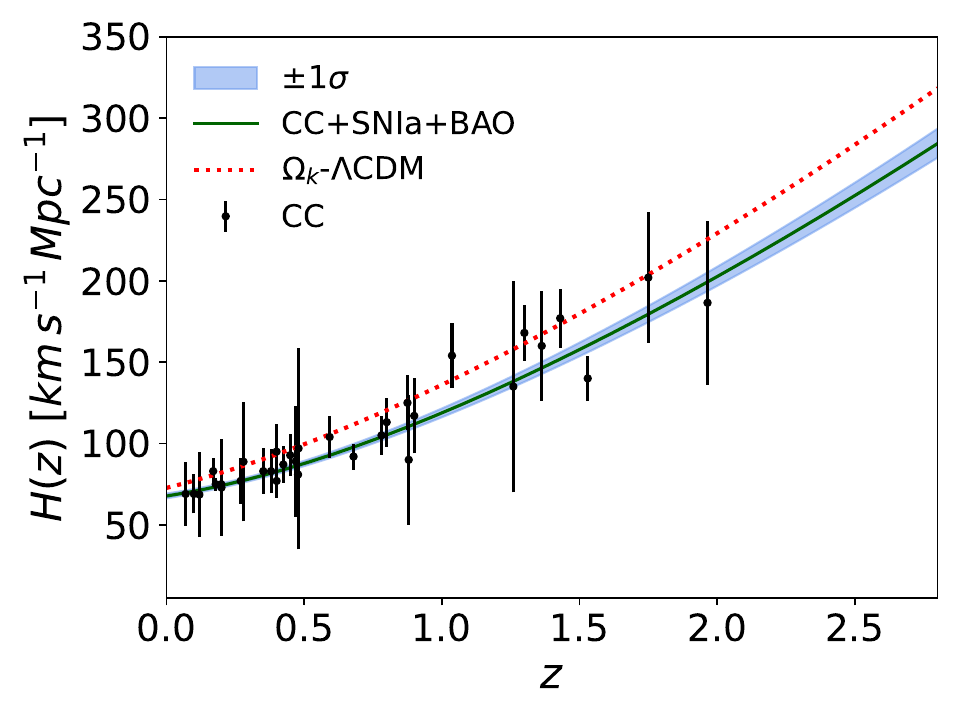}
   \includegraphics[width=0.32\textwidth]{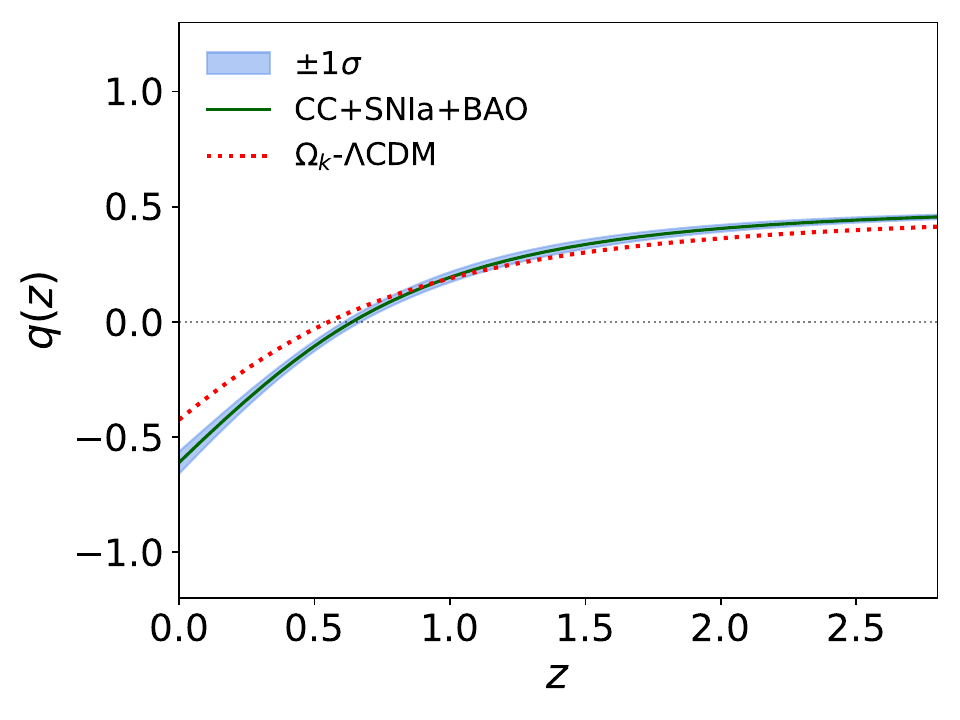}
   \includegraphics[width=0.32\textwidth]{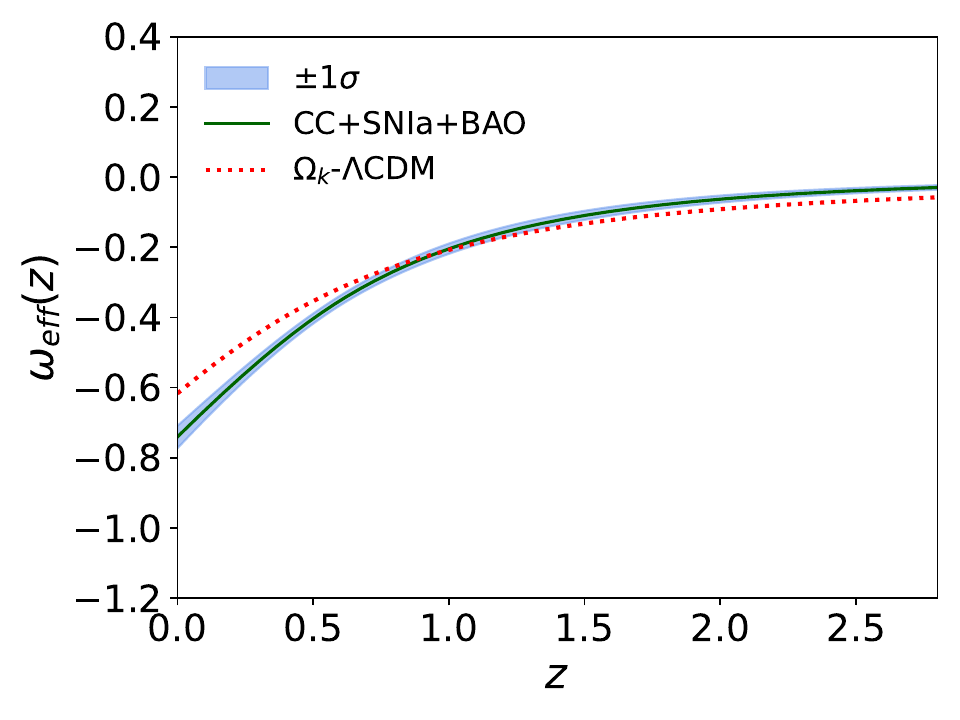} \\
   \includegraphics[width=0.32\textwidth]{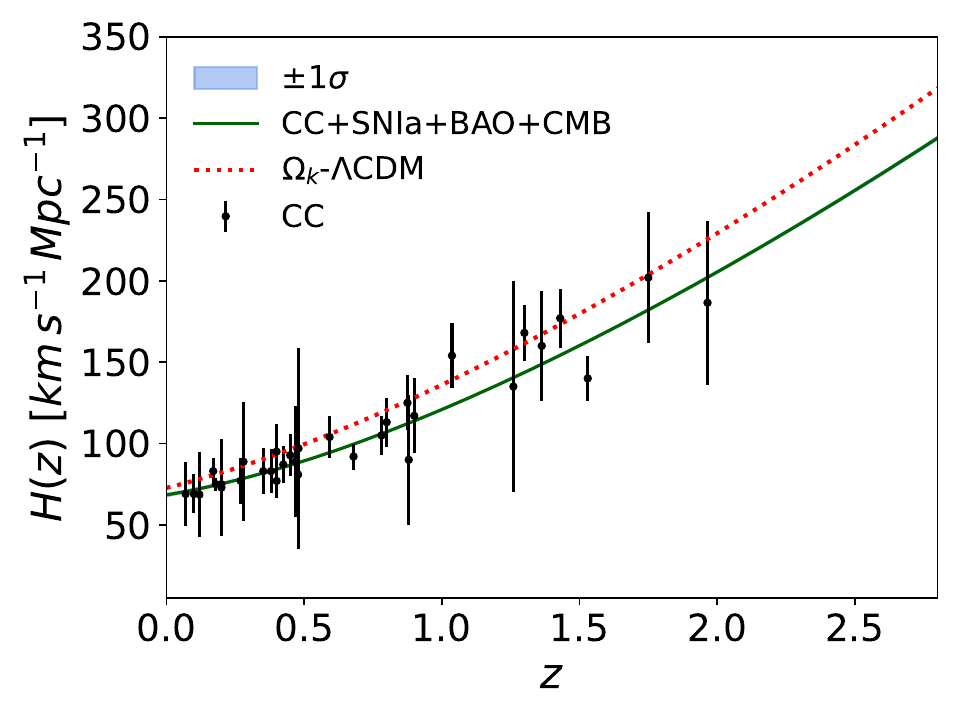}
   \includegraphics[width=0.32\textwidth]{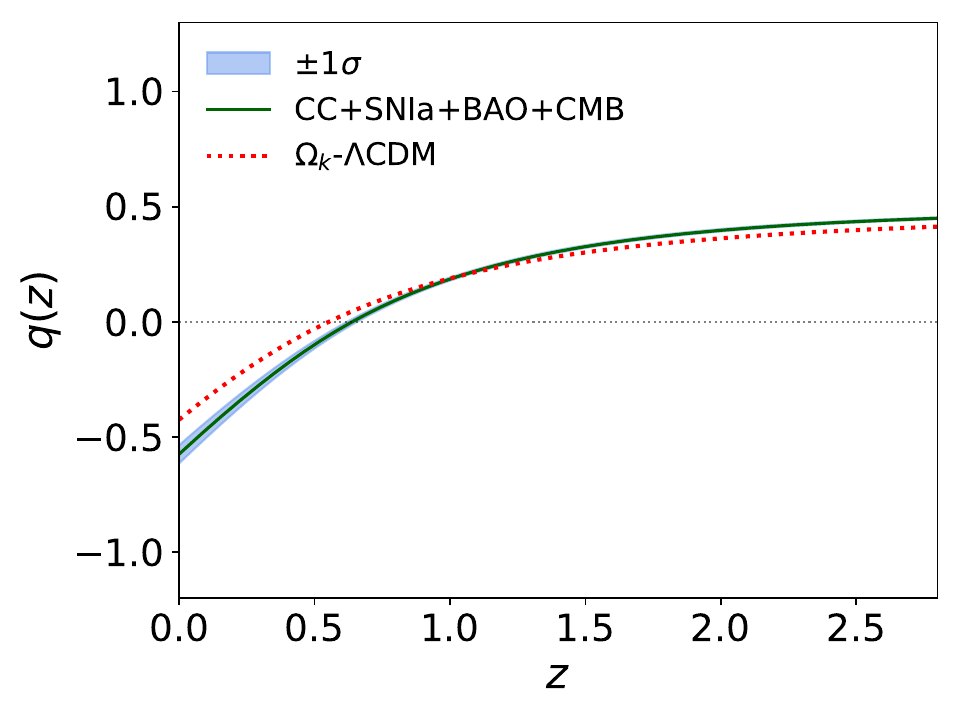}
   \includegraphics[width=0.32\textwidth]{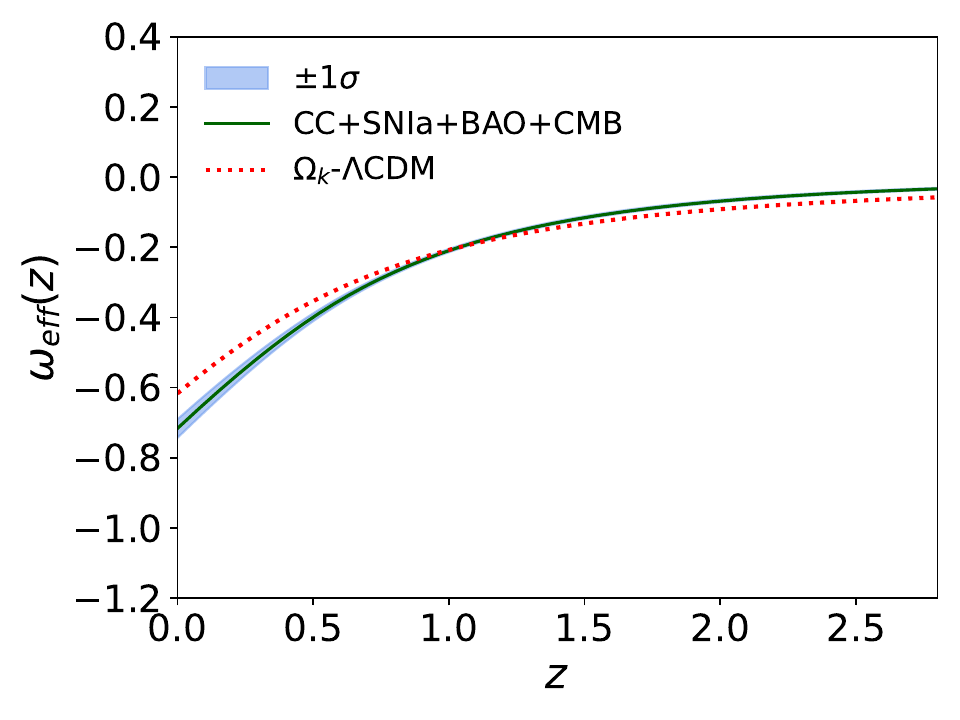}
   \caption{Reconstructions of the expansion history for the GEDE model using CC, SNIa, CC+SNIa, CC+SNIa+BAO, and CC+SNIa+BAO+CMB data combinations (from top to bottom). We show the Hubble parameter $H(z)$ (left panel), the deceleration parameter $q(z)$ (middle panel), and the effective EoS $\omega_{\rm eff}(z)$ (right panel). Solid paths represent the posterior median reconstructions and their $ 1\sigma$ intervals. The reference $\Omega_k$-$\Lambda$CDM prediction is drawn as a red dashed line.}
   \label{fig:cosmo_gede}
\end{figure*}

\begin{figure*}
   \centering
   \includegraphics[width=0.42\textwidth]{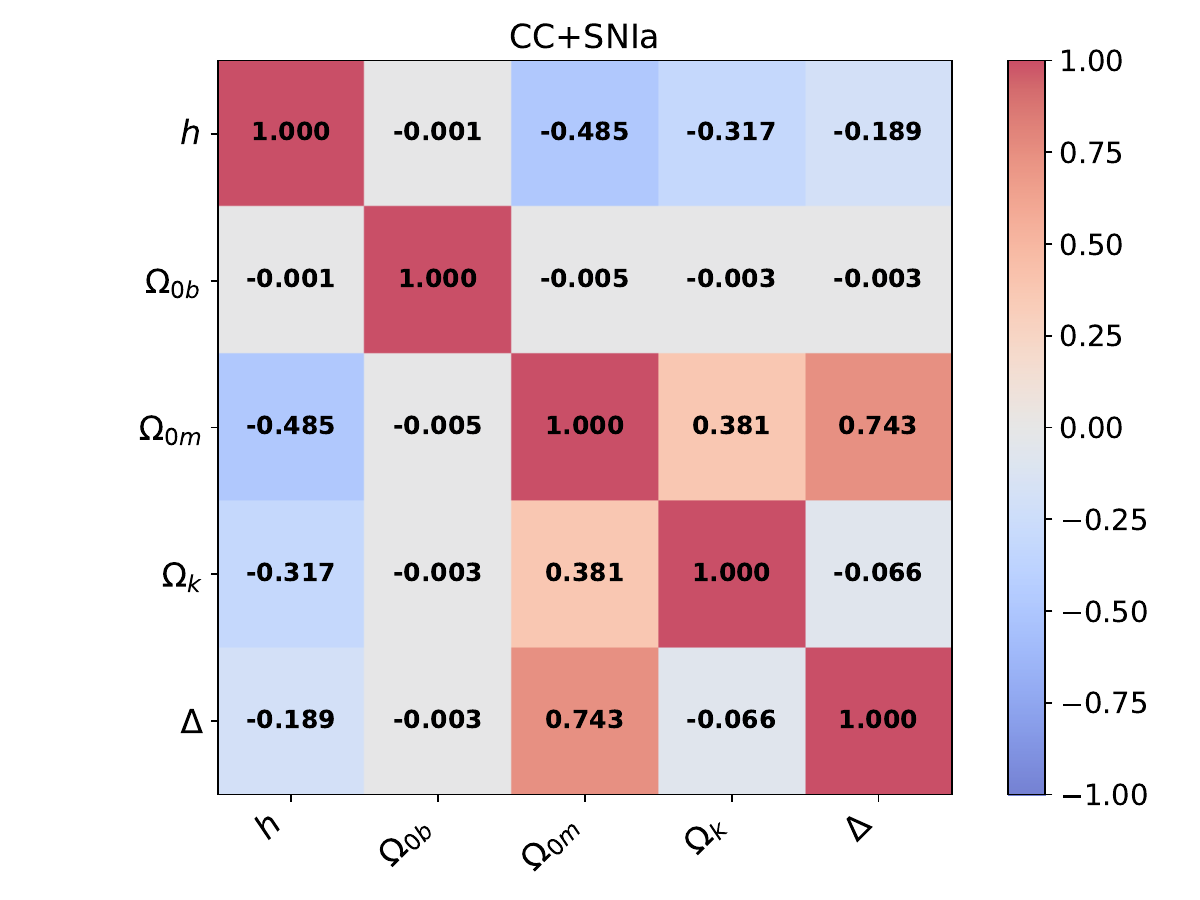}
   \includegraphics[width=0.42\textwidth]{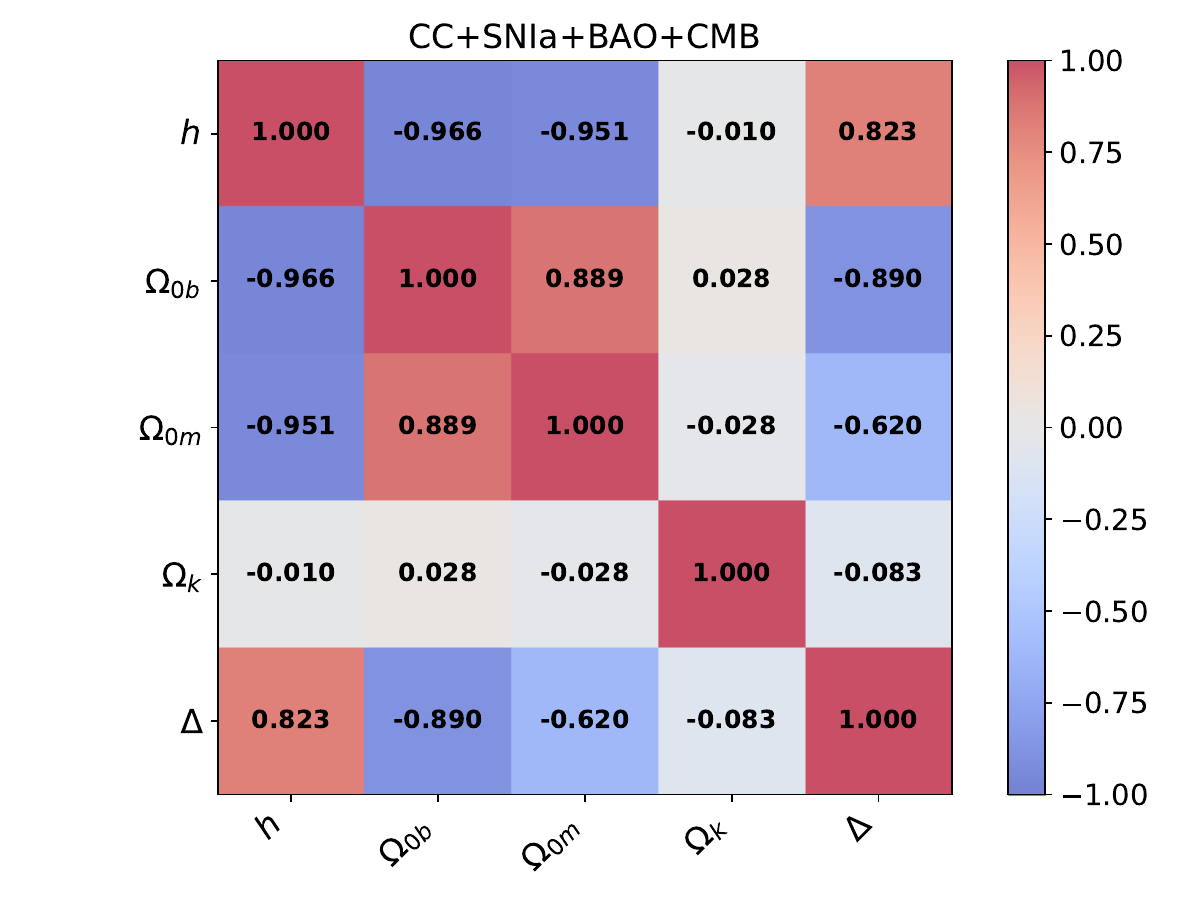}

   \caption{Correlation matrices for the parameters of the GEDE model for CC+SNIa (left panel) and CC+SNIa+BAO+CMB (right panel) datasets.}
   \label{fig:corr_gede}
\end{figure*}

Figure~\ref{fig:corr_gede} presents the correlation matrices for the GEDE model obtained from the low-redshift dataset (CC+SNIa, left panel) and from the full combination including BAO and CMB observations (right panel). For the low-redshift analysis, the baryon density parameter, $\Omega_{0b}$, is essentially uncorrelated with all remaining cosmological parameters, indicating that CC and SNIa data alone provide little sensitivity to the baryonic matter content. Instead, the dominant correlations involve the matter density, with a strong positive correlation between $\Omega_{0m}$ and the GEDE parameter $\Delta$ ($r=0.743$), suggesting that both quantities jointly determine the late-time expansion history, while moderate correlations are observed with $h$ and $\Omega_k$. The situation changes significantly when BAO and CMB measurements are included. High-redshift observations tightly constrain the physical baryon density through the acoustic scale, leading to the emergence of strong correlations between $\Omega_{0b}$ and the remaining cosmological parameters. In particular, $\Omega_{0b}$ becomes strongly anti-correlated with $h$ ($r=-0.966$) and $\Delta$ ($r=-0.890$), while exhibiting a strong positive correlation with $\Omega_{0m}$ ($r=0.889$). At the same time, the curvature parameter remains nearly independent of the other parameters, with correlation coefficients close to zero, indicating that the combined dataset efficiently removes curvature-related degeneracies.

\subsection{GDE}

\begin{figure*}
   \centering
   \includegraphics[width=0.6\textwidth]{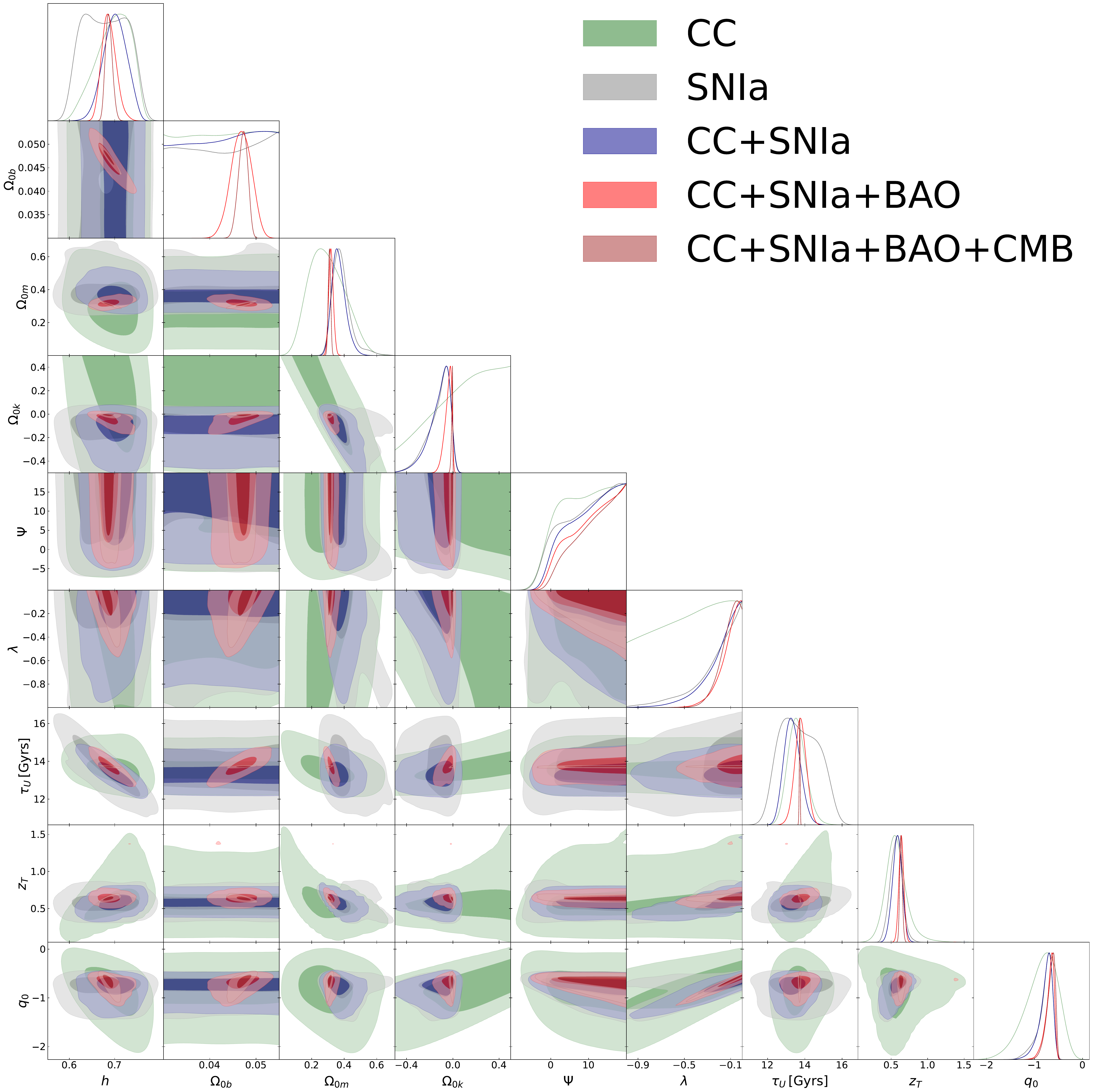}
   \caption{Posterior distributions for the GDE model parameters $\boldsymbol{\Theta}=(h,\Omega_{0m},\Omega_k,\Psi,\lambda)$, obtained from the MCMC analysis. Diagonal panels show the marginalized 1D posteriors, while off-diagonal panels display the joint 2D regions at $1\sigma$  and $3\sigma$ confidence levels.}
   \label{fig:contours_gde}
\end{figure*}

\begin{figure*}
   \centering
   \includegraphics[width=0.32\textwidth]{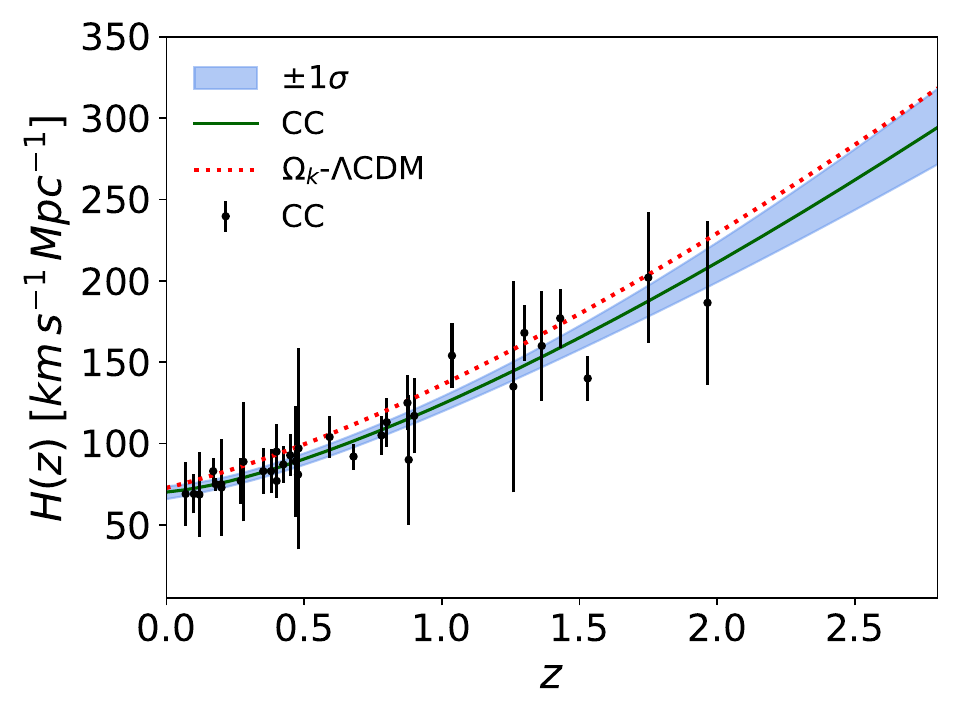}
   \includegraphics[width=0.32\textwidth]{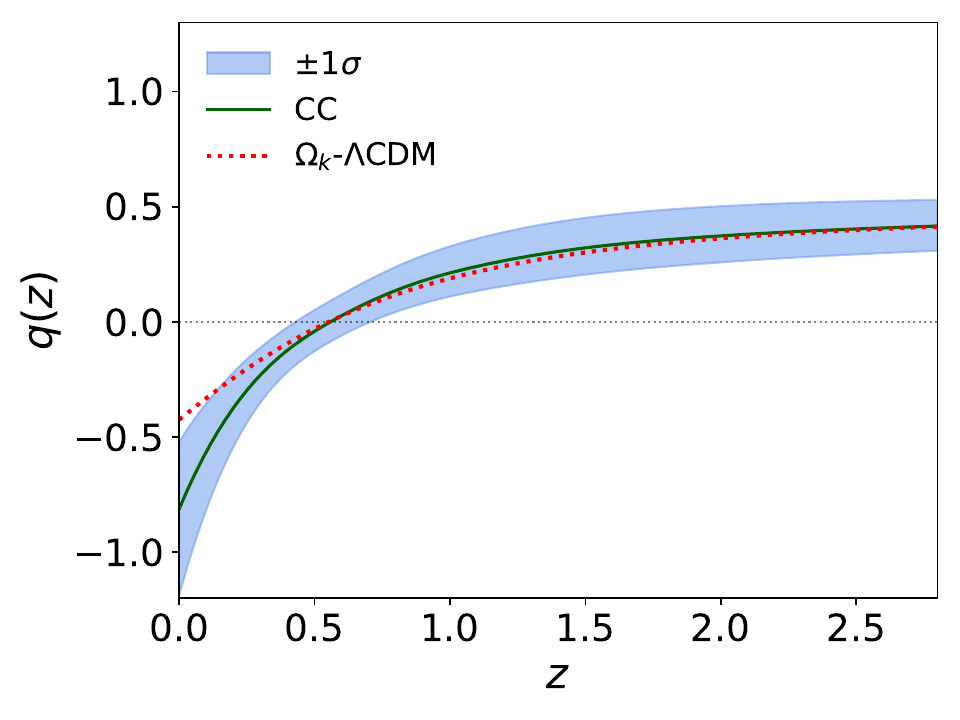}
   \includegraphics[width=0.32\textwidth]{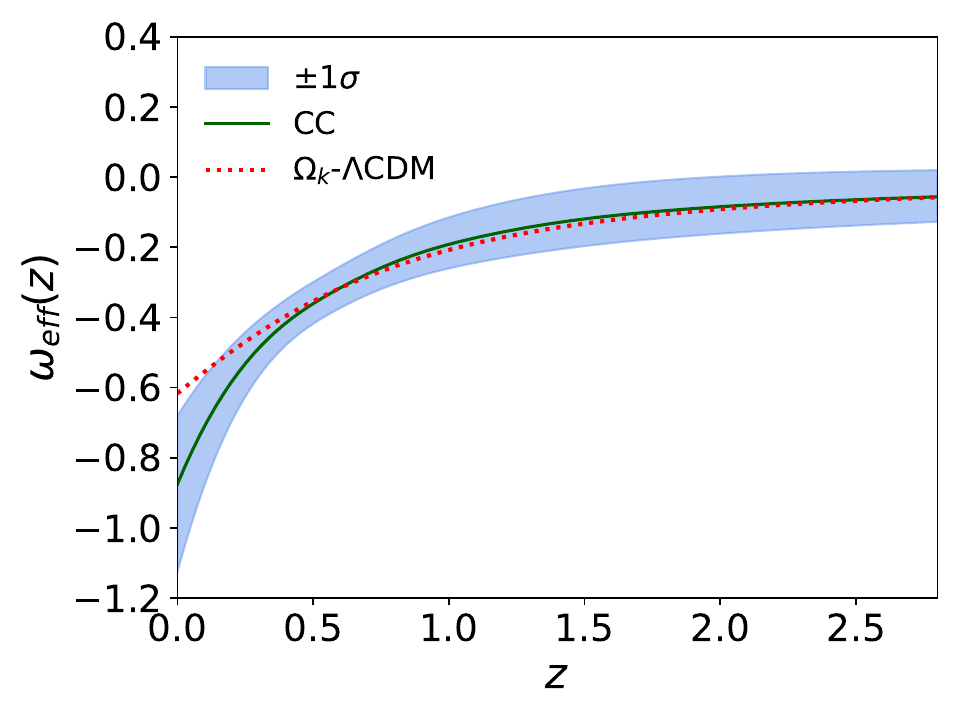}\\
   \includegraphics[width=0.32\textwidth]{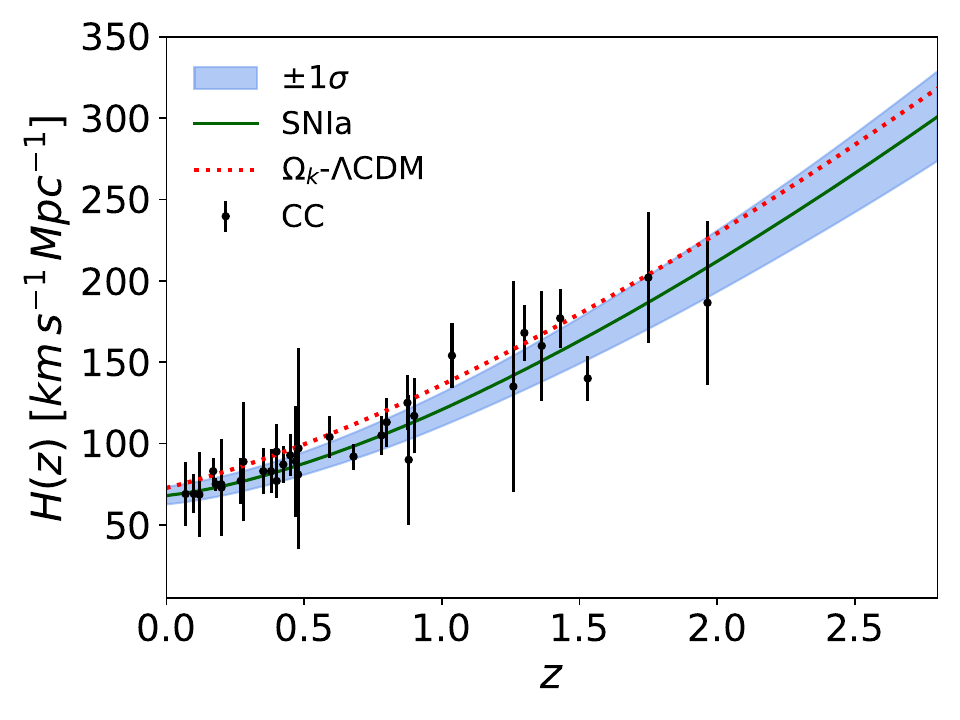}
   \includegraphics[width=0.32\textwidth]{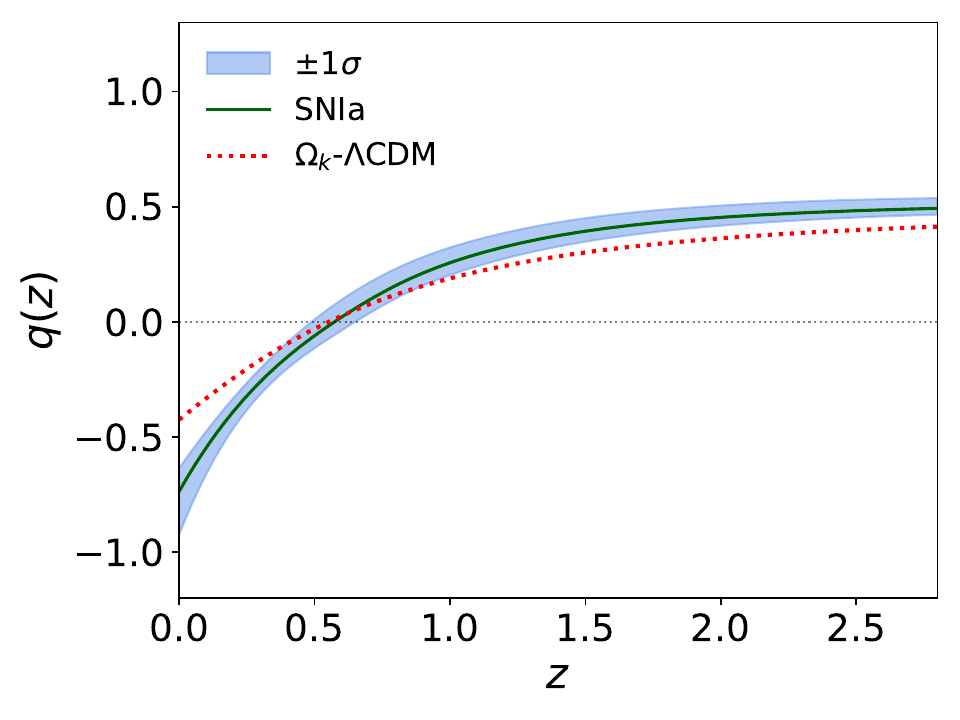}
   \includegraphics[width=0.32\textwidth]{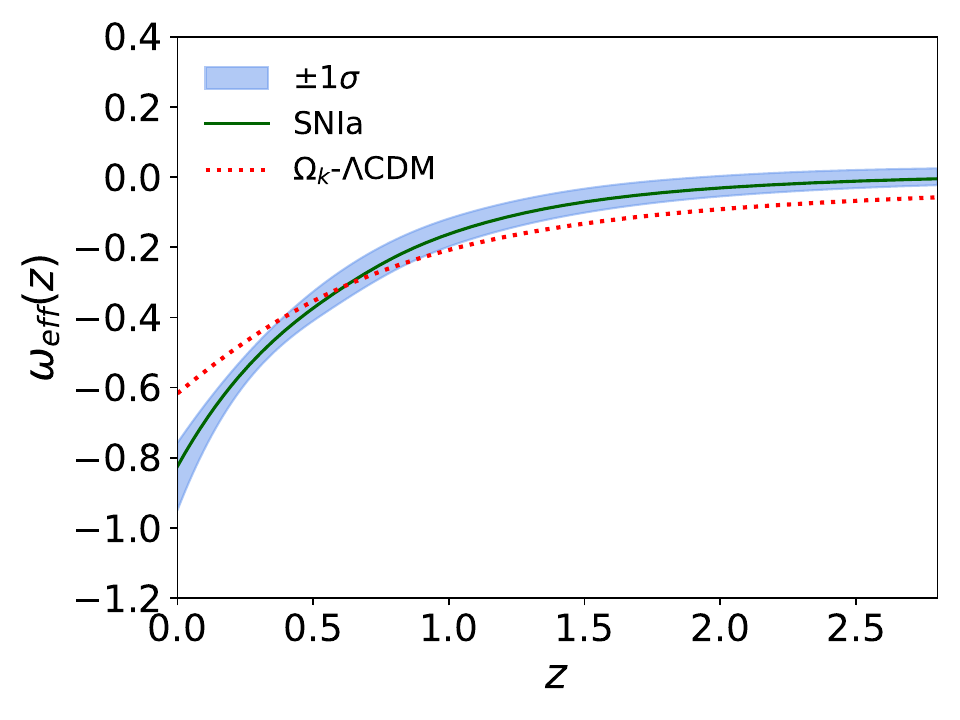} \\
   \includegraphics[width=0.32\textwidth]{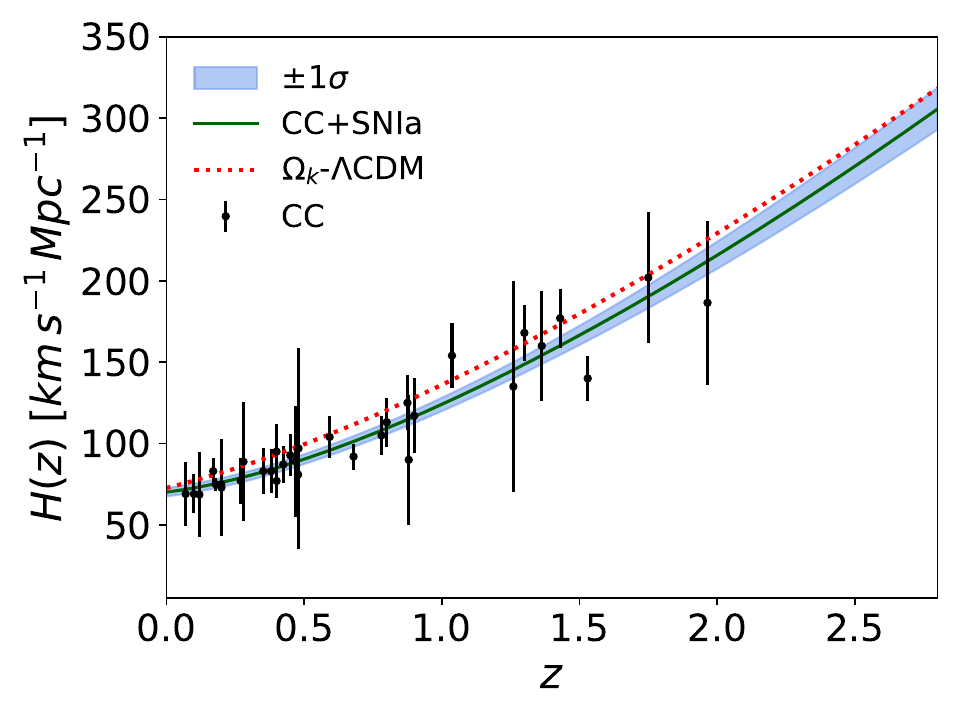}
   \includegraphics[width=0.32\textwidth]{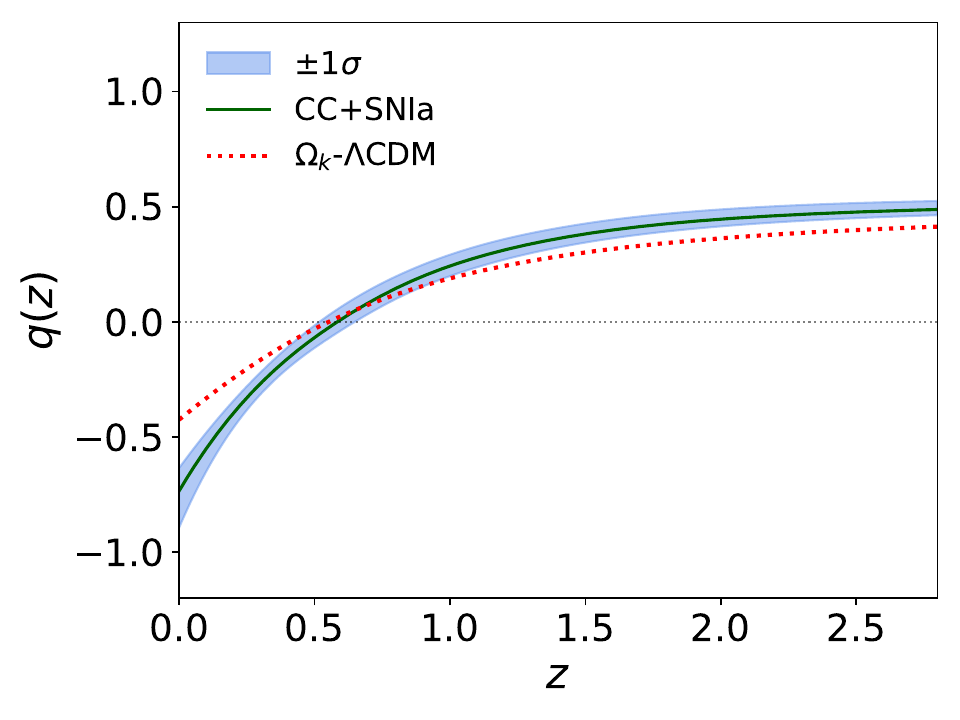}
   \includegraphics[width=0.32\textwidth]{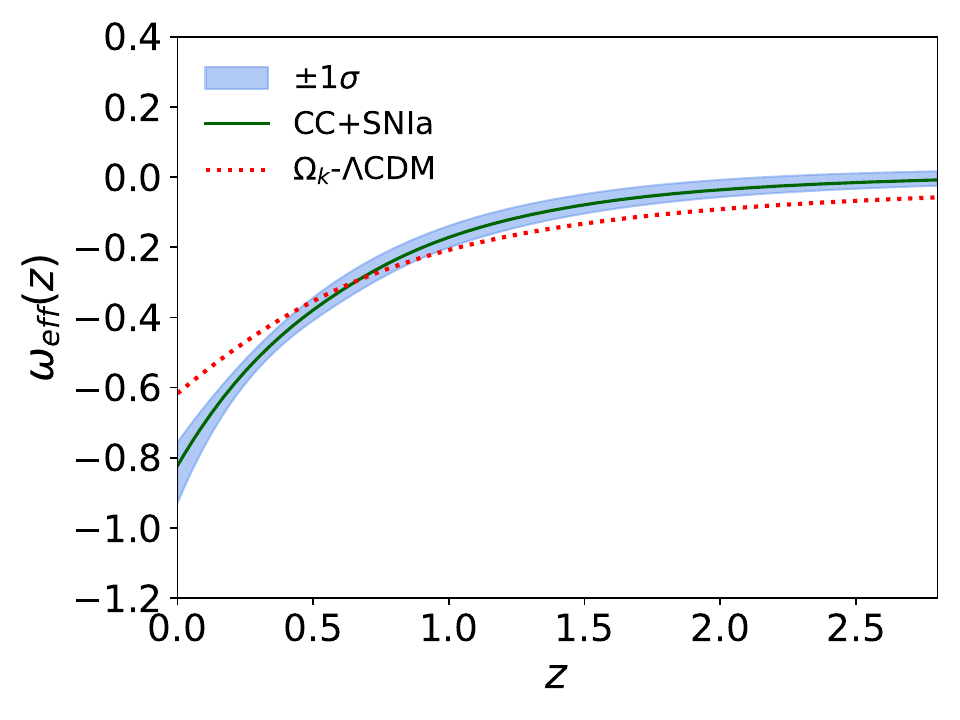} \\
   \includegraphics[width=0.32\textwidth]{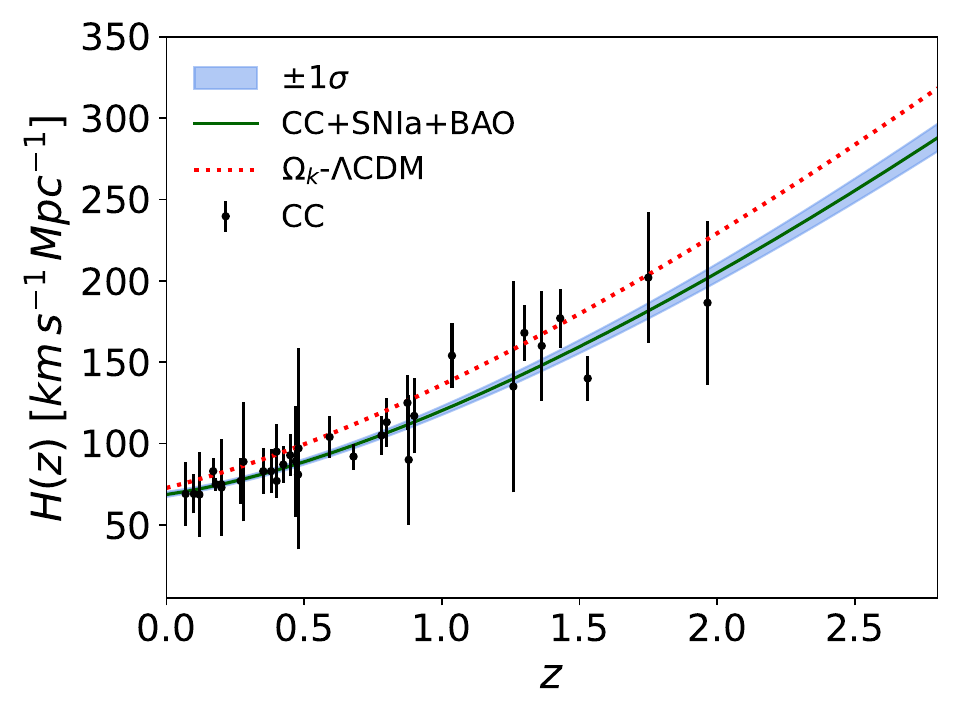}
   \includegraphics[width=0.32\textwidth]{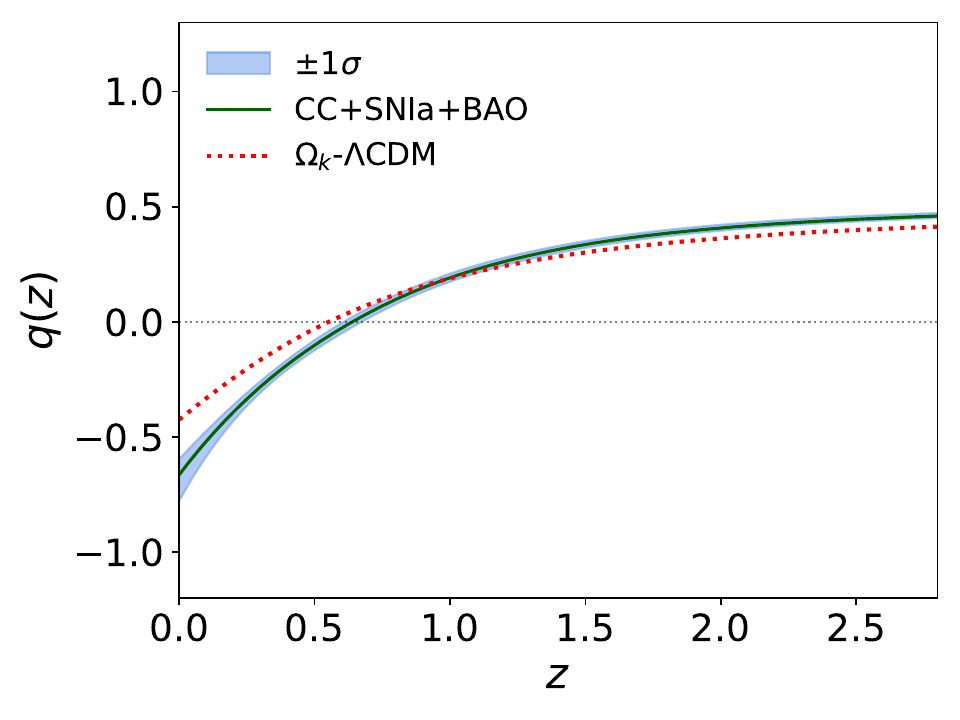}
   \includegraphics[width=0.32\textwidth]{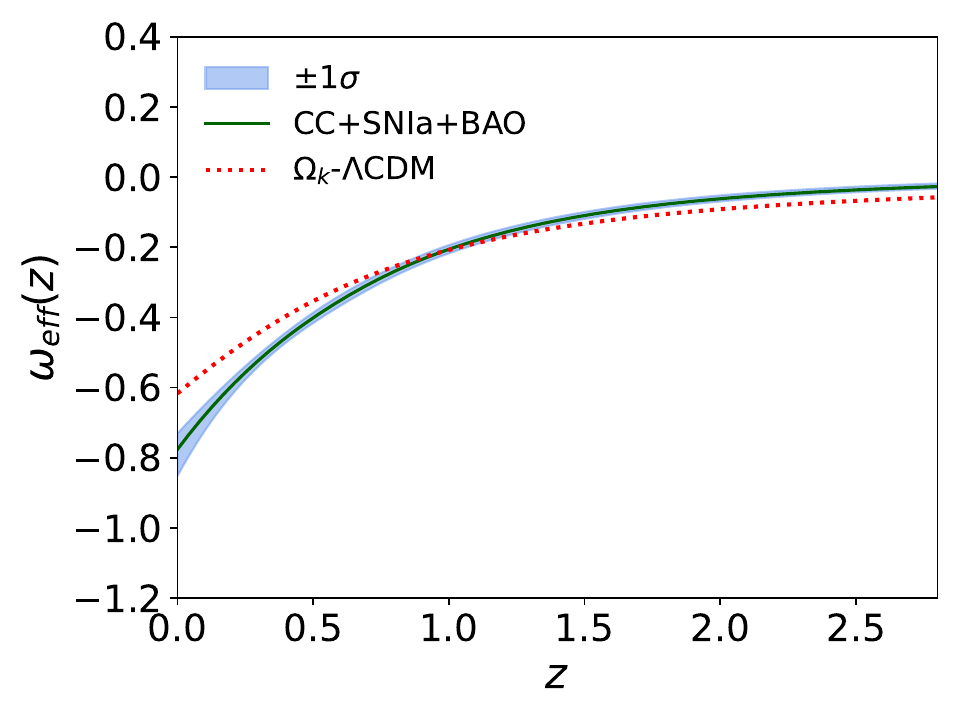} \\
   \includegraphics[width=0.32\textwidth]{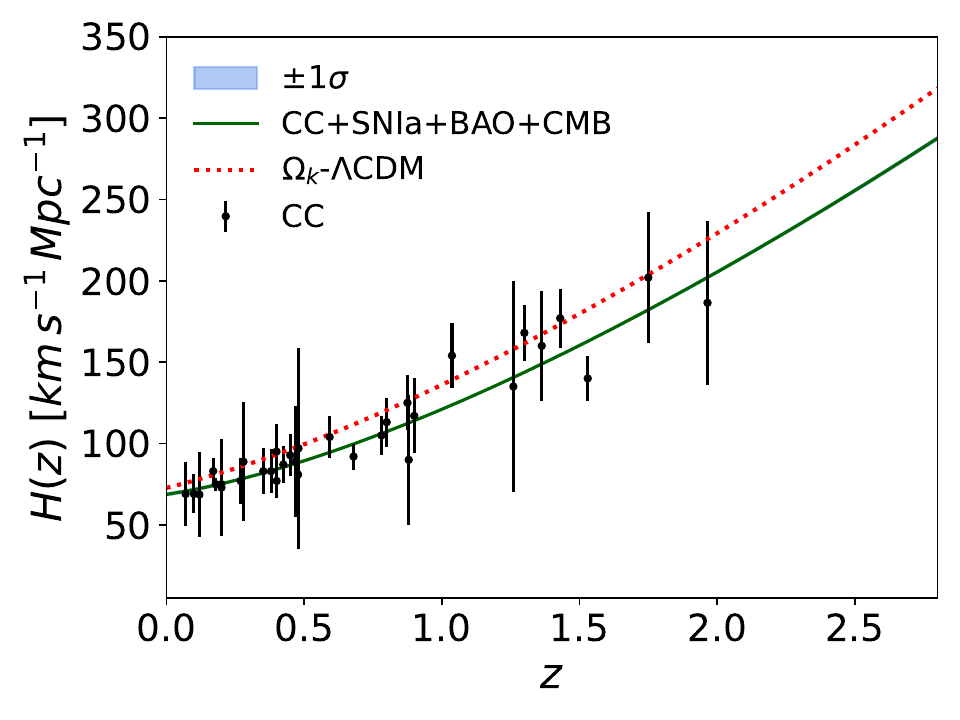}
   \includegraphics[width=0.32\textwidth]{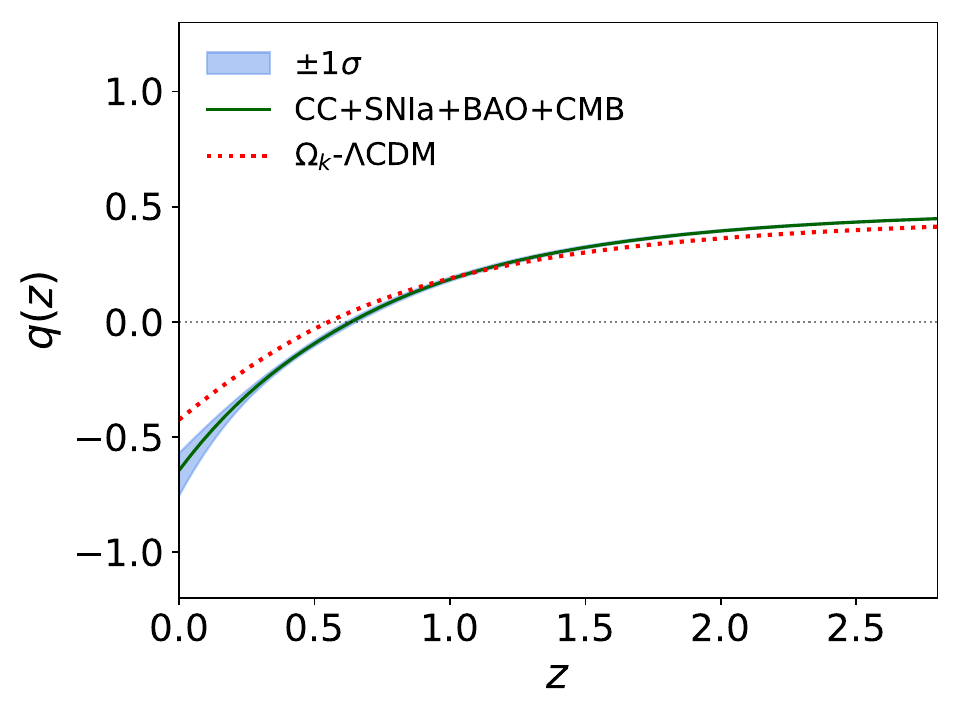}
   \includegraphics[width=0.32\textwidth]{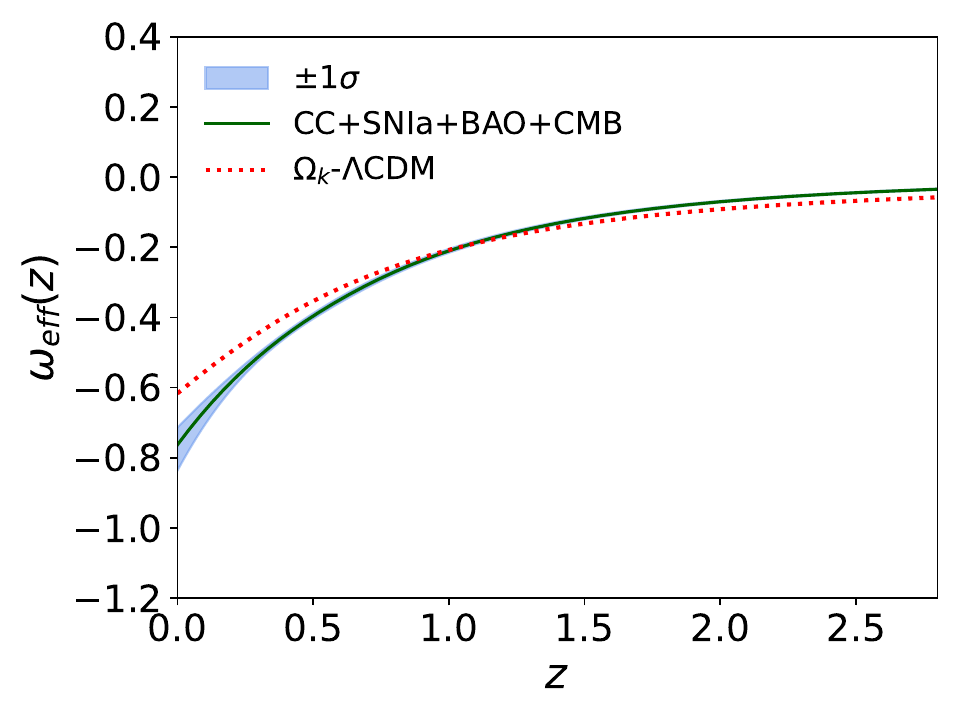}
   \caption{Reconstructions of the expansion history for the GDE model using CC, SNIa, CC+SNIa, CC+SNIa+BAO, and CC+SNIa+BAO+CMB data combinations (from top to bottom). We show the Hubble parameter $H(z)$ (left panel), the deceleration parameter $q(z)$ (middle panel), and the effective EoS $\omega_{\rm eff}(z)$ (right panel). Solid paths represent the posterior median reconstructions and their $ 1\sigma$ intervals. The reference $\Omega_k$-$\Lambda$CDM prediction is drawn as a red dashed line.}
   \label{fig:cosmo_gde}
\end{figure*}

Finally, the GDE model is another proposition where the ${\rm sgn}(1+x)(1+x)^{\alpha}$ function acts like the $\tanh(x)$ function, having a late time DE behavior. The main idea behind this model is that the cosmological constant transitions between a positive lambda ($+\Lambda$) to a negative value of lambda ($-\Lambda$), i.e., De Sitter$\to$ anti-De Sitter space-time. Our MCMC results are summarized in Table \ref{tab:bf_model} and \ref{fig:contours_gde}, with the cosmography shown in Figs. \ref{fig:cosmo_gde}, where again, we observe the non-tension when it is compared with the $\Omega_k$-$\Lambda$CDM. It shows an accelerated phase $z_T=0.635$ (joint) consistent with $\Omega_k$-$\Lambda$CDM and $w_{eff}$, having a tendency to $\approx-0.8$ at $z=0$. Notice also that the model does not exhibit a deceleration phase, even at $z=0$, in contrast to the VC model. The age of the Universe is consistent with $\Omega_k$-$\Lambda$CDM.

\begin{figure*}
   \centering
   \includegraphics[width=0.42\textwidth]{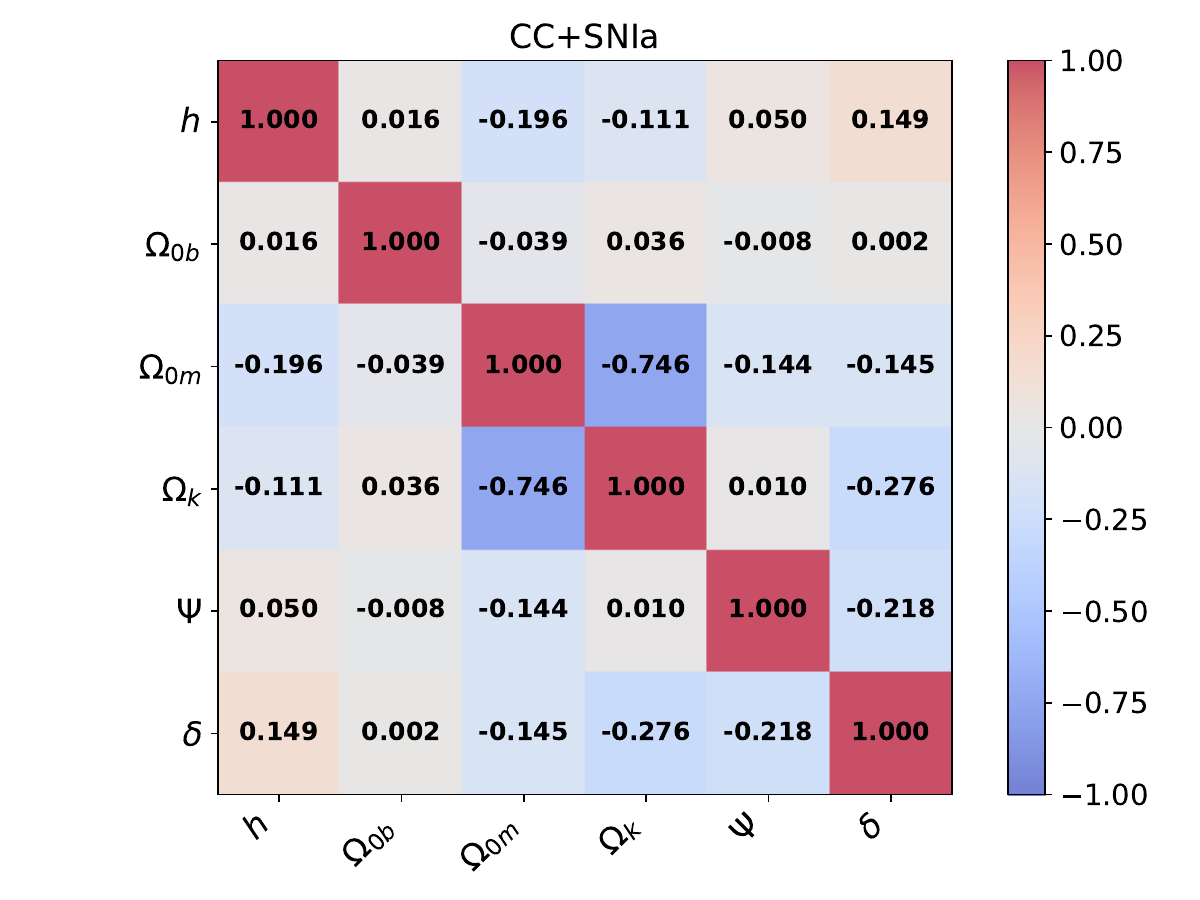}
   \includegraphics[width=0.42\textwidth]{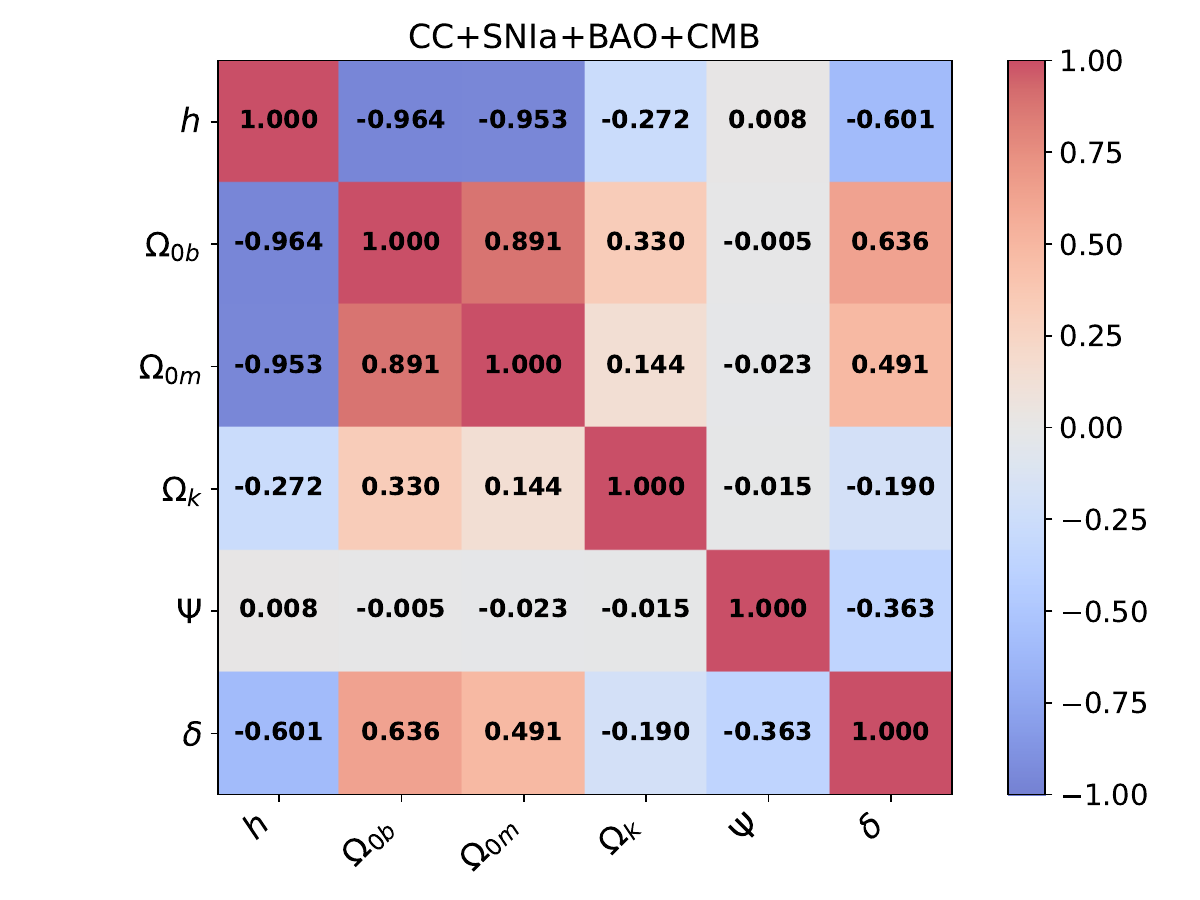}

   \caption{Correlation matrices for the parameters of the GDE model for CC+SNIa (left panel) and CC+SNIa+BAO+CMB (right panel) datasets.}
   \label{fig:corr_gde}
\end{figure*}

Figure~\ref{fig:corr_gde} displays the correlation matrices for the GDE model obtained from the low-redshift dataset (CC+SNIa, left panel) and from the combined CC+SNIa+BAO+CMB analysis (right panel). Using only low-redshift observations, most cosmological parameters exhibit weak correlations, with the exception of the well-known matter--curvature degeneracy, reflected by the strong anti-correlation between $\Omega_{0m}$ and $\Omega_k$ ($r=-0.746$). As in the GEDE model, the baryon density parameter $\Omega_{0b}$ remains essentially unconstrained, showing negligible correlations with all other parameters. Likewise, the GDE-specific parameters $\Psi$ and $\delta$ display only weak to moderate correlations, indicating that the low-redshift expansion history alone is insufficient to tightly constrain the full parameter space of the model. The inclusion of BAO and CMB observations dramatically changes the covariance structure. Strong anti-correlations emerge between $h$ and both $\Omega_{0b}$ ($r=-0.964$) and $\Omega_{0m}$ ($r=-0.953$), while $\Omega_{0b}$ becomes strongly positively correlated with $\Omega_{0m}$ ($r=0.891$), reflecting the joint constraints imposed by the acoustic scale measured at high redshift. In addition, the parameter $\delta$ develops significant correlations with $h$, $\Omega_{0b}$, and $\Omega_{0m}$, indicating that its determination is closely linked to the background expansion constrained by the combined datasets. In contrast, the parameter $\psi$ remains nearly uncorrelated with the standard cosmological parameters in both analyses, suggesting that it represents a largely independent degree of freedom within the current observational sensitivity. Overall, Fig.~\ref{fig:corr_gde} shows that BAO and CMB data are essential for breaking the degeneracies affecting the matter sector while simultaneously revealing the coupling between the GDE-specific parameter $\delta$ and the standard cosmological parameters, whereas $\Psi$ remains only weakly constrained by the available observations.

\subsection{Statistical Model Selection}

We employ the Akaike Information Criterion (AIC) \cite{AIC:1974, Sugiura:1978}, defined as
\begin{equation}
    \mathrm{AIC} \equiv \chi^2_{\rm min} + 2k,
\end{equation}
where $\chi^2_{\rm min}$ corresponds to the minimum chi-square value obtained from the fit and $k$ denotes the number of free fitted parameters of the model. The model preferred by the AIC is the one with the smallest value. To compare competing cosmological scenarios, it is common to define the relative quantity
\begin{equation}
    \Delta \mathrm{AIC} = \mathrm{AIC}_{i} - \mathrm{AIC}_{\rm ref},
\end{equation}
where the reference model is usually taken to be the standard $\Lambda$CDM cosmology. The interpretation of $\Delta$AIC is commonly summarized as follows:
\begin{itemize}
    \item $\Delta$AIC $< 4$: both models are statistically consistent with the data and receive comparable support.
    \item $4 < \Delta$AIC $< 10$: the candidate model is less supported by the data relative to the preferred model.
    \item $\Delta$AIC $> 10$: the candidate model is strongly disfavored by the observations.
\end{itemize}

In addition, we consider the Bayesian Information Criterion (BIC) \cite{schwarz1978}, given by
\begin{equation}
    \mathrm{BIC} \equiv \chi^2_{\rm min} + k \ln(N),
\end{equation}
where $N$ represents the total number of observational data points. In contrast to the AIC, the BIC introduces a stronger penalty for models with larger parameter spaces, especially for large datasets. Therefore, BIC generally favors simpler models. As in the AIC case, the preferred model corresponds to the one with the lowest BIC value. The relative difference is defined as
\begin{equation}
    \Delta \mathrm{BIC} = \mathrm{BIC}_{i} - \mathrm{BIC}_{\rm ref},
\end{equation}
and its interpretation is conventionally given by:
\begin{itemize}
    \item $\Delta$BIC $< 2$: no significant evidence against the candidate model.
    \item $2 < \Delta$BIC $< 6$: positive evidence against the candidate model.
    \item $6 < \Delta$BIC $< 10$: strong evidence against the candidate model.
    \item $\Delta$BIC $> 10$: very strong evidence against the candidate model.
\end{itemize}

\begin{table*}[ht]
\centering
\caption{Comparison of information criteria for different cosmological models. AIC and BIC values are shown, along with their differences relative to the $\Lambda$CDM model.}
\label{tab:model_selection}
\scriptsize
\begin{tabular}{lcccccccccccccccccccc}
\hline
&
\multicolumn{4}{c}{CC} &
\multicolumn{4}{c}{SNIa} &
\multicolumn{4}{c}{CC+SNIa} &
\multicolumn{4}{c}{CC+SNIa+BAO} &
\multicolumn{4}{c}{CC+SNIa+BAO+CMB} \\
\cline{2-21}
Model
& AIC & $\Delta$AIC & BIC & $\Delta$BIC
& AIC & $\Delta$AIC & BIC & $\Delta$BIC
& AIC & $\Delta$AIC & BIC & $\Delta$BIC
& AIC & $\Delta$AIC & BIC & $\Delta$BIC
& AIC & $\Delta$AIC & BIC & $\Delta$BIC \\
\hline

$\Lambda$CDM
& 20.6 & 0.0 & 25.0 & 0.0
& 69.4 & 0.0 & 74.5 & 0.0
& 78.6 & 0.0 & 85.4 & 0.0
& 90.9 & 0.0 & 98.2 & 0.0
& 92.0 & 0.0 & 99.4 & 0.0 \\

VC
& 29.6 & 9.0 & 38.4 & 13.4
& 71.4 & 2.0 & 81.6 & 7.1
& 87.1 & 8.5 & 100.8 & 15.4
& -- & -- & -- & --
& -- & -- & -- & -- \\

PEDE
& 20.5 & -0.1 & 24.9 & -0.1
& 66.2 & -3.2 & 71.2 & -3.3
& 76.1 & -2.5 & 82.9 & -2.5
& 92.0 & 1.1 & 99.3 & 1.1
& 103.7 & 11.7 & 111.0 & 11.6 \\

GEDE
& 22.9 & 2.3 & 28.7 & 3.7
& 56.9 & -12.5 & 63.6 & -10.9
& 71.3 & -7.3 & 80.4 & -5.0
& 89.9 & -1.0 & 99.6 & 1.4
& 94.1 & 2.1 & 103.9 & 4.5 \\

GDE
& 25.0 & 4.4 & 32.3 & 7.3
& 69.8 & 0.4 & 78.2 & 3.7
& 80.6 & 2.0 & 92.0 & 6.6
& 97.8 & 6.9 & 110.0 & 11.8
& 102.8 & 10.8 & 115.1 & 15.7 \\

\hline
\end{tabular}
\end{table*}
\clearpage

Table~\ref{tab:model_selection} summarizes the AIC and BIC for the cosmological models considered in this work. The results indicate that the relative performance of the models depends strongly on the observational dataset. For the low-redshift analyses, both PEDE and GEDE achieve lower AIC and BIC values than the $\Lambda$CDM model for several dataset combinations, indicating that the additional degrees of freedom are justified by the improved fit. In particular, GEDE provides the strongest preference for the SNIa and CC+SNIa datasets, while PEDE performs comparably well and is statistically indistinguishable from $\Lambda$CDM when only CC data are considered.

The situation changes when BAO and CMB observations are included. Although GEDE remains competitive for the CC+SNIa+BAO combination, the inclusion of CMB distance priors significantly favors the simpler $\Lambda$CDM model. In particular, the large values of $\Delta$AIC and $\Delta$BIC obtained for PEDE and GDE indicate that the improvement in the fit is insufficient to compensate for their additional free parameters. By contrast, GEDE exhibits only a modest increase in the information criteria, suggesting that it remains the most competitive extended model when the full dataset is considered. The VC model is consistently disfavored by the low-redshift datasets, especially according to the BIC, which imposes a stronger penalty on model complexity.

In general, the information criteria indicate that while several extended cosmological models can outperform $\Lambda$CDM when only low-redshift background observations are used, the inclusion of high-redshift information substantially reduces their statistical advantage. These results emphasize that the additional parameters introduced by extended dark-energy or curvature models must provide a significant improvement in the likelihood to overcome the penalty associated with increased model complexity. Future analyses based on the full CMB likelihood and perturbation-level observables may help determine whether these additional degrees of freedom remain statistically justified when a more complete description of the cosmological evolution is considered.

\begin{figure*}
   \centering
   \includegraphics[width=0.32\textwidth]{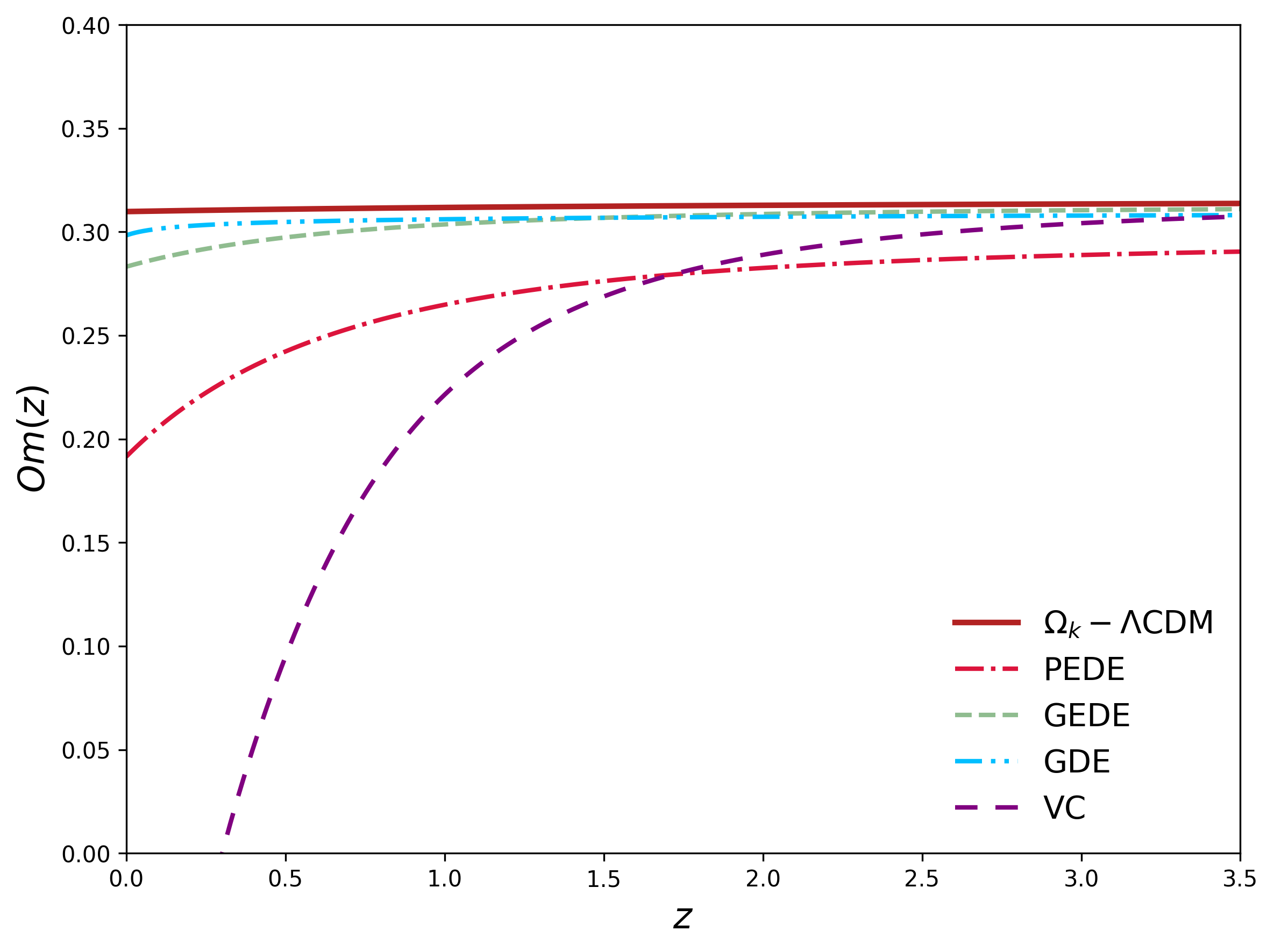}
   \includegraphics[width=0.32\textwidth]{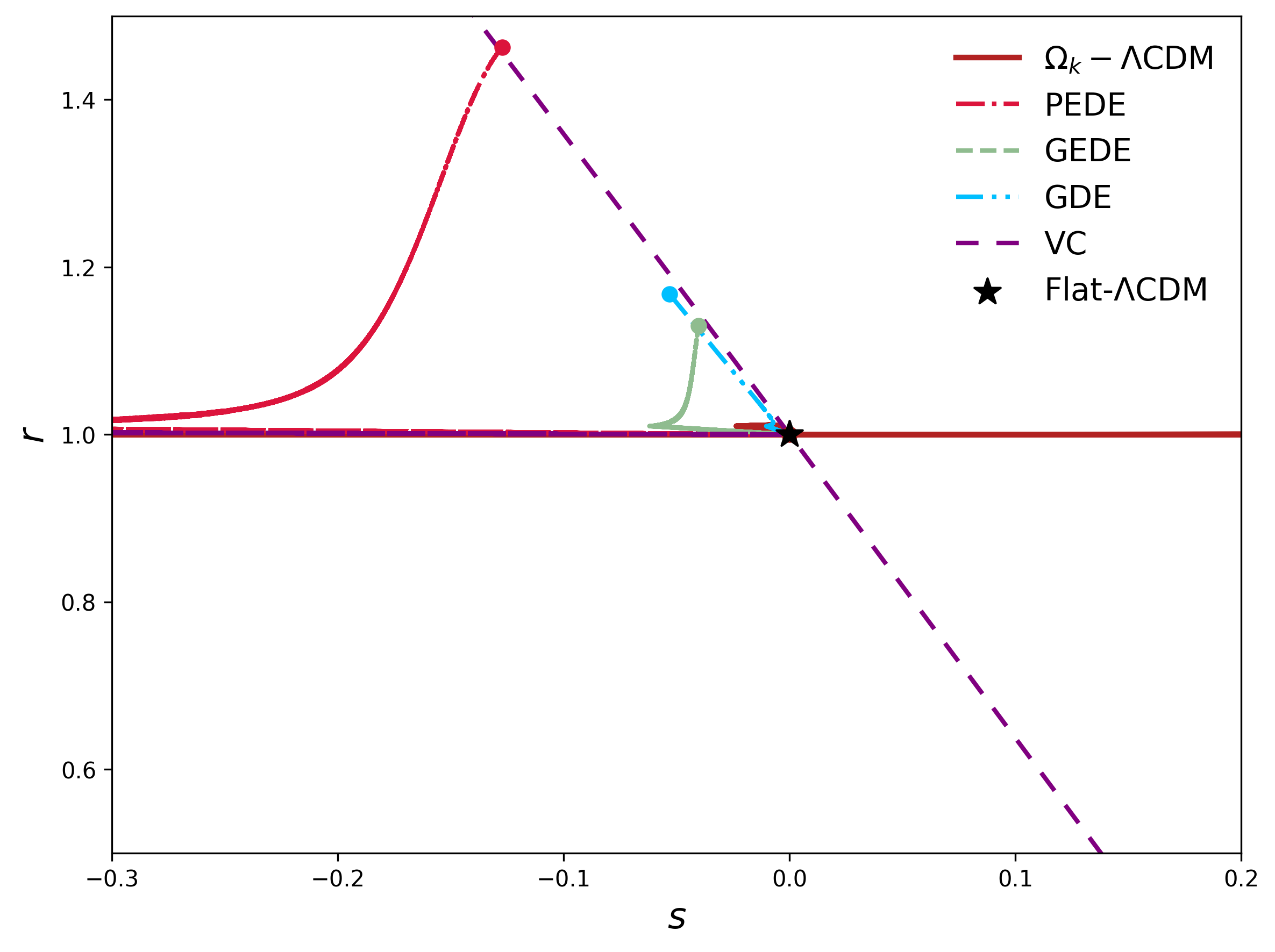}
   \caption{Left panel: Evolution of the $Om(z)$ diagnostic for the four cosmological models considered in this work, together with the non-flat $\Lambda$CDM model. Right panel: Statefinder trajectories for the same models, including both the non-flat and flat $\Lambda$CDM scenarios. The curves correspond to the median parameter values obtained from the combined dataset (CC+SNIa+BAO+CMB) for all models, except for the VC model, whose trajectory is derived from the CC+SNIa dataset only.}
   \label{fig:diagnostic}
\end{figure*}

To further characterize the cosmological models beyond the standard background observables, we consider the $Om(z)$ diagnostic and the Statefinder parameters. Unlike conventional parameter estimation, these diagnostics probe the dynamical evolution of the expansion history and provide a model-independent framework for distinguishing different dark-energy scenarios. In particular, the $Om(z)$ diagnostic is sensitive to departures from the $\Lambda$CDM model, while the Statefinder hierarchy offers a geometric characterization of the cosmic expansion through higher-order derivatives of the scale factor. Together, these complementary diagnostics provide additional discriminatory power and help identify subtle differences among models that may otherwise produce similar background expansion histories. The $Om(z)$ diagnostic is ruled through the formula \cite{Sahni:2002fz}
\begin{equation}
    Om(z)=\frac{\frac{H(z)}{H_0}-1}{(z+1)^3-1},
\end{equation}
where $Om(z)=\Omega_m$ for the $\Lambda$CDM case. Meanwhile, the statefinder is based on a study in the $\lbrace s,r\rbrace$-plane  \cite{Alam:2003sc,Sahni:2002fz} for the four models, including $\Omega_k$-$\Lambda$CDM. The main parameters for the statefinder are defined by the following geometric variables
\begin{eqnarray}
    &&r=j\equiv\frac{\ddot{a}}{aH^3},\nonumber\\&&
    s\equiv\frac{r-1}{3(q-1/2)},
\end{eqnarray}
where $r$ is the traditional jerk. Noticing that $(s,r)=(0,1)$ for the $\Lambda$CDM case (i.e. flat curvature). From Figs. \ref{fig:diagnostic} (left panel) it is possible to observe the behavior of $O_m(z)$ at different redshifts, in particular for $\Omega_k$-$\Lambda$CDM case at $z=0$ the value is the expected and does not exist evolution at least in the region $[0,3.5]$ meanwhile for the others models it is possible to observe a change during the last stages of the Universe evolution. Notice also the VC case where the reduction of $O_m(z)$ is extreme and caused by the jump observed in $q(z)$, for example.
From the statefinder analysis observed in \ref{fig:diagnostic} (right panel), the dynamics caused by dark energy in the four models (despite VC does not contains DE) are totally clear when we compare with $\Omega_k$-$\Lambda$CDM and $\Lambda$CDM. Again, notice the completely different behavior of VC, caused by the change of curvature in the last stages. This notorious behavior is, in essence, due to the VC model not exhibiting homogeneity symmetry.

\section{Conclusions and Discussions} \label{section:CD}

This paper is dedicated to comparing models of emergent dark energy, in particular those with a hyperbolic tangent or a similar function that acts only at late times in the Universe's evolution ($z\approx0$). The idea is to include the curvature term even in $\Lambda$CDM to allow a clear comparison, because the main feature of the VC model is the change in curvature. Moreover, we center and standardize our analysis using low-redshift datasets, in particular CC and SNIa together with BAO and CMB datasets, except for the VC model constrained only with low-redshift observations.
First, the unique model that exhibits a beginning of deceleration at $z=0$ is the VC, consistent with recent DESI observations \cite{desicollaboration2024desi}; the other emergent dark energy models do not decelerate at $z=0$. An important point of the VC model is that $H(z)=0$ at $z\to-1$ according to \cite{Esteban-Gutierrez:2024rpz}, which is also an important difference in comparison with the $\Lambda$CDM model.
Note also that PEDE, GEDE, and GDE exhibit a more aggressive acceleration that reaches values near $\sim-0.8$, according to the analysis of $w_{eff}$. In these last three cases, the age of the Universe is consistent with the $\Omega_k$-$\Lambda$CDM, but it is slightly lower or higher, which can have important repercussions when compared with the oldest stars in our Universe \cite{OldestStar}. Nevertheless, for the VC, the age of the Universe is larger than $\Omega_k$-$\Lambda$CDM, alleviating the tension with the older stars observed. For example, in \cite{OldestStar}, the VC universe is $ \sim2.4$ Gyrs older than the oldest star observed; meanwhile, for $\Omega_k$-$\Lambda$CDM, the Universe is $\sim0.74$ Gyr younger. Similarly, for \cite{deAndres:2024vmr}, VC universe is $\sim9.84$ Gyrs younger, meanwhile for $\Omega_k$-$\Lambda$CDM it is $\sim12.98$Gyrs younger. Thus, the VC model resolves or alleviates the tension with the oldest stars.

From Fig. \ref{fig:ode_comparison}, we observe a comparison only by the DE density parameter. In essence, the difference lies because the universe's acceleration in VC is caused by a subtle change in curvature, meanwhile in the other models the density parameters are associated with a DE, for example, a scalar field for PEDE and GEDE, and a sign change of $\Lambda$ in GDE. This behavior is also related to the observed deceleration in VC, while for the other models the acceleration is eternal. Moreover, in Fig. \ref{fig:eos_comparison}, we reconstruct different DE equations of state and compare them with the results obtained by DESI \cite{DESI:2025zgx}. Notoriously, all models always live in the phantom region, except VC, which lives in the quintessence region, eventually transiting to the phantom region, reaching a minimum, and starting its deceleration. GDE shows a singularity\footnote{A singularity in GDE does not cause problems in our constraints because the CC maximum redshift is $z<1.965$, meanwhile, for SNIa the maximum redshift is $z<2.26$. Additionally, in the numerical analysis, a singularity is observed in Fig. \ref{fig:eos_comparison}; however, it does not cause conflicts when compared with the other models, and the dynamics are clearly observed.} near $z=3.5$ previously reported in \cite{Akarsu:2019hmw}, having a tendency to a cosmological constant for $z\leq0$. Due to the dispersion in the DESI observations, it is complicated to decide on the best model. However, notice that the CPL model remains the best candidate for DE (light green line). 

Statistical analysis based on the AIC and BIC criteria (Table~\ref{tab:model_selection}) indicates that the standard $\Omega_k$-$\Lambda$CDM model remains the preferred cosmological scenario when both goodness of fit and model complexity are considered. In particular, $\Omega_k$-$\Lambda$CDM consistently yields the lowest BIC values across all datasets, suggesting that current observations do not provide sufficient evidence to justify more complex cosmological extensions.

Among alternative models, PEDE and GEDE remain observationally competitive, especially for the CC data set, where PEDE shows marginally lower AIC and BIC values than $\Omega_k$-$\Lambda$CDM, although these differences are not statistically significant. For the SNIa and combined CC+SNIa+BAO+CMB datasets, PEDE, GEDE, and GDE exhibit moderate $\Delta$AIC values, indicating that they cannot be conclusively ruled out. However, the stronger penalization imposed by BIC disfavors the more parameter-rich models, particularly GDE.

The VC model is clearly the least favored scenario, with moderate values of $\Delta$AIC and large values of $\Delta$BIC for all datasets, indicating strong to very strong evidence against it. Overall, while some phenomenological dark energy models remain viable alternatives, current data continue to favor the simpler $\Omega_k$-$\Lambda$CDM framework. However, we need to take into account that the VC model breaks the symmetry of homogeneity and, interestingly, causes an accelerated phase; meanwhile, for the others, a DE component is necessary. Additionally, this lack of symmetry generates more free parameters. Thus, a conclusive verdict about the VC model needs to take into account the fact that the number of symmetries is not the same. Therefore, it is more precise to compare models with the same number of symmetries in order to have the same penalization for all the models during the comparison.

The correlation analysis reveals that the parameter degeneracies are strongly model dependent and evolve significantly with the inclusion of complementary cosmological observations. While low-redshift data (CC+SNIa) are sufficient to constrain the background expansion, they leave several parameters weakly constrained, particularly those associated with the extended dark-energy or curvature sectors. The addition of BAO and CMB observations substantially reduces these degeneracies by tightly constraining the matter and baryon densities through the acoustic scale, although new correlations naturally emerge among the standard cosmological parameters and the additional model degrees of freedom. In contrast, the VC model exhibits its dominant degeneracies within the intrinsic curvature sector, with only weak correlations with the standard cosmological parameters.

Overall, these results highlight both the complementarity and the limitations of background cosmological probes. Since several extended cosmological models can produce very similar expansion histories, background observables alone cannot completely break the remaining parameter degeneracies. Future analyses based on the full CMB likelihood, together with perturbation-level observables such as weak gravitational lensing, redshift-space distortions, galaxy clustering, and growth-rate measurements, are expected to provide significantly tighter constraints and a more robust assessment of the viability of these cosmological models.

In summary, the PEDE model is one of the most favored emergent DE models, sharing the same degrees of freedom as $\Omega_k$-$\Lambda$CDM and also offering the advantage that DE can be understood as a scalar field. However, GEDE and GDE can not be totally discarded, even the VC model, which is the most disfavored, has the key characteristic of allowing a decelerated phase to occur close to redshifts $z=0$, which is compatible with recent DESI results \cite{desicollaboration2024desi}. Additionally, it is the unique model that allows $\Omega_{k}^1=-0.003^{+0.006}_{-0.004}$, fitting better with Planck results ($\Omega_k=0.001\pm0.002$) \cite{Planck:2018}. Moreover, the $Om(z)$ diagnostic and statefinder analysis show, first, the change in matter evolution, with different results compared with $\Omega_k$-$\Lambda$CDM, particularly for VC, which shows an interesting tendency to zero in the value of $Om(z)$ which corresponds to the bump observed in the reconstruction of $q(z)$ and $w_{eff}(z)$. For the statefinder, it is notorious that the dynamics associated with the four models differ when compared with the standard paradigm (flat $\Lambda$CDM a point and $\Omega_k$-$\Lambda$CDM a horizontal line in $r$). Remarkably, the analysis gives us an idea of how the dark energy evolves in the different models and its fingerprints among them. Future studies that generate a final decision require a deep study that involves structure growth, weak lensing, among others. It is also necessary in future studies to explore whether the models maintain positivity of the effective energy density, stability against perturbations, or absence of ghost instabilities. However, these will be presented elsewhere. Thus, we need more precise observations to elucidate whether the Universe is decelerating in its final phase. This result could help us finally decide which model is the most competitive and best fits the observations.

\begin{acknowledgments}
We thank anonymous referee for thoughtful remarks and suggestions. A.H.A. thanks the support from Luis Aguilar, 
Alejandro de Le\'on, Carlos Flores, and Jair Garc\'ia of the Laboratorio 
Nacional de Visualizaci\'on Cient\'ifica Avanzada. M.A.G.-A. acknowledges support from catedra Marcos Moshinsky, SECIHTI for the support with the National Research System (SNII) grant and the project 0056 from Universidad Iberoamericana: Nuestro Universo en Aceleraci\'on, energ\'ia oscura o modificaciones a la relatividad general. V.M. acknowledges partial support from Centro de Astrof\'{\i}sica de Valpara\'{\i}so CIDI 21. The numerical analysis was also carried out by {\it Numerical Integration for Cosmological Theory and Experiments in High-energy Astrophysics} (Nicte Ha) cluster at IBERO University, acquired through c\'atedra MM support.  A.H.A and M.A.G.-A acknowledge partial support from project ANID Vinculaci\'on Internacional FOVI220144.
\end{acknowledgments}

\bibliography{main}

\end{document}